\documentclass[12pt]{article}
\pdfoutput=1
\usepackage{amsmath,amsfonts,amsthm,amssymb,slashed,graphicx}
\usepackage{xcolor}
\usepackage{graphicx}
\usepackage{epstopdf}
\usepackage{array,booktabs}
\usepackage{hyperref,bookmark}
\usepackage{tabulary}
\usepackage{multirow}
\usepackage{array,arydshln}
\usepackage{bbm} 
\usepackage{cite}
\usepackage{mathtools} 

\hypersetup{linktoc = all}  
\hypersetup{pdfborderstyle={/S/U/W 0.5}}

\numberwithin{equation}{section}
\def\l{\lambda}
\def\n{\nu}

\def\vare{\varepsilon}

\newcommand{\beq}{\begin{equation}}
\newcommand{\eeq}{\end{equation}}

\newcommand{\address}[1]{\vbox{\center\em#1}}
\renewcommand{\title}[1]{\vbox{\center\LARGE{#1}}\vspace{5mm}}

\let\Im\undefined
\DeclareMathOperator{\Im}{Im}

\DeclareMathOperator{\Tr}{Tr}
\DeclareMathOperator{\Pexp}{Pexp}
\newcommand{\res}{\mathop{\operatorname{res}}\limits}
\newcommand{\rf}[1]{\eqref{#1}}
\declareslashed{}{/}{-.08}{0}{V} 

\makeatletter

\newcommand*{\letterdef@}{}
\newcommand*{\letterdef}[3]{%
  \def\letterdef@##1{\expandafter\newcommand\csname #1\endcsname{#2{##1}}}%
  \@tfor\@tempa :=#3\do{\expandafter\letterdef@\expandafter{\@tempa}}}
\makeatother
\letterdef{b#1} {\mathbb} {CHPRZ}    
\letterdef{c#1} {\mathcal}{ABCDEFGHIJKLMNOPQRSTUVWXYZ} 
\letterdef{rm#1}{\mathrm} {dDmM} 
\letterdef{#1frak}{\mathfrak}{t}

\newcommand{\antiNF}{\widetilde{N}_F} 

\def\changed#1#2\to#3\done{#3}

\begin{document}
\bibliographystyle{utphys}


\begin{titlepage}
\setcounter{page}{0}
%
%
%
%
%

 \hfill {\tt DAMTP-2012-31}\\
  \vspace{10mm}

\centerline{\LARGE{Exact Results in \texorpdfstring{$D=2$}{D=2} Supersymmetric Gauge Theories}}
\hypersetup{pdftitle={Exact Results in D=2 Supersymmetric Gauge Theories}}
\vspace{10mm}

\begin{center}
\renewcommand{\thefootnote}{$\alph{footnote}$}
Nima Doroud,\footnote{\href{mailto:ndoroud@perimeterinstitute.ca}{\tt ndoroud@perimeterinstitute.ca}} Jaume Gomis,\footnote{\href{mailto:jgomis@perimeterinstitute.ca}
{\tt jgomis@perimeterinstitute.ca}} Bruno Le Floch,\footnote{\href{mailto:bruno@le-floch.fr}
{\tt bruno@le-floch.fr}} and Sungjay Lee\,\footnote{\href{mailto:S.Lee@damtp.cam.ac.uk}
{\tt S.Lee@damtp.cam.ac.uk}}

\vspace{10mm}

\address{${}^{a,b,c}$Perimeter Institute for Theoretical Physics,\\
Waterloo, Ontario, N2L 2Y5, Canada}
\address{${}^{a}$Department of Physics, University of Waterloo,\\
Waterloo, Ontario N2L 3G1, Canada}
\address{${}^{c}$\'Ecole Normale Sup\'erieure,\\
Paris, 75005, France}
\address{${}^{d}$DAMTP, Centre for Mathematical Sciences,\\ Cambridge University,
Cambridge CB3 0WA, United Kingdom}

\renewcommand{\thefootnote}{\arabic{footnote}}
\setcounter{footnote}{0}

\end{center}


\abstract{
\medskip\medskip
\normalsize{
\noindent
We compute exactly the partition function of two dimensional $\cN=(2,2)$ gauge theories on $S^2$ and show that it admits two dual  descriptions:  either as an integral over  the Coulomb branch  or  as a sum over
vortex and anti-vortex excitations on the Higgs branches of the theory. We further demonstrate that
correlation functions in two dimensional  Liouville/Toda CFT compute the $S^2$ partition function for a class of $\cN=(2,2)$ gauge theories, thereby uncovering novel modular properties in two dimensional gauge theories.
Some of these gauge theories flow in the infrared to Calabi-Yau sigma models -- such as the conifold -- and the topology changing flop transition  is realized as crossing symmetry in Liouville/Toda CFT.
   Evidence for Seiberg duality in two dimensions is exhibited by demonstrating that the partition function of conjectured Seiberg dual pairs are the same.
}}

\vspace{10mm}

\noindent
\vfill

\end{titlepage}

\tableofcontents


\newpage

\section{Introduction}
\label{sec:intro}

It has long been recognized that many of the dynamical and quantum properties  of four dimensional gauge theories are mirrored in two dimensional quantum field theories. This includes -- among the wealth of phenomena that a
four dimensional gauge theory can exhibit -- the remarkable and not yet completely understood physics of confinement and dynamical generation of a mass gap.
Instantons, which mediate non-perturbative effects in four dimensional gauge theories, are also present in two dimensional field theories, and play a central role in determining the quantum properties of these theories.
While the dynamics of two dimensional gauge theories is tamer than in four dimensions, few exact results for correlation functions are available. In most examples, such computations  heavily  rely on  integrability.
Furthermore, given that two dimensional theories share many of the beautiful phenomena present in four dimensions, it is a desirable goal to attain exact results in two dimensional quantum field theories.

In this paper we obtain exact results in two dimensional $\cN=(2,2)$ supersymmetric gauge theories on $S^2$.  These results are obtained using the powerful machinery of supersymmetric localization
\cite{Witten:1988ze, Witten:1991zz, Pestun:2007rz}. We uncover that the partition function of these theories admit two seemingly different representations.\footnote{This can be enriched with the insertion of
supersymmetric Wilson loop operators.}  In one, the partition function is written as an integral (and discrete sum) over vector multiplet field configurations.  This yields the Coulomb branch representation of the
partition function
\[
 Z_{\text{Coulomb}} (m, \tau) = \sum_{B} \int_{\tfrak} \rmd a\,  Z_{\text{cl}}(a, B, \tau)\, Z_{\text{one-loop}}(a, B, m)\,.
\]
$B$ is the quantized flux on   $S^2$,  $a$ the Coulomb branch parameter, $m$ denotes the masses of the matter fields and $\tau$ are the complexified gauge theory parameters
\[
\tau=\frac{\vartheta}{2\pi}+i \xi\,,
\]
where $\xi$ and $\vartheta$ are the Fayet-Iliopoulos (FI) parameter and topological angle associated to each $U(1)$ factor in the gauge group. Expressions for  $ Z_{\text{cl}}(a, B, \tau)$ and  $Z_{\text{one-loop}}(a, B, m)$
are given in section~\ref{sec:classical}.

In the other representation, the path integral is given as a discrete sum over Higgs branches of the product of the vortex partition function \cite{Shadchin:2006yz}  at the north pole and the anti-vortex partition function at
the south pole.  This gives the Higgs branch representation of the partition function
\[
Z_{\text{Higgs}} (m, \tau) = \hskip-12pt\sum_{v \in \text{Higgs vacua }}  \hskip-12pt Z_{\text{cl}}(v, 0, \tau)\, \res_{a=v}\left[ Z_{\text{one-loop}}(a, 0, m)\right]
Z_{\text{vortex}} (v, m, e^{2\pi i\tau}) Z_{\text{anti-vortex}}(v, m, e^{-2\pi i\bar \tau})\,.
\]
In this formula the residue of the pole of $Z_{\text{one-loop}}(a, 0, m)$ at the location of each Higgs branch must be taken.\footnote{A Higgs branch is a solution to the equation $a+m=0$.}  Equivalently, this expression can
be written in a holomorphically factorized form as a sum of the ``norm'' of the vortex partition function
\[
Z_{\text{Higgs}} (m, \tau) = \hskip-12pt \sum_{v \in \text{Higgs vacua}} \hskip-12pt Z_{\text{cl}}(v, 0, \tau)\, \res_{a=v}\left[ Z_{\text{one-loop}}(a, 0, m)\right] \left|Z_{\text{vortex}} (v, m, e^{2\pi i\tau})\right|^2\,.
\]

 \begin{figure}[t]
\begin{center}
 \includegraphics[scale=0.7]{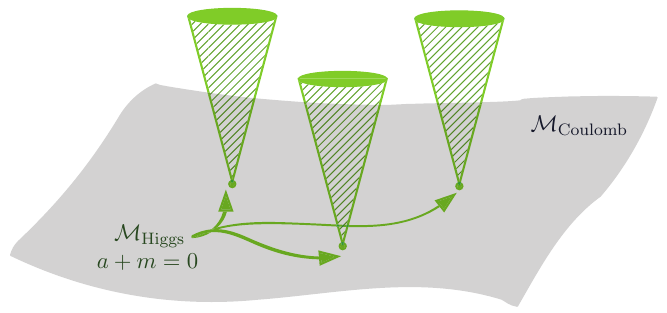}
 \end{center}
 \caption{Higgs vacua. Vortices and anti-vortices on these vacua contribute to $Z_{\text{Higgs}}(m, \tau)$}
\end{figure}

Despite  that the expressions for the Coulomb and Higgs branch representations are rather distinct and involve different degrees of freedom, we show that the two yield identical, dual representations of the partition function
of $\cN=(2,2)$ gauge theories on $S^2$
\[
Z=Z_{\text{Coulomb}}=Z_{\text{Higgs}}\,.
\]
We have explicitly shown this equivalence for SQCD, with $U(N)$ gauge group and $N_F$ fundamental and $\antiNF$ anti-fundamental chiral multiplets. The factorization of the Coulomb branch integral is akin to the one found by
Pasquetti \cite{Pasquetti:2011fj} and Krattenthaler et al. \cite{Krattenthaler:2011da} in evaluating the partition function of three dimensional $\cN=2$ abelian gauge theories on the squashed $S^3$ \cite{Hama:2011ea} and $S^1\times S^2$.\footnote{Other related works on localization include
\cite{Kim:2009wb,Kapustin:2009kz,Jafferis:2010un,Hama:2010av,Imamura:2011su}.}

The fact that a correlation function in a supersymmetric gauge theory may admit multiple representations can be understood to be a consequence of the different choices of supercharge and/or deformation terms available when
performing supersymmetric localization.  Different choices may lead to integration over different supersymmetric configurations,  but the localization argument guarantees that all (reasonable) choices must ultimately yield
the same correlation function.\footnote{In particular we obtain the Coulomb branch representation of the partition function using two different choices of supercharge.} See section~\ref{sec:conclusion} for a more detailed
discussion. Our choice of localization supercharge has the elegant feature of
giving rise to supersymmetry equations which interpolate between vortex equations at the north pole and anti-vortex equations at the south pole while also allowing for configurations on the Coulomb branch.

\begin{figure}[ht]
  \begin{center}
    \includegraphics[scale=1]{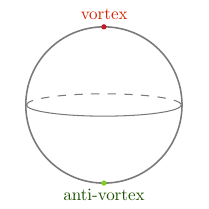}
  \end{center}
  \caption{Vortex and anti-vortex configurations in the Higgs branch}
\end{figure}

We demonstrate that the partition function of certain two dimensional $\cN=(2,2)$ gauge theories on $S^2$ admits a dual description in terms of correlation functions in two dimensional Liouville/Toda CFT.  This is akin to
the AGT correspondence \cite{Alday:2009aq} between the partition function of four dimensional $\cN=2$ gauge theories on $S^4$ and correlators in these two dimensional CFTs.  The key difference is that the correlators in
Liouville/Toda CFT that capture the two dimensional gauge theory partition function on $S^2$ involve the insertion of degenerate vertex operators of the Virasoro or $W$-algebra at suitable punctures on the Riemann surface.
These insertions have the sought after property of restricting the sum over intermediate states to a discrete sum of conformal blocks, which precisely capture the sum over Higgs vacua in the Higgs branch representation of
the partition function.  Pleasingly, $Z_{\text{Higgs}}$  exactly reproduces the sum over conformal blocks with the precise modular invariant
Liouville/Toda measure by summing over vortices and anti-vortices over all Higgs vacua.

The simplest instance of this correspondence is SQED, described by a $U(1)$ vector multiplet and $N_F$ electron and $N_F$ positron chiral multiplets.  The partition function of SQED corresponds to the $A_{N_F-1}$ Toda CFT
on the four-punctured sphere with the insertion of two non-degenerate, a semi-degenerate and a fully degenerate puncture:

\vskip+10pt

\begin{figure}[ht]
\begin{center}
 \includegraphics[scale=0.7]{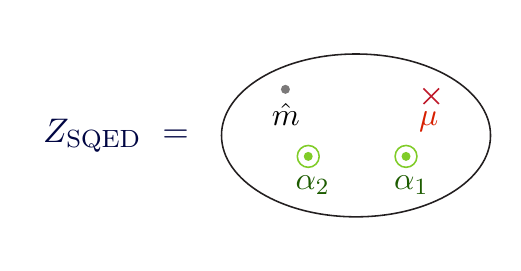}
 \end{center}
 \caption{SQED partition function as Toda CFT correlator}
\end{figure}

\noindent
The fact that two dimensional $\cN=(2,2)$ gauge theories on $S^2$ admit a Liouville/Toda CFT description with degenerate fields is consistent with the observation that certain half-BPS surface operators in four dimensional
$\cN=2$ gauge theories on $S^4$ are realized by the insertion of a degenerate field \cite{Alday:2009fs}.

The correspondence we establish with Liouville/Toda CFT implies that two dimensional $\cN=(2,2)$ gauge theories enjoy rather interesting modular properties with respect to the complexified gauge theory parameters $\tau$.
This is a direct consequence of modular invariance, which implies that CFT correlators are independent of the choice of factorization channel (or pants decomposition) used to represent a correlator as a sum over
intermediate states. The moduli of the punctured Riemann surface on which modular duality acts correspond to the vortex fugacity parameters
\[
z=e^{2\pi i \tau}\,.
\]
It is rather interesting that the partition function of two dimensional $\cN=(2,2)$ gauge theories on $S^2$ assembles into a modular invariant object.

Another important motivation to study two dimensional $\cN=(2,2)$ gauge theories is string theory.  As shown in \cite{Witten:1993yc}, the Higgs branch of such a gauge theory flows in the infrared to a two dimensional
$\cN=(2,2)$ supersymmetric non-linear sigma model with a K\"ahler target space.  Moreover, with a suitable choice of matter content and gauge group, the gauge theory flows to an $\cN=(2,2)$ superconformal field theory,
which provides the worldsheet description of string theory on a Calabi-Yau manifold.  One can hope that the exact formulae for the partition function of these  gauge theories will provide a novel way to compute worldsheet
instantons in the corresponding Calabi-Yau manifolds, as well as shed new light  into the dynamics of these phenomenologically appealing string theory backgrounds.

The ultraviolet gauge theory description of these string theory backgrounds provides  a qualitative characterization of the ``phase'' structure as the  K\"ahler moduli of the Calabi-Yau manifold are changed by studying the
gauge dynamics as a function of the complexified gauge theory parameters $\tau$ \cite{Witten:1993yc}. An interesting topology changing transition -- the so called flop transition -- occurs in some models as the sign of the
FI parameter is reversed $\xi\rightarrow -\xi$. The string dynamics in the two phases connected by a flop transition are expected to be related by analytic continuation in $\tau$. Our exact results for the partition function
of $\cN=(2,2)$ SQED -- which includes the conifold for $N_F=2$ and higher dimensional Calabi-Yau manifolds for  $N_F>2$ -- demonstrate that the results for $\xi>0$ and $\xi<0$ are indeed related by analytic continuation.
Given the representation of the partition function of SQED in terms of a Toda CFT correlator on the four-punctured sphere, the analytic continuation describing the flop transition admits an elegant realization as
crossing symmetry in Toda CFT
\[
\text {flop transition}\longleftrightarrow \text{crossing symmetry}\,.
\]
Furthermore, our exact results demonstrate that the geometric singularity as we move from $\xi>0$ to $\xi<0$ across the singular point $\xi=0$ can be avoided by turning on a nonzero topological angle $\vartheta$, as
anticipated in \cite{Witten:1993yc,Aspinwall:1993yb}.

Our findings are used to provide quantitative evidence for Seiberg duality \cite{Seiberg:1994pq} in two dimensions by comparing the partition functions of putative dual theories in various limits and finding exact agreement.
Seiberg  duality in two dimensional $\cN=(2,2)$ gauge theories \cite{Hori:2006dk} relates theories with $N_F>N$ fundamental chiral multiplets, trivial superpotential and gauge groups
\[
SU(N)\longleftrightarrow SU(N_F-N)\,.
\]
The conjectured duality was put forward in \cite{Hori:2006dk} to give a physical realization of R{\o}dland's conjecture stating that two Calabi-Yau manifolds appear as distinct large volume limits of the same
K\"ahler moduli space.  Our results, therefore, provide further evidence for this conjecture.

The plan of the rest of the paper is as follows.
In section~\ref{sec:gauge} we explicitly write down for gauge theories on $S^2$ the $\cN=(2,2)$ supersymmetry transformations of the vector and chiral multiplet fields and the associated supersymmetric action.
In section~\ref{sec:local} we specify a particular supercharge with which we perform the localization computation.  We derive the partial differential equations that determine the space of supersymmetric field configurations
corresponding to our choice of supercharge and show that the system of equations we get smoothly interpolates between the vortex equations at the north pole and the anti-vortex equations at the south pole.  A vanishing theorem
finding the most general smooth, supersymmetric solutions to our system of supersymmetry equations is proven. We find that smooth solutions are parametrized by vector multiplet fields and correspond to Coulomb phase
configurations, while singular localized vortices and anti-vortices, which exist in the Higgs phase, may appear at the north and south poles of the $S^2$.
In section~\ref{sec:classical} we localize the path integral by choosing a specific deformation term and show that only Coulomb branch configurations can contribute if we consider the saddle point equations of the combined
action in the limit that the coefficient of the deformation term goes to infinity.  This yields the Coulomb branch representation of the partition function.  Quite remarkably, the integral and sum over the Coulomb branch
configurations can be carried out for arbitrary choices of gauge group $G$ and matter representation.  The resulting expression can be written as a finite sum of the product of a function with its complex conjugate.
We identify this expression as the sum over Higgs vacua of the product of the vortex partition function at the north pole with the anti-vortex partition function at the south pole.
In section~\ref{sec:instanton} we argue, by first looking at the saddle point equations for a different deformation term, that the Coulomb branch configurations are lifted and that
vortex and anti-vortex configurations at the poles are the true saddle points of the path integral in this other limit.  This yields the Higgs branch representation of the partition function.  This way of computing the
path integral gives a first principles derivation of the result obtained by brute force evaluation of the Coulomb branch representation of the partition function.
The identification of the partition function of certain two dimensional $\cN=(2,2)$ gauge theories with Liouville/Toda correlation functions is uncovered in section~\ref{sec:Toda}, and some of their consequences explored.
In section~\ref{sec:Seiberg} we provide quantitative evidence for Seiberg duality in two dimensions by matching the partition function of Seiberg dual pairs in various limits.
We conclude in section~\ref{sec:conclusion} with a  discussion of our findings and future directions. The appendices contain some detailed computations used in the bulk of the paper.

 Note added:  While this work was being completed, we became aware of related
work \cite{Benini:2012ui}, which has some overlap with this paper.

\section{Two Dimensional \texorpdfstring{$\cN=(2,2)$}{N=(2,2)} Gauge Theories on \texorpdfstring{$ S^2 $}{S\texttwosuperior}}
\label{sec:gauge}

In this section we explicitly construct the Lagrangian of $\cN=(2,2)$ supersymmetric gauge theories on $S^2$.
The basic multiplets of two dimensional $\cN=(2,2)$ supersymmetry are the vector multiplet and the chiral multiplet, which arise by dimensional reduction to two dimensions of the familiar four dimensional $\cN=1$
supersymmetry multiplets.
The field content is therefore
\begin{equation}
\begin{aligned}
\text{vector multiplet: } & (A_i,\sigma_1,\sigma_2,\lambda, \bar \lambda, \rmD)\\
\text{chiral multiplet: } & (\phi ,\bar \phi ,\psi, \bar \psi,F,\bar F)\,.
\end{aligned}
\end{equation}
The fields $(\lambda, \bar\lambda, \psi, \bar \psi)$ are two component complex Dirac spinors,\footnote{Our conventions for spinors are listed in appendix~\ref{app:conven}.}
 $(\phi ,\bar \phi,F,\bar F)$ are complex scalar fields while $(\sigma_1, \sigma_2, \rmD)$ are real scalar fields.\footnote{The reality of the auxiliary field $\rmD$ is altered when coupled with matter fields. }
The fields in the vector multiplet transform in the adjoint representation of the gauge group $G$ while
the chiral multiplet fields transform in a representation $\bf R$ of $G$. The field content of an arbitrary $\cN=(2,2)$ supersymmetric gauge theory admitting a Lagrangian description is captured by these multiplets by
letting $G$ be a product gauge group and $\bf R$ a reducible representation.

While it is well known how to construct the Lagrangian of $\cN=(2,2)$ supersymmetric gauge theories in $\bR^2$ (i.e.\@ flat space), constructing supersymmetric theories on $S^2$ requires some thought, as $S^2$ does not
admit covariantly constant spinors. Indeed, we must first characterize the $\cN=(2,2)$ supersymmetry algebra on $S^2$.
This is the subalgebra of the two dimensional $\cN=(2,2)$ superconformal algebra on $S^2$ that generates the isometries of $S^2$, but none of the  conformal transformations of $S^2$.
The $\cN=(2,2)$ supersymmetry algebra on $S^2$ thus defined obeys the (anti)commutation relations of the $SU(2|1)$ superalgebra\footnote{See appendix~\ref{app:sca} for details.}
\begin{equation}
  \begin{aligned}
    [J_{m},J_{n}] &= i \epsilon_{mnp} J_p
    \qquad
    [J_{m},Q_\alpha] = -\frac{1}{2} \gamma_{m}{}^{\beta}{}_{\alpha} Q_\beta
    \qquad
    [J_{m},S_ \alpha] =-\frac{1}{2} \gamma_{m}{}^{\beta}{}_{\alpha} S_\beta
    \\
    \{S_ \alpha,Q_\beta\} &= \gamma^{m}_{\alpha \beta} J_{m} - \frac{1}{2} C_ {\alpha\beta}R
    \qquad
    [R,Q_\alpha] = -Q_\alpha
    \qquad
    [R,S_\alpha] =S_\alpha\,.
  \end{aligned}
  \label{SU(2|1)}
\end{equation}
 \smallskip
The supercharges $Q_\alpha$ and $S_\alpha$ are two dimensional Dirac spinors generating the supersymmetry transformations, $J_m$ are the $SU(2)$ charges generating the isometries of $S^2$ while $R$ is a
$U(1)$ $R$-symmetry charge. This supersymmetry algebra is the $S^2$  counterpart of the $\cN=(2,2)$ super-Poincar\'e algebra in flat space.

Constructing a supersymmetric Lagrangian on $S^2$ requires finding  supersymmetry transformations on the vector and chiral multiplet fields that represent the $SU(2|1)$ algebra. We construct these by restricting
the  $\cN=(2,2)$ superconformal transformations to those corresponding to the $SU(2|1)$ subalgebra. The $\cN=(2,2)$ superconformal transformations on the fields are easily obtained by combining the
$\cN=(2,2)$ super-Poincar\'e  transformations in flat space (with the flat metric  replaced by an arbitrary metric),  with additional terms  that are uniquely fixed by demanding that the supersymmetry transformations
are covariant under Weyl transformations.\footnotemark[7]
Given the $SU(2|1)$ supersymmetry transformations on the vector and chiral multiplet fields constructed this way and shown below, it is straightforward to construct the corresponding $SU(2|1)$ invariant Lagrangian.
The supersymmetry transformations and action may equivalently be obtained by ``twisted'' dimensional reduction from  three dimensional $\cN=2$ gauge theories on $S^1\times S^2$, considered in \cite{Imamura:2011su}.

Without further ado, we write down the most general renormalizable  $\cN=(2,2)$ supersymmetric action of an arbitrary gauge theory on $S^2$
\begin{equation}
S=S_{\text{v.m.}}+S_{\text{top}}+S_{\text{FI}}+S_{\text{c.m.}}+S_{\text{mass}}+S_{\text{W}}\,.
\label{action}
\end{equation}
The vector multiplet action is given by
\begin{equation}
\label{vm_action}
S_{\text{v.m.}} = \frac{1}{2g^{2}} \int \rmd^{2}x\sqrt{h}\ \Tr
\left\{
V_{i}V^{i}+V_{3}V^{3}+\rmD^{2}+i\lambda\left(\slashed{D}\bar{\lambda}-\left[\sigma_{1},\bar{\lambda}\right]-i\left[\sigma_{2},\gamma^{\hat 3}\bar{\lambda}\right]\right)
\right\}\,,
\end{equation}
where
\begin{equation}
\begin{aligned}
V^{i} &= \varepsilon^{ij}D_{j}\sigma_{2}+D^{i}\sigma_{1}\,,
\\
V^{3} &= \frac{1}{2}\varepsilon^{ij}F_{ij}+i\left[\sigma_{1},\sigma_{2}\right]+\frac{1}{r}\sigma_{1}\,.
\label{Bogo}
\end{aligned}
\end{equation}
The bosonic part of the action can also be written as
\begin{equation}
 \frac{1}{2g^{2}} \int \rmd^{2}x\sqrt{h}\ \Tr
\left\{\left( F_{\hat{1}\hat{2}} +\frac{1}{r}\sigma_{1}\right)^2+\left(D_i\sigma_1\right)^2+\left(D_i\sigma_2\right)^2-[\sigma_1,\sigma_2]^2+\rmD^{2}\right\}\,.
\label{vm_actionb}
\end{equation}
In the vector multiplet action $g$ denotes the super-renormalizable gauge coupling\footnote{For a  product gauge group, there is an independent gauge coupling for each factor in the gauge group.}, $h$ is the
round metric on $S^2$ and $r$ is its radius.

For each  $U(1)$ factor in $G$, the gauge field action in two dimensions can be enriched by the addition of the topological term
\begin{equation}
S_{\text{top}}=-i\frac{\vartheta}{2\pi} \int \Tr F\,,
\end{equation}
and
of a supersymmetric Fayet-Iliopoulos (FI) $\rmD$-term on $S^2$
\begin{equation}
S_{\text{FI}}=-i\hskip+1pt \xi \int \rmd^{2}x\sqrt{h}\  \Tr \left( \rmD -\frac{\sigma_{2}}{r} \right)\,.
\end{equation}
The couplings $\vartheta$ and $\xi$ are classically marginal, and can be  combined into a complex gauge coupling
\begin{equation}
\tau=\frac{\vartheta}{2\pi}+i\xi
\label{taucoup}
\end{equation}
for each $U(1)$ factor in the gauge group. Quantum mechanically, the coupling $\tau$ depends on the energy scale, and can be traded with the dynamically generated, renormalization group invariant
scale $\Lambda$.\footnote{The dynamical scale is given by $\Lambda^{b_0}=\mu^{b_0} e^{2\pi i \tau(\mu)}$, where $\beta(\xi)\equiv\frac{b_0}{2\pi}$ and $\mu$ is the floating scale.} We will return to this dynamical
transmutation in section~\ref{sec:classical}.

The action for the chiral multiplet coupled to the vector multiplet  is\footnote{The representation matrices of $G$ in the representation $\bf R$, which we do not write explicitly to avoid clutter, intertwine the
vector multiplet
and chiral multiplet fields in the usual way.}
\begin{equation}
\begin{aligned}
S_{\text{c.m.}}=\int \rmd^{2}x\sqrt{h}\
\bigg\{\bar{\phi}&\left( -D_{i}^{2}+\sigma_{1}^{2}+\sigma_{2}^{2} + i\rmD + i\frac{q-1}{r}\sigma_{2}-\frac{q^{2}-2q}{4r^{2}}\right)\phi+
\bar{F}F\\ &-i\bar{\psi}\left(\slashed{D}-\sigma_{1}-i\sigma_{2}\gamma^{\hat 3}+\frac{q}{2r}\gamma^{\hat 3}\right)\psi
+i\bar{\psi}\lambda\phi-i\bar{\phi}\bar{\lambda}\psi
\bigg\}\,.
\end{aligned}
\label{chiralmult}
\end{equation}
Here $q$ denotes the $U(1)$ $R$-charge of the chiral multiplet, which takes the value $q=0$ for  the  canonical chiral multiplet.\footnote{$q$ also determines the Weyl weight of the fields in the chiral multiplet.
The Weyl weight of a field can be read from the commutator of two superconformal transformations (see appendix~\ref{app:sca}), which represents the two dimensional $\cN=(2,2)$ superconformal algebra on the fields.}
In a theory with flavour symmetry $G_F$, the $U(1)$ $R$-charges take values in the Cartan subalgebra of $G_F$ (see discussion below).

In two dimensions, it is possible to turn on in a  supersymmetric  way twisted masses for the chiral multiplet. These supersymmetric mass terms are obtained by first weakly gauging the flavour symmetry group $G_F$
acting on the theory, coupling the matter fields to a vector multiplet for $G_F$, and then turning on a supersymmetric background expectation value for the fields in that vector multiplet. For $\cN=(2,2)$ gauge
theories on $S^2$, unbroken $SU(2|1)$ supersymmetry (see equations \rf{dhlambda} and \rf{dhblambda}) implies that the mass parameters are given by  a constant background expectation value for the scalar field
$\sigma_2$ in the vector multiplet for $G_F$. This can be taken in the Cartan subalgebra of the flavour symmetry group $G_F$.
Therefore, the supersymmetric twisted mass terms on $S^2$ are obtained by substituting
\begin{equation}
\sigma_2\rightarrow \sigma_2+m
\label{shiftt}
\end{equation}
in \rf{chiralmult}, with $m$ in the Cartan subalgebra of $G_F$
\begin{equation}
S_{\text{mass}}=\int \rmd^{2}x\sqrt{h}\ \bigg\{\bar\phi \left(m^2+2m \sigma_2 +i\frac{q-1}{r}m\right)\phi-\bar{\psi}m \gamma^{\hat 3} \psi\bigg\}\,.
\label{mass}
\end{equation}
Likewise, the $U(1)$ $R$-charge parameters $q$ introduced  in \rf{chiralmult} can be obtained by turning on an  imaginary expectation value for the scalar field $\sigma_2$ in the vector multiplet for $G_F$.
The corresponding supersymmetric terms in the action
are obtained by shifting the  action   in \rf{chiralmult} for  $q=0$ by
\begin{equation}
\sigma_2\rightarrow \sigma_2+\frac{i}{2r} q\,.
\end{equation}
 The flavour symmetry $G_F$ is determined by the representation $\bf R$ under which the chiral multiplet transforms and by the choice of superpotential, as this can break the group of transformations rotating the
 chiral multiplets down to the actual $G_F$ symmetry of the theory.
 If $\bf R$ contains $N_F$ copies of an irreducible representation $\mathbf{r}$ and  the theory has a trivial superpotential, then the theory has $U(N_F)$ as part of its flavour symmetry group and gives rise to $N_F$
 twisted mass parameters $m=(m_1,\ldots,m_{N_F})$ and $N_F$ $U(1)$ $R$-charges $q=(q_1,\ldots,q_{N_F})$. Occasionally, we will find it convenient to combine these parameters into the holomorphic combination
 \begin{equation}
 \rmM_I=m_I+\frac{i}{2r} q_I\,.
 \label{holomm}
 \end{equation}

Finally, we can add in a supersymmetric way a superpotential for the chiral multiplet
\begin{equation}
S_{\text{W}} = \int \rmd^{2}x\sqrt{h} \ \bigg\{F_W+\bar F_{\bar W}\bigg\}\,,
\label{superp}
\end{equation}
whenever the total $U(1)$ $R$-charge of the superpotential is $-q_W=-2$.
$F_W$ is the gauge invariant auxiliary component of the superpotential chiral multiplet.\footnote{In terms of the $\phi$ chiral multiplet,
$F_W={\frac{\partial W}{\partial \phi}} F-{\frac{1}{2}}{\frac{\partial^2 W}{\partial\phi^2}}\psi\psi$. Invariance of \rf{superp} under supersymmetry when $q_W=2$ follows from equations \rf{dhF} and \rf{dhFb}.}
Under these conditions, the Lagrangian in \rf{superp} transforms into a total derivative under the $SU(2|1)$ supersymmetry transformations below.

A few brief remarks about the $\cN=(2,2)$ gauge theories in $S^2$ thus constructed are in order.  The action (and supersymmetry transformations) can be organized in a power series expansion in $1/r$, starting with the
covariantized $\cN=(2,2)$ gauge theory action in flat space. The action is deformed by terms  of order $1/r$ and  $1/r^2$, with terms proportional to $1/r$  not being reflection positive.  These features are consistent
with the general arguments in \cite{Festuccia:2011ws}.
The theory on $S^2$ breaks the classical\footnote{This classical symmetry of the flat space theory, being chiral, can be anomalous.} $U(1)_A$ $R$-symmetry of the corresponding  $\cN=(2,2)$ gauge theory in flat space.
This can be observed in the asymmetry between the scalar fields $\sigma_1$ and $\sigma_2$ in the action on $S^2$, which are otherwise rotated into each other by the $U(1)_A$ symmetry of  the flat space theory.
This asymmetry is also manifested in the twisted masses $m$ being real on $S^2$, while they are complex in flat space.\footnote{Where twisted masses correspond to background values of  $\sigma_1, \sigma_2$ in the vector
multiplet for $G_F$.} The real twisted masses $m$ on $S^2$, however, combine  with the $U(1)$ $R$-charges $q$ into the holomorphic parameters $\rmM=m+\frac{i}{2r} q$ introduced in \rf{holomm}.

The gauge theory action we have written down is invariant under the $SU(2|1)$ supersymmetry algebra. The supersymmetry transformations are parametrized by conformal Killing spinors\footnote{Thus named since the
defining equation $\nabla_{i}\epsilon=\gamma_i\tilde\epsilon$ is conformally invariant.}  $\epsilon$ and $\bar\epsilon$ on $S^2$. These can be taken to obey
\begin{equation}
\begin{aligned}
\nabla_{i}\epsilon &= +\frac{1}{2r}\gamma_{i}\gamma^{\hat 3}\epsilon\,
\\
\nabla_{i}\bar{\epsilon} &= -\frac{1}{2r}\gamma_{i}\gamma^{\hat 3}\bar{\epsilon}\,,
\end{aligned}
\label{killing}
\end{equation}
where $\epsilon$ and $\bar \epsilon$ are complex Dirac spinors in two dimensions and $r$ is the radius of the $S^2$.
The spinors $\epsilon_{\alpha}$ and $\bar \epsilon_{\alpha}$ are the supersymmetry parameters associated to the supercharges $Q_\alpha$ and $S_\alpha$ respectively.
More details  about the supersymmetry transformations can be found in appendix~\ref{app:sca}.

As mentioned earlier, the explicit supersymmetry transformations can be found by restricting the $\cN=(2,2)$ superconformal transformations to the $SU(2|1)$ subalgebra.  The $SU(2|1)$  supersymmetry transformations
of the vector multiplet fields are
\begin{align}
\label{dhlambda}
   {\delta} \lambda &=
    \left(iV_{m}\gamma^{m}-\rmD\right)\epsilon
\\
\label{dhblambda}
   {\delta} \bar{\lambda} &=
    \left(i\bar{V}_{m}\gamma^{m}+\rmD\right)\bar{\epsilon}
\\
\label{dhA}
  {\delta} A_{i} &=
    -\frac{i}{2}\left(\bar{\epsilon}\gamma_{i}\lambda+\epsilon\gamma_{i}\bar{\lambda}\right)
\\
\label{dhsigma1}
  {\delta} \sigma_{1} &=
    \frac{1}{2}\left(\bar{\epsilon}\lambda-\epsilon\bar{\lambda}\right)
\\
\label{dhsigma2}
   {\delta} \sigma_{2} &=
    -\frac{i}{2}\left(\bar{\epsilon}\gamma_{\hat 3}\lambda+\epsilon\gamma_{\hat 3}\bar{\lambda}\right)
\\
\label{dhD}
\begin{split}
   {\delta} \rmD &=
    -\frac{i}{2}\bar{\epsilon}\left(\slashed{D}\lambda+\left[\sigma_{1},\lambda\right]-i\left[\sigma_{2},\gamma^{\hat 3}\lambda\right]\right)
    \\
    &\quad + \frac{i}{2}\epsilon\left(\slashed{D}\bar{\lambda}-\left[\sigma_{1},\bar{\lambda}\right]-i\left[\sigma_{2},\gamma^{\hat 3}\bar{\lambda}\right]\right)\,,
\end{split}
\end{align}
with $V_m$ and  $\bar V_m$ defined by
\begin{equation}
  \begin{aligned}
    V^{i} &= \varepsilon^{ij}D_{j}\sigma_{2}+D^{i}\sigma_{1}\,,
    &
    V^{3} &= \frac{1}{2}\varepsilon^{ij}F_{ij}+i\left[\sigma_{1},\sigma_{2}\right]+\frac{1}{r}\sigma_{1}
    \\
    \bar{V}^{i} &= \varepsilon^{ij}D_{j}\sigma_{2}-D^{i}\sigma_{1} \,,
    &
    \bar{V}^{3} &= \frac{1}{2}\varepsilon^{ij}F_{ij}-i\left[\sigma_{1},\sigma_{2}\right]+\frac{1}{r}\sigma_{1}\,.
  \end{aligned}
   \label{Bogob}
\end{equation}

The transformations of the massless chiral multiplet fields are
\begin{align}
\label{dhphi}
  {\delta} \phi&= \bar{\epsilon}\psi
\\
\label{dhphib}
  {\delta} \bar{\phi}&= \epsilon\bar{\psi}
\\
\label{dhpsi}
  {\delta} \psi&=
    i\left(\slashed{D}\phi+\sigma_{1}\phi-i\sigma_{2}\phi\gamma^{\hat 3}+\frac{q}{2r}\phi\gamma^{\hat 3}\right)\epsilon+\bar{\epsilon}F
\\
\label{dhpsib}
   {\delta} \bar{\psi}&=
    i\left(\slashed{D}\bar{\phi}+\bar{\phi}\sigma_{1}+i\bar{\phi}\sigma_{2}\gamma^{\hat 3}-\frac{q}{2r}\bar{\phi}\gamma^{\hat 3}\right)\bar{\epsilon}+\epsilon \bar{F}
\\
\label{dhF}
   {\delta} F&=
    -i\left(D_{i}\psi\gamma^{i}+\sigma_{1}\psi-i\sigma_{2}\psi\gamma^{\hat 3}+\lambda\phi+\frac{q}{2r}\psi\gamma^{\hat 3}\right)\epsilon
\\
\label{dhFb}
   {\delta} \bar{F}&=
    -i\left(D_{i}\bar{\psi}\gamma^{i}+\bar{\psi}\sigma_{1}+i\bar{\psi}\sigma_{2}\gamma^{\hat 3}-\bar{\phi}\bar{\lambda}-\frac{q}{2r}\bar{\psi}\gamma^{\hat 3}\right)\bar{\epsilon}\,.
\end{align}
The supersymmetry transformations of the theory with twisted masses are obtained from equations (\ref{dhphi}--\ref{dhFb}) by shifting $\sigma_2\rightarrow \sigma_2+m$ as in \rf{shiftt}.

With these transformations, the $SU(2|1)$ supersymmetry algebra \rf{SU(2|1)} is realized  off-shell on the vector multiplet and chiral multiplets fields.
Splitting $\delta\equiv \delta_\epsilon+\delta_{\bar\epsilon}$, we find that this representation of $SU(2|1)$ on the fields obeys
\begin{equation}
[\delta_\epsilon,\delta_\epsilon]=0\qquad\qquad  [\delta_{\bar\epsilon},\delta_{\bar\epsilon}]=0\,,
\end{equation}
and\footnote{The explicit form of the commutator of supersymmetry transformations on the vector multiplet and chiral multiplet fields can be found in appendix~\ref{app:sca}.}
\begin{equation}
[\delta_\epsilon,\delta_{\bar\epsilon}]=\delta_{SU(2)}(\xi) + \delta_{R}(\alpha)+\delta_{G}(\Lambda)+\delta_{G_F}(\Lambda_m)\,,
\end{equation}
thus generating an infinitesimal $SU(2)\times R\times G\times G_F$ transformation.  When localizing the path integral of $\cN=(2,2)$ gauge theories on $S^2$, we will choose a particular supercharge $\cQ$ in $SU(2|1)$.
The $SU(2)\times R\times G\times G_F$ transformation it generates will play an important role in our computation of the partition function.

The $SU(2)$ isometry transformation induced by the commutator of supersymmetry transformations is parametrized by the Killing vector field\footnote{The fact that $\xi$ is a Killing vector, that it obeys
$\nabla^i\xi^j+\nabla^j\xi^i=0$, is a consequence of the choice of conformal Killing spinors in \rf{killing}. As desired, it does not generate conformal transformations of $S^2$.}
\begin{equation}
\xi^i=-i\bar{\epsilon}\gamma^{i}\epsilon\,.
\end{equation}
It acts on the bosonic fields via the usual Lie derivative and on the fermions via the Lie-Lorentz derivative
\begin{equation}
\label{LLd}
\cL_\xi \equiv \xi^i\,\nabla_i + \frac{1}{4} \nabla_{i} \,\xi_{j}\,\gamma^{ij} \,.
\end{equation}

The $U(1)$ $R$-symmetry transformation generated by the commutator of the supersymmetry transformations is parametrized by
\begin{equation}
\alpha = -\frac{1}{2r}\bar{\epsilon}\gamma^{\hat{3}}\epsilon\,.
\end{equation}
It acts on the fields by multiplication by the corresponding charge.
The $U(1)$ $R$-symmetry charges of the various fields, supercharges and parameters are given by:
\medskip
\begin{center}
  \begin{tabular}{*{16}{>{\(}c<{\)}}}
    \toprule
    \multicolumn{4}{c}{supersymmetry} &
    \multicolumn{6}{c}{vector multiplet} &
    \multicolumn{6}{c}{chiral multiplet}
    \\
    \cmidrule(lr{1mm}){1-4}\cmidrule(lr{1mm}){5-10}\cmidrule(lr{1mm}){11-16}
    \epsilon & \bar{\epsilon} & Q & S &
    A_\mu & \sigma_1 & \sigma_2 & \lambda & \bar{\lambda} & D &
    \phi & \psi & F & \bar{\phi} & \bar{\psi} & \bar{F}
    \\
    1 & -1 & -1 & 1 &
    0 & 0 & 0 & 1 & -1 & 0 &
    -q & -(q-1) & -(q-2) & q & q-1 & q-2 \\
    \bottomrule
  \end{tabular}
\end{center}
\medskip
Since the action of $R$ on the fields is non-chiral, this classical symmetry is not spoiled by quantum anomalies and is an exact symmetry of the $\cN=(2,2)$ gauge theories we have constructed.

The commutator of two supersymmetry transformations generates a field dependent gauge transformation, taking values in  the Lie algebra of the gauge group $G$. The induced gauge transformation is labeled by the gauge parameter
\begin{equation}
\Lambda =(\bar{\epsilon}\epsilon)\sigma_{1}-i(\bar{\epsilon}\gamma^{\hat 3}\epsilon)\sigma_{2}+  \xi^i A_i\,,
\end{equation}
which acts on the various fields by the standard gauge redundancy transformation laws. On the gauge field it acts by
\begin{equation}
\delta_\Lambda A_i=D_i\Lambda
\end{equation}
while on a field $\varphi$ it acts by
\begin{equation}
\delta_\Lambda\varphi = i \Lambda \cdot \varphi \,,
\end{equation}
where \(\Lambda\) acts on $\varphi$ in the corresponding representation of $G$.

Finally, in the presence of twisted masses $m$, a $G_F$ flavour symmetry rotation on the chiral multiplet fields is generated by $ [\delta_\epsilon,\delta_{\bar\epsilon}]$.
The induced flavour symmetry transformation acts on the chiral multiplet fields in the fundamental representation of $G_F$, and is parametrized  by
\begin{equation}
\Lambda_m=-i(\bar{\epsilon}\gamma^{\hat 3}\epsilon)m\,,
\end{equation}
with $m$  taking values in the Cartan subalgebra of $G_F$. It acts trivially on the vector multiplet fields.

\section{Localization of the Path Integral}
\label{sec:local}

In this paper  our goal is to  perform  the exact computation of the partition function of $\cN=(2,2)$ gauge theories on $S^2$. The powerful tool that
allow us to achieve this goal is supersymmetric localization.

The central idea of supersymmetric localization \cite{Witten:1991mk} is that the path integral -- possibly decorated with the insertion of observables or boundary conditions invariant under a supercharge $\cQ$ --
 localizes to the $\cQ$-invariant field configurations.  If the orbit of $\cQ$ in the space of fields
 is non-trivial,\footnote{By definition of $\cQ$-invariance of the path integral, the space of fields admits the action of $\cQ$.}  then  the path integral vanishes upon integrating over the associated
 Grassman collective coordinate. Therefore,  the non-vanishing contributions to the path integral can only arise from the trivial orbits, i.e.\@ the fixed points of supersymmetry.
  These fixed point field configurations are the solutions to the supersymmetry variation equations generated by the supercharge $\cQ$, which we denote by
\begin{equation}
\delta_\cQ\,\text{fermions}=0\,.
\label{PDE}
\end{equation}
In the path integral we must  integrate over   the moduli space of solutions of the partial differential equations implied by
supersymmetry fixed point equations \rf{PDE}.

Under favorable asymptotic behavior, integration by parts implies that the result of the path integral does not depend on the deformation of  the original supersymmetric Lagrangian by a
$\cQ$-exact term\footnote{$\cQ\cdot V$ denotes the supersymmetry transformation of $V$ generated by $\cQ$ (see also \rf{deforme}).}
\begin{equation}
\cL\rightarrow \cL+t\, \cQ\cdot V\,,
\label{deformm}
\end{equation}
as long as $V$ is invariant under the bosonic transformations generated by $\cQ^2$. Obtaining a  sensible path integral requires that the action is nondegenerate and that the path integral is convergent in the presence of
the deformation term $\cQ\cdot V$.

In the $t\rightarrow \infty$ limit, the semiclassical approximation with respect to $\hbar_{\text{eff}}\equiv 1/t$ is exact. In this  limit,  only the saddle points of $\cQ\cdot V$ can contribute and, moreover,
the path integral is dominated by the saddle points with vanishing action. However, of all the saddle points of $\cQ\cdot V$, only the $\cQ$-supersymmetric field configurations
give a non-vanishing contribution. Therefore, we must integrate over the intersection of supersymmetric field configurations and saddle points of $\cQ\cdot V$. We denote this intersection by $\cF$.

Using the saddle point approximation,  the path integral in the  $t\rightarrow \infty$ limit can be calculated by restricting  the original Lagrangian $\cL$  to
$\cF$,\footnote{The deformation term $\cQ\cdot V$ vanishes on $\cF$ since it is a linear combination of the supersymmetry equations.} integrating out the quadratic fluctuations of all the fields in the
deformation $\cQ\cdot V$ expanded around a point in $\cF$, and integrating the combined expression over $\cF$.\footnote{The original Lagrangian $\cL$ is irrelevant for the localization one-loop analysis.}
 Of course, even though the path integral is one-loop exact with respect to $t$, it yields exact results with respect to the original  coupling constants and parameters of the theory.

The final result of the localization computation does not depend on the choice of deformation $\cQ\cdot V$.  One may add to $\cQ\cdot V$ another $\cQ$-exact term, and the result of the path integral
will not change as long as the new $\cQ$-exact  term  is non-degenerate,  and no new supersymmetric saddle points are introduced that can flow from infinity. This can be accomplished by choosing the
deformation term such that  it  does not change the asymptotic behavior of the potential in the space of fields. We will take advantage of this freedom and choose a deformation term  $\cQ\cdot V$
that makes computations most tractable.

Since our aim is to localize the path integral of gauge theories, some care has to be taken to localize the gauge fixed theory. This requires combining in a suitable way the deformed action $\cQ\cdot V$
and gauge fixing terms $\cL_{\text{g.f.}}$ into a $\hat\cQ=\cQ+Q_{\text{BRST}}$ exact term $\hat \cQ\cdot \hat V$, where $\hat V=V+V_{\text{ghost}}$. This refinement, while technically important, does not modify
the fact that the gauge fixed path integral localizes to $\cF$. The inclusion of the gauge fixing term, however, plays an important role in the evaluation of the one-loop determinants in the directions
normal  to  $\cF$.

\subsection{Choice of Supercharge}
\label{sec:supercharge}

In this section we choose a particular supersymmetry generator  $\cQ$ in the $SU(2|1)$ supersymmetry algebra with which to localize the
path integral of $\cN=(2,2)$ gauge theories on $S^2$.  We consider\footnote{In section~\ref{sec:classical} we also analyze   localization of the path integral with respect to both $Q_1$ and $Q_2$.
The analysis leads directly to the Coulomb branch representation of the partition function. On the other hand, this other choice does not allow non-trivial field configurations in the Higgs branch,
and therefore cannot give rise to the Higgs branch representation of the partition function.}
\begin{equation}
\cQ= S_{1} + Q_{2}\,.
\label{choice}
\end{equation} This supercharge  generates an $SU(1|1)$ subalgebra of $SU(2|1)$, given by
\begin{equation}
\cQ^2=J+\frac{R}{2}  \hspace{50pt} \left[J+\frac{R}{2},\cQ\right]=0\,,
\label{SUSSalg}
\end{equation}
where $J$ is the charge corresponding to  a $U(1)$ subgroup of the $SU(2)$ isometry group of  the  $S^2$ while $R$ is the $R$-symmetry generator in $SU(2|1)$.
In terms of embedding coordinates where $S^2$ is parametrized by
\begin{equation}
X_1^2+X_2^2+X_3^2=r^2\,,
\label{embeddd}
\end{equation}
 $J$ acts under an infinitesimal transformation, as follows
\begin{equation}
\begin{aligned}
 X_1\rightarrow &X_1-\varepsilon X_2\\
  X_2\rightarrow &X_2+\varepsilon X_1\,.
\end{aligned}
\label{rotacio}
\end{equation}
Geometrically, the action of $J$ has two antipodal fixed points on $S^2$, which can be used to define the north and south poles of $S^2$. These are located at $(0,0,r)$ and $(0,0,-r)$ in the
embedding coordinates \rf{embeddd}. In terms of the coordinates of the round metric on $S^2$
\begin{equation}
 \rmd s^2=r^2\left(\rmd\theta^2+\sin^2\theta \rmd\varphi^2\right)
 \label{round}
 \end{equation}
the corresponding Killing vector is
\begin{equation}
 i\frac{\partial}{\partial \varphi}\,,
\end{equation}
with the north and south poles corresponding to $\theta=0$ and $\theta=\pi$ respectively. The supersymmetry algebra \rf{SUSSalg} is the same used in \cite{Pestun:2007rz} in the computation of the
partition function of four dimensional $\cN=2$ gauge theories on $S^4$.

In order to derive the supersymmetry fixed point equations \rf{PDE} generated by the supercharge $\cQ$, first we need to construct the
conformal Killing spinors associated to it, which we denote by $\epsilon_{\cQ}$ and $\bar{\epsilon}_{\cQ}$. The  conformal Killing spinors on $S^2$  obeying  \rf{killing}
are explicitly
given by\footnote{In the vielbein basis  $e^{\hat 1}=r \rmd\theta$ and  $e^{\hat 2}=r \sin\theta  \rmd\varphi$.
For details, please refer to appendix~\ref{app:susyv}.}
   \begin{equation}
\begin{aligned}
\epsilon & =
  \exp\left(-\frac{i\theta}{2}\gamma^{\hat{2}}\right)\exp\left(\frac{i\varphi}{2}\gamma^{\hat{3}}\right)\epsilon_{\circ}\,
\\
\bar{\epsilon} &=
  \exp\left(+\frac{i\theta}{2}\gamma^{\hat{2}}\right)\exp\left(\frac{i\varphi}{2}\gamma^{\hat{3}}\right)\bar{\epsilon}_{\circ}\,,
\end{aligned}
\label{killspin}
\end{equation}
where $\epsilon_{\circ}$ and $\bar{\epsilon}_{\circ}$ are constant, complex Dirac spinors.  The conformal Killing spinors  $\epsilon_{\cQ}$ and $\bar{\epsilon}_{\cQ}$ are given by \rf{killspin},
with
$\epsilon_{\circ}$ and $\bar{\epsilon}_{\circ}$  being chiral spinors of opposite
chirality, that is
\begin{equation}
\begin{aligned}
\gamma^{\hat 3}\epsilon_{\circ}&=+\epsilon_{\circ}\,
\\
\gamma^{\hat 3}\bar{\epsilon}_{\circ}&=-\bar{\epsilon}_{\circ}\,.
\end{aligned}
\end{equation}
Therefore, explicitly
 \begin{equation}
\begin{aligned}
\epsilon_{\cQ} & =e^{i\varphi/2}
  \exp\left(-\frac{i\theta}{2}\gamma^{\hat{2}}\right)\epsilon_{\circ}\,
\\
\bar{\epsilon}_{\cQ} &=e^{-i\varphi/2}
  \exp\left(+\frac{i\theta}{2}\gamma^{\hat{2}}\right)\bar{\epsilon}_{\circ}\,.
\end{aligned}
\label{killspinQ}
\end{equation}
We note that at  the north and the south poles of the $S^2$  the conformal Killing spinors $\epsilon_{\cQ}$ and $\bar{\epsilon}_{\cQ}$
have definite chirality, and that the chirality  at the north pole  is opposite to that at the south pole
\begin{equation}
\begin{aligned}
\gamma^{\hat 3}\epsilon_{\cQ}(N)&=\epsilon_{\cQ}(N)\qquad \qquad \hskip+9pt\gamma^{\hat 3}\epsilon_{\cQ}(S)=-\epsilon_{\cQ}(S)\\
\gamma^{\hat 3}\bar{\epsilon}_{\cQ}(N)&=-\bar{\epsilon}_{\cQ}(N)\qquad \qquad \gamma^{\hat 3}\bar{\epsilon}_{\cQ}(S)=\bar{\epsilon}_{\cQ}(S)\,.
\end{aligned}
\end{equation}
As we shall see, the fact that $\cQ$
is  chiral at the poles  implies that the corresponding chiral field configurations  --  vortices localized at the north pole and anti-vortices  at the  south pole --
may contribute  to the partition function of $\cN=(2,2)$ gauge theories on $S^2$.

We note that the circular Wilson loop operator supported on a latitude angle $\theta_0$
\begin{equation}
\label{WilsonLoop}
W_{\theta_{\circ}} = \Tr \Pexp  \oint_{\theta_{\circ}}  \left[-i  A_i \rmd x^i +i r (\sigma_{1}\cos\theta_{\circ} - i \sigma_2)\rmd\varphi \right]
\end{equation}
is invariant   under the action of $\cQ$. Therefore the expectation value of these operators can be computed when localizing with respect to the supercharge~$\cQ$.

Given our choice of supercharge $\cQ$, we can explicitly determine the infinitesimal $J\times R\times G\times G_F$ transformation that $\cQ^2$ generates when acting on the fields. The spinor bilinears
constructed from $\epsilon_{\cQ}$ and $\bar{\epsilon}_{\cQ}$ in section~\ref{sec:gauge} evaluate to\footnote{By fixing the overall normalization
$\bar{\epsilon}_{\circ}\epsilon_{\circ}=i$.}
\begin{equation}
\begin{aligned}
  \bar{\epsilon}_{\cQ}\epsilon_{\cQ} &= i \cos\theta\
  \\[5pt]
  \bar{\epsilon}_{\cQ}\gamma^{\hat 3}\epsilon_{\cQ} &= i
\end{aligned}
  \hspace{50pt}
\begin{aligned}
  \xi &=-\frac{i}{r} \partial_{\varphi}
  \\
  \alpha &= -\frac{i}{2r} \,.
\end{aligned}
\end{equation}
Therefore, in view of \rf{rotacio}, $\cQ^2$ generates $J+R/2$, i.e.\@ a simultaneous infinitesimal rotation and $R$-symmetry transformation with parameter
 \begin{equation}
\varepsilon=\frac{1}{r}\,,
 \end{equation}
and a gauge transformation with gauge parameter
 \begin{equation}
 \Lambda = i\cos\theta\sigma_1 + \sigma_2 - \frac{i}{r} A_2\,.
 \end{equation}
On the chiral multiplet fields, $\cQ^2$ also induces a $G_F$ flavour symmetry rotation parametrized by the twisted masses $m$.

\subsection{Localization Equations}
\label{sec:sphere}

Here we present the key steps in the derivation of the set of partial differential equations that characterize the vector multiplet and chiral multiplet field configurations that are invariant under the action of $\cQ$.
The details of the derivation are omitted here and can be found in appendix~\ref{app:susyv}.

We must identify the partial differential equations implied by \rf{PDE}
\begin{align}
  \delta_{{\cQ}} \lambda&= \delta_{{\cQ}}  \bar\lambda=0 \label{fixedpts}\\
  \delta_{{\cQ}} \psi&= \delta_{{\cQ}}  \bar\psi=0\, ,  \label{fixedptsb}
\end{align}
where $\delta_\cQ \equiv \delta_{\epsilon_\cQ}+\delta_{\bar\epsilon_\cQ}$\,,  from the explicit supersymmetry transformations given in equations (\ref{dhlambda},~\ref{dhblambda}) and
(\ref{dhpsi},~\ref{dhpsib}) for the choice of  conformal Killing spinors  $\epsilon_{\cQ}$ and $\bar \epsilon_{\cQ}$ in \rf{killspinQ}. The moduli space of
solutions to these equations, once intersected with the saddle points of our choice of $\cQ$-exact deformation term, determines the space of field configurations that need to be integrated over in the path integral.

Given a choice of deformation term, in order for the path integral to converge we need to impose reality conditions on the fields. These reality conditions restrict the contour of path integration so that the integrand
falls off sufficiently fast in the asymptotic region in the space of field configurations. The residual freedom in the choice of contour i.e.\@ deformations of the contour which do not change the asymptotic behavior
of the integrand, is then used to make sure that the contour of integration includes the saddle points of the deformed action.

We are interested in deformation terms that do not alter the asymptotic behavior of the original action \rf{action}. We may therefore extract the reality conditions by requiring the original path integral
for some effective couplings to be convergent.

From the kinetic terms in the bosonic part of the action \rf{action} we conclude that the scalar fields $\sigma_1,\sigma_2$ and the connection $A_{i}$ in the vector multiplet are hermitian while the chiral multiplet
complex scalars $\phi$ and $\bar{\phi}$ satisfy $\bar{\phi}=\phi^{\dagger}$. Next we note that the path integration over the chiral multiplet auxiliary fields $F,\,\bar{F}$ is just a Gaussian integral and we simply require $\bar{F}=F^{\dagger}$. Provided that a $\cQ$-exact deformation term contains following terms
\begin{align}
  \cQ \cdot V = - \frac{1}{2 g_\text{eff}^2} {\rm D}^2 - i {\rm D} (\phi \bar{\phi} - \xi_\text{eff} \mathbbm{1} ) + \cdots\,,
  \label{effFI}
\end{align}
one should choose the contour of integration for the auxiliary field $\rm D$ such that
$\rmD + ig_{\text{eff}}^{2}(\phi \bar{\phi} - \xi_{\text{eff}} \mathbbm{1})$ is hermitian for the convergence of the path integral.
In other words
\begin{align}
  \operatorname{Im}\rmD + g_{\text{eff}}^{2}(\phi \bar{\phi} - \xi_{\text{eff}} \mathbbm{1}) = 0 \,, \nonumber
\end{align}
where the explicit form of the coupling constants $g_{\text{eff}}^{2}$ and $\xi_{\text{eff}}$ are determined
by choice of $\cQ$-exact deformation terms (after taking $t\rightarrow \infty$).

The supersymmetry fixed point equations for the vector multiplet fields \rf{fixedpts}  are given by
\begin{align}
\label{vmeq1}
  D_{\hat{2}} \sigma_1 = D_{\hskip+0.5pt\hat{i}} \sigma_2 & = 0 &
  D_{\hat{1}} \sigma_1 + g_{\text{eff}}^{2}(\phi \bar{\phi} - \xi_{\text{eff}} \mathbbm{1}) \sin \theta & = 0
   \\
\label{vmeq2}
  \operatorname{Re}\rmD = [ \sigma_1 , \sigma_2 ] & = 0 &
  F_{\hat{1}\hat{2}}  + \frac{\sigma_1}{r} + g_{\text{eff}}^{2}(\phi \bar{\phi} - \xi_{\text{eff}} \mathbbm{1}) \cos\theta & = 0\,,
\end{align}
while the supersymmetry equations for the chiral multiplet fields \rf{fixedptsb} reduce to
\begin{align}
\label{cmeq1}
 \cos \frac{\theta}{2} \left(  D_{\hat{1}} + i D_{\hat{2}} \right) \phi + \sin \frac{\theta}{2} \left( \sigma_{1} - \frac{q}{2r} \right) \phi &=0 &
 F &= 0
\\
\label{cmeq2}
 \sin \frac{\theta}{2} \left(  D_{\hat{1}} - i D_{\hat{2}}\right) \phi + \cos \frac{\theta}{2} \left(  \sigma_{1} + \frac{q}{2r} \right) \phi &=0 &
 \left(\sigma_{2}+m\right)\phi &= 0\,.
\end{align}
These differential  equations on $S^2$ are a supersymmetric extension of  classic differential  equations in physics.
Our equations interpolate between BPS vortex equations at the north pole ($\theta=0$)
\begin{equation}
  \begin{aligned}
    \left( D_{\hat{1}} + i D_{\hat{2}} \right) \phi&=0 & \qquad \qquad D_{\hat{i}}\left(\sigma_1+i\sigma_2\right)&= 0 \\
    F_{\hat{1}\hat{2}}  + \frac{\sigma_1}{r} + g_{\text{eff}}^{2} (\phi \bar{\phi} - \xi_{\text{eff}} \mathbbm{1} ) &= 0  & \operatorname{Re}\rmD = [\sigma_1,\sigma_2]&=0\\
    \left(  \sigma_{1} + \frac{q}{2r} \right) \phi &=0 & \left(\sigma_{2}+m\right)\phi &= 0\,,
  \end{aligned}
\end{equation}
and BPS anti-vortex equations at the south pole ($\theta=\pi$)
\begin{equation}
  \begin{aligned}
    \left( D_{\hat{1}} - i D_{\hat{2}} \right) \phi&=0 & \qquad \qquad D_{\hat{i}}\left(\sigma_1+i\sigma_2\right)&= 0 \\
    F_{\hat{1}\hat{2}}  + \frac{\sigma_1}{r} - g_{\text{eff}}^{2} (\phi \bar{\phi} - \xi_{\text{eff}} \mathbbm{1} ) &= 0  & \operatorname{Re}\rmD = [\sigma_1,\sigma_2]&=0\\
    \left(  \sigma_{1} - \frac{q}{2r} \right) \phi &=0 & \left(\sigma_{2}+m\right)\phi &= 0\,.
  \end{aligned}
\end{equation}
This system of differential equations is akin to the one found in \cite{Gomis:2011pf} in the localization
computation of four dimensional $\cN=2$ gauge theories on $S^4$.
We return later to the study of the  supersymmetry equations at the poles, which play  a crucial
role in our analysis, yielding the Higgs branch representation of  the gauge theory partition function on $S^2$.

\subsection{Vanishing Theorem}
\label{sec:solutions}

As explained previously, the path integral localizes to the space $\cF$ of supersymmetric field configurations which are also saddle points of the localizing deformation term. In this section, we consider
the supersymmetry equations in the absence of effective FI parameters and we write down the most general \emph{smooth} solutions to the supersymmetry equations for
generic values of the $R$-charges. These solutions are parametrized by the expectation value of fields in the vector multiplet, thus, we denote this space of solutions by $\cF_{\text{Coul}}$.
In section~\ref{sec:classical} we localize the path integral to $\cF_{\text{Coul}}$ and derive the Coulomb branch representation of the partition function.

With $\xi_{\text{eff}}=0$ and for generic $R$-charges, the most general smooth solution to the equations \rf{vmeq1},\rf{vmeq2},\rf{cmeq1} and \rf{cmeq2} is given by\footnote{A detailed derivation of this
result is presented in appendix~\ref{app:susyv}.}
\begin{equation}
  \begin{aligned}
    A & = \frac{B}{2}\left(\kappa-\cos\theta\right) \rmd \varphi \qquad &
    \qquad \sigma_1 & = -\frac{B}{2r} &
    \qquad\qquad \phi & = 0
    \\
    \rmD & = 0 &
    \sigma_2 & = a &
    F & = 0 \,,
  \end{aligned}
  \label{saddlept}
\end{equation}
where $a$ and $B$ are constant commuting matrices which live in the gauge Lie algebra and its Cartan subalgebra respectively. The matrix $B$ is further restricted by the first Chern class quantization to
have integer eigenvalues. The constant $\kappa$ parametrizes a pure gauge background which is necessary in any coordinate patch which includes one of the poles and can be gauged away in the coordinate patch which excludes
the poles.

It is interesting to note that if the $R$-charge is tuned to be a negative integer or zero, then there are nontrivial solutions of the form
\begin{equation}
\phi=e^{\frac{i}{2}(\kappa B-q)\varphi}\frac{ (\sin\frac{\theta}{2})^{\frac{B-q}{2}} }{ (\cos\frac{\theta}{2})^{\frac{B+q}{2}} }\phi_{\circ}
\end{equation}
with $\phi_{\circ}$ being a constant in the kernel of $a+m$.  Imposing regularity at the poles restricts
the allowed value of $q$ and $B$ as follows: $q + |B|$ must be even and non-positive
integers. In such a case, the above field configuration can be written in terms of the magnetic flux $B$
monopole scalar harmonics $Y^{\frac{B}{2}}_{j,m}$
as
\begin{equation}
\phi=Y^{\frac{B}{2}}_{-\frac{q}{2},-\frac{q}{2}}\phi_{\circ}\,.
\label{zeromodes}
\end{equation}
It is worth mentioning that these field configurations are also supersymmetric configurations    in the localization computation of the partition function of three dimensional
$\cN=2$ gauge theories on $S^1\times S^2$ \cite{Imamura:2011su}, which computes the superconformal index of these theories.
In our computations,  we can ignore these discrete, tuned solutions to the supersymmetry equations: for   theories flowing   to superconformal theories in the infrared, unitarity  constrains the
$R$-charges to be non-negative. Furthermore, as will be explained     in section~\ref{sec:classical}, these solutions are not saddle points of the localized path integral.

We note that even though our  choice of $\cQ$ breaks the $SU(2)$ symmetry of $S^2$, the $\cQ$-invariant field configurations \rf{saddlept} are $SU(2)$ invariant. Later on, we take an alternative approach
in which the Coulomb branch is lifted and the saddle point equations admit singular solutions at the poles thereby breaking the $SU(2)$ symmetry.
We will consider the physics behind singular solutions localized at the north and south poles of $S^2$ in section~\ref{sec:instanton}.

\section{Coulomb Branch}
\label{sec:classical}

In order to evaluate the path integral of an \(\cN=(2,2)\) gauge theory on \(S^2\) using supersymmetric localization, we must choose a deformation of the original supersymmetric Lagrangian by a
$\cQ$-exact term \rf{deformm}
\begin{equation}
  \cL \rightarrow \cL + t\, \delta_{\cQ}V\,.
  \label{deforme}
\end{equation}
The deformation term
$ \delta_{\cQ}V$ defines the measure of integration through the associated one-loop determinant. In this section we calculate the contribution to the path integral due to the smooth field configurations
\rf{saddlept}. This yields the Coulomb branch representation of the path integral, as an integral over the Coulomb branch saddle points  \(\cF_{\text{Coul}}\).

A calculation shows that the vector multiplet action \rf{vm_action} and the chiral multiplet action \rf{chiralmult} are \(\cQ\)-exact with respect to our choice of supercharge \rf{choice}.  Specifically,
\begin{equation}
  ( \bar\epsilon_\cQ \gamma^{\hat 3} \epsilon_\cQ) \, g^2\, \cL_{\text{v.m.}}
  = \delta_{\cQ}\delta_{\bar\epsilon_\cQ}\Tr\left(
    \frac{1}{2} \bar\lambda \gamma^{\hat 3}\lambda
    - 2i\rmD\sigma_2 + \frac{i}{r} \sigma_2^2\right)\,,
    \label{defvm}
\end{equation}
and
\begin{equation}
  - ( \bar\epsilon_\cQ \gamma^{\hat 3} \epsilon_\cQ) \, \left(\cL_{\text{c.m.}}+\cL_{\text{mass}}\right)
  = \delta_{\cQ}\delta_{\bar\epsilon_\cQ} \Tr\left(
    \bar\psi \gamma^{\hat 3}\psi
    - 2\bar\phi\left( \sigma_2+m + i\frac{q}{2r}\right)\phi
    + \frac{i}{r}\bar\phi\phi
  \right)\,,
   \label{defcm}
\end{equation}
where $\delta_\cQ \equiv \delta_{\epsilon_\cQ}+\delta_{\bar\epsilon_\cQ}$.
This implies  that correlation functions of $\cQ$-closed observables in an $\cN=(2,2)$ gauge theory on $S^2$ are independent of $g$, the Yang-Mills coupling constant.  Despite being $g$ independent,
these correlators  are   nontrivial functions of the renormalized FI parameter $\xi_{\text{ren}}$ for each $U(1)$ factor in the gauge group, and of the twisted masses $m$.

We now turn to the choice of deformation term $\delta_{\cQ}V$. The most canonical choice would be to take
\begin{equation}
V_{\text{can}}=\left(\delta_\cQ \lambda\right)^\dagger \lambda+\left(\delta_\cQ \bar \lambda\right)^\dagger \bar \lambda+
\left(\delta_\cQ \psi\right)^\dagger \psi+\left(\delta_\cQ \bar \psi\right)^\dagger \bar \psi\,.
\end{equation}
 For this choice, the bosonic part of the  deformation term  $\delta_{\cQ}V_{\text{can}}$ is manifestly non-negative. It is therefore guaranteed that
all $\cQ$-invariant field configurations are the saddle points of $\delta_{\cQ}V_{\text{can}}$ with minimal (zero) action.  The disadvantage of such a deformation term is that the resulting action
$\delta_{\cQ}V_{\text{can}}$ does not necessarily preserve the $SU(2)$ symmetries of $S^2$, thus technically complicating the computation of the one-loop determinants in the
directions transverse to the $\cQ$-invariant field configurations. But as we argued in section~\ref{sec:local}, the result is largely insensitive to the choice of deformation, as long as it is non-degenerate
and  does not change the asymptotics of the potential in the space of fields. Therefore, we will instead use as the deformation term   the technically simpler, $SU(2)$ symmetric, vector multiplet and chiral
multiplet actions \(\delta_{\cQ}V=\cL_{\text{v.m.}} + \cL_{\text{c.m.}} + \cL_{\text{mass}}\). Contrarily to the canonical choice $\delta_{\cQ}V_{\text{can}}$, the saddle points of $\delta_{\cQ} V$ do not
coincide with the supersymmetric configurations and thus fully localize the path integral to the intersection.
For this choice of deformation, the effective FI parameter in \rf{effFI} vanishes $\xi_{\text{eff}}=0$.

It is straightforward to show that all Coulomb branch field configurations in  \(\cF_{\text{Coul}}\) are saddle points of  $\delta_{\cQ} V$ and must be integrated over. However, the solutions to the vortex and
anti-vortex equations we found at the poles are not saddle points of $\delta_{\cQ} V$.  This can be demonstrated using both the supersymmetry and the saddle point equations at the poles as
follows.\footnote{With some more effort it is possible to prove using only the equation of motion for $\rmD$ that the vortex and anti-vortex configurations are not saddle points of the action in the
limit in which the coefficient of the deformation term $\delta_{\cQ} V$ goes to infinity.} Since we are taking the masses to be non-degenerate, it follows from the   equations
\begin{equation}
(\sigma_2 + m_I) \phi_I = 0
\end{equation}
that any pair of distinct non-vanishing vectors $\phi_{I}$ and $\phi_{J}$ have to be independent. In addition, the above equation combined with the covariant constancy of $\sigma_2$ and its equation of motion imply
\begin{equation}
\label{eoms2eq0}
  \sum_I (q_I-1)\phi_I\bar{\phi}_I =0\,,
\end{equation}
while the equation of motion for $\rmD$ yields
\begin{equation}
\label{eomDeq0}
i\rmD-\sum_{I}\phi_{I}\bar{\phi}_{I}=0\,.
\end{equation}
However, since all non-vanishing $\phi_{I}$ are independent, we can conclude\footnote{This step requires us to assume that none of the $R$-charges is $1$.}  from \rf{eoms2eq0}  that $\phi_{I}\bar{\phi}_{I}$
vanishes for each $I$.  It therefore excludes the aforementioned supersymmetric solutions \rf{zeromodes} with fine-tuned values of $q$ from the set of saddle points. Combined with  \rf{eomDeq0}, it also sets
$\rmD=0$. Plugging this result in the supersymmetry equations fixes $F=-\sigma_{1}/r = B/2r^2$ and $\sigma_{2}=a$ and we recover the Coulomb branch field configurations spanning $\cF_{\text{Coul}}$, thus
eliminating the vortex and anti-vortex configurations.

The conclusion that the  path integral can be   written as a integral over just $\cF_{\text{Coul}}$ can also be derived  as follows. As we remarked earlier, the path integral does not depend on the choice of
supercharge $\cQ$ used in the localization computation. Therefore, we may instead try to localize the partition function with respect to the supercharges $Q_1$ and $Q_2$. This, however, requires  finding  a
deformation term which is $Q_1$ and $Q_2$ exact. Such a deformation term is provided by the following terms in the action
\begin{equation}
\cL_{\text{v.m.}} + \cL_{\text{c.m.}} + \cL_{\text{mass}}= \delta_{\epsilon_1} \delta_{\epsilon_2}V'\,,
\label{defnew}
\end{equation}
 with $V'=1/2 \Tr(\lambda \lambda) + \bar{\phi} F$, which are exact with respect to both supercharges since $[\delta_{\epsilon_1}, \delta_{\epsilon_2}]=0$.  In this approach the path integral localizes to the
 $Q_1$ and $Q_2$ invariant field configurations, which are the solutions to the equations
\begin{equation}
\label{susydeltadelta}
\begin{aligned}
\delta_{\epsilon_1} \lambda = \delta_{\epsilon_2} \lambda &= 0 &
\delta_{\epsilon_1} \psi = \delta_{\epsilon_2} \psi &=0 \\
\delta_{\epsilon_1} \bar{\lambda} = \delta_{\epsilon_2}\bar{\lambda} &= 0 \qquad &
\delta_{\epsilon_1} \bar{\psi} = \delta_{\epsilon_2}\bar{\psi} &= 0\,.
\end{aligned}
\end{equation}
These equations directly lead\footnote{Supersymmetry  implies that $V_1=V_2=V_3=\rmD=0$. The fact that the solutions to these equations are the Coulomb branch field configurations \rf{saddlept} follows
by using the equality of actions in \rf{vm_action} and \rf{vm_actionb}, derived  by integrating by parts. Non-trivial chiral multiplet configuration are manifestly non-supersymmetric.} to the Coulomb branch
field configurations \rf{saddlept} parametrizing  $\cF_{\text{Coul}}$ while immediately rendering the  vortex and anti-vortex configurations non-super\-symmetric. Note that this conclusion is reached by
considering the supersymmetry equations alone, contrary to   localization with respect to $\cQ$, where the saddle point equations of $\delta_{\cQ}V$ also  need to be invoked to show that vortex and
anti-vortex configurations do not contribute. Since the saddle points and deformation term \rf{defnew} are  precisely the same as the one for $\cQ$,
  this guarantees that we obtain  the same Coulomb branch representation of the path integral. A drawback of localizing with respect to $Q_1$ and $Q_2$ is
  that we cannot study the expectation value of the circular Wilson loop \rf{WilsonLoop}  since it is not $Q_1$ and $Q_2$ invariant.

In section~\ref{sec:instanton} we will obtain the payoff of using the supercharge $\cQ$. As we have shown in section~\ref{sec:local}, supersymmetry leads to the
  vortex and anti-vortex equations  at the poles. In that section, we will argue that localizing the path integral $\cQ$ in a different
  limit yields the Higgs branch representation of the partition function.

\subsection{Integral Representation of the Partition Function}

We   now can  write down the expression of the partition function as an integral over the Coulomb branch field configurations $\cF_{\text{Coul}}$.
The Coulomb branch representation of  the partition function is thus given by\footnote{The partition function has an anomalous dependence on the radius $r$ of the $S^2$ due to the conformal anomaly in two
dimensions. We do not retain this factor throughout our formulae, which  can be extracted from our one-loop determinants.}
\begin{equation}
Z_{\text{Coulomb}}(m,\tau)= \sum_{B} \int_{\tfrak} \rmd a\, Z_{\text{cl}}(a,B,\tau) \, Z_{\text{one-loop}}(a,B,m)\,,
\label{semicl}
\end{equation}
where the integral over \(a\) has been  reduced  to the Cartan subalgebra $\tfrak$ of \(G\).  The first factor arises from evaluating the renormalized gauge theory action on the smooth supersymmetric field
configurations \rf{saddlept}
\begin{equation}
Z_{\text{cl}}(a,B,\tau)= e^{-4\pi ir\xi_{\text{ren}} \Tr a+i\vartheta \Tr B}\,,
\end{equation}
 and the one-loop determinant $Z_{\text{one-loop}}(a,B,m)$ specifies the measure of integration over $a$, which is determined by the
 deformation term $\delta_{\cQ}V$.

Some care has been taken to ensure that the computation, including the regularization of the one-loop determinants $Z_{\text{one-loop}}(a,B,m)$, is $\cQ$-invariant. Even though the FI parameter $\xi$ is
classically marginal, it runs quantum mechanically according to the renormalization group equation
\begin{equation}
\xi(\mu)=\xi+{\frac{1}{2\pi}}\sum_{j} Q_j\ln\left(\frac{\mu}{M_{\text{UV}}} \right)={\frac{1}{2\pi}}\sum_{j} Q_j\ln\left(\frac{\mu}{\Lambda}\right) \,,
\label{running}
\end{equation}
where $Q_j$ is the charge of the $j$-th chiral multiplet under the $U(1)$ gauge group corresponding to $\xi$, $M_{\text{UV}}$ is the ultraviolet cutoff,  $\mu$ is the floating scale and $\Lambda$ is the
renormalization group invariant scale. A simple way of performing this renormalization in a  $\cQ$-invariant way, is to enrich the theory one is interested in with an ``expectator'' chiral multiplet of
mass $M$ and charge $-Q=-\sum_jQ_j$, so that in the enriched theory the FI parameter does not run. Now,  to extract the result for the theory of interest, we take the answer of the finite theory in the
limit where $M$ is very large, thereby decoupling the expectator chiral multiplet. This procedure results in a  $\cQ$-invariant ultraviolet cutoff $M$ for the theory under study. As shown in appendix~\ref{FIrunning}, taking $M$ large in the one-loop determinant  \rf{chiralloop} for the expectator chiral multiplet precisely reproduces the running of the FI parameter \rf{running} with $M_{\text{UV}}=M$
and $\mu=\varepsilon={1/r}$. That is,  the renormalized coupling obtained in this way is  evaluated at the inverse radius of the $S^2$, which is the infrared scale of $S^2$
\begin{equation}
\xi_{\text{ren}}\equiv\left.\xi\left(\mu={1/r}\right)\right|_{M_{\text{UV}}=M}=\xi + \frac{1}{2\pi} \sum_{i}Q_{i}\ln\left(\frac{\varepsilon}{M}\right)\,.
\end{equation}

The one-loop factor in the  localization computation $Z_{\text{one-loop}}(a,B,m)$ takes the form
\begin{equation}
  Z_{\text{one-loop}}(a,B,m) = Z^{\text{v.m.}}_{\text{one-loop}}(a,B) \cdot Z^{\text{c.m.}}_{\text{one-loop}}(a,B,m) \cdot \cJ(a,B) \,,
\end{equation}
where the Jacobian factor \(\cJ(a,B)\) accounts for the reduction of  the integral over all \(a\)  such that $[a,B]=0$ to an integral over the Cartan subalgebra $\tfrak$.  The magnetic flux \(B\) over the
\(S^2\) breaks the gauge symmetry \(G\) down to a subgroup \(H_B=\{g\in G \mid gBg^{-1}=B\}\). Therefore,  the associated Jacobian factor is
\begin{equation}
\cJ(a,B) = \frac{1}{|\cW(H_B)|} \prod_{\substack{\alpha\in\Delta^+\\ \alpha\cdot B  = 0}} (\alpha\cdot a)^2\,,
\end{equation}
where $\alpha\in\Delta^+$ are positive roots of the Lie algebra of $G$ and $|\cW(H_B)|$ is the order of the Weyl group of $H_B$.

The one-loop determinants
for  our choice of deformation term $\delta_{\cQ}V$,  which is  the sum of \rf{defvm} and \rf{defcm},  are computed in appendix~\ref{app:one-loop}. For a chiral multiplet
in a reducible representation ${\bf R}=\oplus_I {\bf r}_I$ we obtain
\begin{align}
  Z_{\text{one-loop}}^{\text{c.m.}}(a,B,m)
  & = \prod_{I}\prod_{w_I\in{\bf r}_I}
  \changed{10}
  (-i)^{w_I\cdot B} (-1)^{|w_I\cdot B|/2}
  \to
  (-1)^{(w_I\cdot B+|w_I\cdot B|)/2}
  \done
  \frac {\Gamma\left(\frac{q_I}{2} -ir({w_I}\cdot a +m_{I}) + \frac{|{w_I}\cdot B|}{2}\right)} {\Gamma\left(1 - \frac{q_I}{2}
      + ir({w_I}\cdot a+m_{I}) + \frac{|{w_I}\cdot B|}{2}\right)} ,
  \label{chiralloop}
\end{align}
 where
\(w_I\) are the weights of the representation \({\bf r}_I\) and \(\Gamma(x)\) is the Euler gamma function.  The twisted masses and $R$-charges $m_{I}$ and $q_I$ of the chiral multiplets, which take values
in the Cartan subalgebra of the flavour symmetry $G_F$, combine into the holomorphic combination \(M = m + \frac{i}{2r} q\) introduced in \rf{holomm}.

For the vector multiplet contribution we obtain
\begin{align}
 Z^{\text{v.m.}}_{\text{one-loop}}(a,B)
  & =\prod_{\substack{\alpha\in\Delta^+\\ \alpha\cdot B\neq 0 }}
  \changed{20}
  \left[\left(\frac{\alpha\cdot B}{2r}\right)^{2} +\left(\alpha\cdot a\right)^2\right]
  \to
  \left[(-1)^{\alpha\cdot B}\left(\left(\frac{\alpha\cdot B}{2r}\right)^{2} +\left(\alpha\cdot a\right)^2\right)\right]
  \done
  \,.
\end{align}
We note that the Jacobian factor and the vector multiplet determinant combine nicely into an unconstrained product over the positive roots of the Lie algebra
\begin{equation}
  Z^{\text{v.m.}}_{\text{one-loop}}(a,B) \cdot J(a,B)
  =\frac{1}{|\cW(H_B)|}\prod_{\alpha\in\Delta^+}
  \changed{20}
  \left[\left(\frac{\alpha\cdot B}{2r}\right)^{2} +\left(\alpha\cdot a\right)^2 \right]
  \to
  \left[(-1)^{\alpha\cdot B}\left(\left(\frac{\alpha\cdot B}{2r}\right)^{2} +\left(\alpha\cdot a\right)^2 \right)\right]
  \done
  \,.
\end{equation}
\changed{-10}(paragraph added)
\to
The sign is $\prod_{\alpha\in\Delta^+} (-1)^{\alpha\cdot B}=e^{2\pi i\rho\cdot B}$ in terms of the half-sum of positive roots~$\rho$.\footnote{The sign $(-1)^{2\rho\cdot B}$, pointed out in~\cite{Hori:2013ika,Honda:2013uca}, was missing in an earlier version of the paper.  For simply-connected gauge groups the sign is trivial: for such groups $\rho$~is an integer weight and GNO quantization of~$B$ ensures then that $2\rho\cdot B\in 2\mathbb{Z}$ rather than~$\mathbb{Z}$.  For $U(N)$ gauge group the sign is equivalent to $\vartheta\to\vartheta+(N-1)\pi$.}
\done

The Coulomb branch representation of  the partition function of an $\cN=(2,2)$ gauge theory on $S^2$ is thus given by
\begin{align}
  Z_{\text{Coulomb}}(m,\tau)
  & = \sum_{B}\frac{1}{|\cW(H_B)|} \int_{\tfrak} \rmd a\,  e^{-4\pi i \xi_{\text{ren}} r \Tr a+i\vartheta \Tr B
\changed{10}+0
\to
+2\pi i\rho\cdot B
\done
}
    \prod_{\alpha\in\Delta^+} \left[\left(\frac{\alpha\cdot B}{2r}\right)^{2} +\left(\alpha\cdot a\right)^2 \right]
  \nonumber\\
  & \qquad
    \times\prod_{I, w_I} \left[
    \changed{10}
    (-i)^{w_I\cdot B} (-1)^{|w_I\cdot B|/2}
    \to
    (-1)^{(w_I\cdot B+|w_I\cdot B|)/2}
    \done
    \frac {\Gamma\left(-ir({w_I}\cdot a +\rmM_{I}) + \frac{|w_I\cdot B|}{2}\right)} {\Gamma\left(1  + ir({w_I}\cdot a+\rmM_{I}) +
    \frac{|w_I\cdot B|}{2}\right)} \right] \, .
  \label{semicolumb}
\end{align}
The expectation value of the circular Wilson loop \rf{WilsonLoop} is obtained by enriching the integrand in \rf{semicolumb} with the  insertion of
\begin{equation}
 \Tr e^{2\pi a -i\pi B}\,.
 \label{insertlooop}
\end{equation}

\subsection{Factorization of the Partition Function}
\label{sec:factorization}

We show in this subsection that the Coulomb branch representation of the partition function \rf{semicolumb} can be written as a discrete sum, whose summand factorizes into the product of two functions.
A related factorization was  found previously by Pasquetti \cite{Pasquetti:2011fj} when evaluating the partition function of three dimensional  $\cN=2$ abelian gauge theories on the squashed
$S^3$.\footnote{The partition function of   three dimensional gauge theories on $S^2\times S^1$ can also be factorized \cite{pasquetti}.}

We recognize the expression we obtain as the sum over Higgs vacua of the product of the vortex partition function due to vortices at the north pole  with  the  anti-vortex partition function due to
the anti-vortices at the south pole. This result is interpreted  in section~\ref{sec:instanton} as a direct path integral evaluation  of the partition function, where the path integral is argued to
localize on vortices and anti-vortices in the Higgs branch.

Let us consider for definiteness the case of two dimensional $\cN=(2,2)$ SQCD. This theory has  \(G = U(N)\) gauge group and  \(N_F\) fundamental chiral multiplets and \(\antiNF\) anti-fundamental
chiral multiplets.  The partition function \rf{semicolumb} of this theory is\footnote{Without loss of generality we set $r=1$ to unclutter formulas. It can easily be restored by dimensional analysis.}
\begin{equation}
  \label{ZUN-Coulomb}
  \begin{aligned}
    & \! Z_{\text{SQCD}}^{U(N)} 
    = \frac{1}{N!} \! \sum_{B_1, \ldots, B_N \in \mathbb{Z}} \int \! \rmd a_1 \cdots \rmd a_N \Bigg\{ e^{-4\pi i \xi  \Tr a} e^{i
\changed{10}
\vartheta
\to
(\vartheta+(N-1)\pi)
\done
\Tr B} \prod_{i < j} \Biggl[ (a_i - a_j)^2 + \biggl(\frac{B_i - B_j}{2}\biggr)^2 \Biggr]
    \\
    & \qquad \cdot \prod_{s=1}^{N_F} \prod_{i=1}^{N} \frac{(-1)^{\frac{|B_i|+B_i}{2}} \Gamma(-ia_i - i\rmM_s + |B_i| / 2)} {\Gamma(1 + ia_i + i\rmM_s + |B_i| / 2)}
      \prod_{s=1}^{\antiNF} \prod_{i=1}^{N} \frac{(-1)^{\frac{|B_i|-B_i}{2}} \Gamma(ia_i - i\widetilde{\rmM}_s + |B_i| / 2)} {\Gamma(1 - ia_i + i\widetilde{\rmM}_s + |B_i| / 2)} \Bigg\} \, .
  \end{aligned}
\end{equation}
In the large $a$ limit, the integrand is of order \(|a|^{N(N-1) + N \sum_I (q_I - 1)}\), hence this \(N\)-dimensional integral is convergent as long as
\begin{equation}
  \label{U(N)-q-condition}
  \sum_{s = 1}^{N_F} q_s + \sum_{s = 1}^{\antiNF} \widetilde{q}_s < N_F + \antiNF - N \, .
\end{equation}
In the cases where \(N_F > \antiNF\), or \(N_F = \antiNF\) and \(\xi > 0\), the contour can be closed towards \(ia_i \to + \infty\), enclosing poles of the fundamental multiplets' one-loop determinants; the contour must be chosen to enclose poles of the anti-fundamental multiplets' one-loop determinants in cases where \(N_F < \antiNF\), or \(N_F = \antiNF\) and \(\xi < 0\).
Assuming that all $R$-charges are positive, or deforming the integration contour to ensure that we enclose the same set of poles, this expresses the Coulomb branch integral as a sum of the residues at combined poles
\begin{equation}
  ia_i = -i\rmM_{p_i} + n_i + \frac{|B_i|}{2} \quad \text{for all \(1\leq i\leq N\),}
\end{equation}
with \(1\leq p_1, \ldots, p_N\leq N_F\) and \(n_1, \ldots, n_N \geq 0\) labelling the poles.  The resulting ratios of Gamma functions in the integrand can be recast in terms of Pochhammer raising factorials
\((x)_n = x(x+1) \cdots (x+n-1)\) as
\begin{equation} \label{eq:GammaPoch}
  \frac{\Gamma(i\rmM_{p_i} -i\rmM_s - n_i)} {\Gamma(1 + i\rmM_s-i\rmM_{p_i} + |B_i| + n_i)}
  = \frac{\gamma(i\rmM_{p_i} -i\rmM_s) (-1)^{n_i}}{(1+i\rmM_s-i\rmM_{p_i})_{n_i} (1+i\rmM_s-i\rmM_{p_i})_{n_i + |B_i|}} \, ,
\end{equation}
where
\begin{equation}
  \gamma(x) = \frac{\Gamma(x)}{\Gamma(1 - x)} \,,
\end{equation}
and similarly for the ratios of Gamma functions coming from the anti-fundamental chiral multiplets.

The symmetry between \(n_i\) and \(n_i + |B_i|\) in \rf{eq:GammaPoch} leads us to introduce new coordinates
\begin{equation}
  k^\pm_i = n_i + [B_i]^\pm = n_i + |B_i| / 2 \pm B_i / 2 \geq 0
\end{equation}
on the summation lattice, such that \(\{n_i, n_i+|B_i|\} = \{k^\pm_i\}\).  In section~\ref{sec:instanton}, the \(N\) integers \(k^+_i\) will be   interpreted as labelling vortices located at the north pole,
and \(k^-_i\) anti-vortices at the south pole.  More precisely,   \(k^\pm_i\) measures  the amount of vortex and anti-vortex charge carried by the $i$-th Cartan generator in $U(N)$: note that the flux \(B_i = k^+_i - k^-_i\).

This change of coordinates decouples the sums over \(k^+\geq 0\) and \(k^-\geq 0\) and yields the following expression after converting signs to a shift in the theta angle
\begin{equation}
  \label{Coulomb-factorized}
  \! \begin{aligned}
    & Z_{\text{SQCD}}^{U(N)} 
    = \frac{(2\pi)^N}{N!} \sum_{p_1,\ldots,p_N=1}^{N_F} \Bigg[ e^{4\pi \xi \sum_j i \rmM_{p_j}}
    \prod_{i=1}^{N} \frac{ \prod_{s=1}^{\antiNF} \gamma( -i\widetilde{\rmM}_s -i\rmM_{p_i}) } { \prod_{s\neq p_i}^{N_F} \gamma(1 + i\rmM_s -i\rmM_{p_i}) }
    \\
    & \times \!
    \sum_{k^+_i \geq 0} \! \Bigg[ e^{(2\pi i\tau+i\pi
\changed{10}
N_F
\to
(N_F+N-1)
\done
) \sum_i k^+_i} \prod^N_{i < j} \left( \rmM_{p_j} - \rmM_{p_i} + ik^+_j -ik^+_i\right)
      \prod_{i=1}^{N} \frac{\prod_{s=1}^{\antiNF} ( -i\widetilde{\rmM}_s -i\rmM_{p_i} )_{k^+_i}} { \prod_{s=1}^{N_F} (1 + i\rmM_s-i\rmM_{p_i})_{k^+_i}} \Bigg]
    \\
    & \times \!
    \sum_{k^-_i \geq 0} \! \Bigg[ e^{(-2\pi i\bar{\tau}+i\pi
\changed{10}
\antiNF
\to
(\antiNF+N-1)
\done
) \sum_i k^-_i} \prod^N_{i < j} \left( \rmM_{p_j} - \rmM_{p_i} + ik^-_j -ik^-_i\right)
      \prod_{i=1}^{N} \frac{\prod_{s=1}^{\antiNF} ( -i\widetilde{\rmM}_s -i\rmM_{p_i} )_{k^-_i} } { \prod_{s=1}^{N_F} (1 + i\rmM_s-i\rmM_{p_i})_{k^-_i} } \Bigg] \Bigg] .
  \end{aligned}
\end{equation}
Terms with \(p_a = p_b\) for some \(a \neq b \leq N\) vanish, because the sum over \(k^+\) is then antisymmetric under the exchange of \(k^+_a\) and \(k^+_b\).  We can thus normalize the series as
\begin{align}
  f(\{p_i\},\rmM,z)
  &= \sum_{k_i \geq 0} \Bigg[ z^{\sum_i k_i} \prod^N_{i < j} \frac{i\rmM_{p_j} -i\rmM_{p_i} + k_i - k_j}{i\rmM_{p_j} -i\rmM_{p_i}}
    \frac{\prod_{s=1}^{\antiNF} \prod_{i=1}^{N} ( -i\widetilde{\rmM}_s -i\rmM_{p_i} )_{k_i}} { \prod_{s=1}^{N_F} \prod_{i=1}^{N} (1 + i\rmM_s-i\rmM_{p_i})_{k_i}} \Bigg]
  \nonumber\\
  &= \sum_{k_i \geq 0} \Bigg[ \frac{z^{\sum_i k_i}}{\prod_i k_i!}
    \frac{\prod_{s=1}^{\antiNF} \prod_{i=1}^{N} ( -i\rmM_{p_i} -i\widetilde{\rmM}_s )_{k_i}} {\prod^N_{i \neq j}(i\rmM_{p_j} -i\rmM_{p_i} - k_j)_{k_i} \prod_{s\not\in\{p\}}^{N_F}
	\prod_{i=1}^{N} (1 + i\rmM_s -i\rmM_{p_i})_{k_i}} \Bigg] \,,
  \label{fpi}
\end{align}
which as we will see in the next section, corresponds to the vortex partition function
studied in \cite{Shadchin:2006yz}, with $z=\exp(2\pi i \tau)$
\changed{8}()
\to
(up to a sign)
\done
playing the role of the vortex fugacity.
Note that this series converges for all \(z\) (all \(\xi\)) if \(N_F > \antiNF\), and for \(|z| < 1\) (that is, \(\xi > 0\)) if \(N_F = \antiNF\), consistent with the constraints required by our choice of contour.
All in all, the partition function factorizes as
\begin{equation}
  Z_{\text{SQCD}}^{U(N)} 
  = 
  \sum_{\mathclap{\substack{v_i = - \rmM_{p_i}\\1 \leq p_1< \ldots < p_N\leq N_F}}}
    Z_{\text{cl}} (v, 0, \tau)
    \, \res_{a = v} Z_{\text{one-loop}} (a, 0, \rmM)
    \, f(\{p_i\}, \rmM, (-1)^{
\changed{10}
N_F
\to
N_F+N-1
\done
} z)
    \, f(\{p_i\}, \rmM, (-1)^{
\changed{10}
\antiNF
\to
\antiNF+N-1
\done
} \bar z)
\end{equation}
up to a constant factor, with
\begin{equation}
 \res_{a_i = - \rmM_{p_i}} Z_{\text{one-loop}} (a, 0, \rmM)
 =
 \prod_{i=1}^{N}
 \frac{ \prod_{s=1}^{\antiNF} \gamma( -i\widetilde{\rmM}_s -i\rmM_{p_i})}{ \prod_{s\not\in\{p\}}^{N_F} \gamma(1 + i\rmM_s -i\rmM_{p_i}) }\,.
\end{equation}
In the next section we obtain this result directly by localizing the path integral to Higgs branch configurations with vortices and anti-vortices.
In the matching, some care must be taken when comparing  the mass parameters of the gauge theory on the sphere with the parameters describing the theory in the
$\Omega$-background used to evaluate the vortex partition function.

The final expression we find is  reminiscent of the discrete sums   of the product of holomorphic and anti-holomorphic conformal blocks that appear in correlators of the \(A_{N_F-1}\) Toda CFT in the presence of
completely degenerate fields. A precise matching between the partition function of $\cN=(2,2)$ gauge theories on $S^2$ and correlators in Toda is provided in the abelian case in section~\ref{sec:Toda}, and in the
case of \(U(N)\) in \cite{Gomis:2014eya}.

Note that this factorization result applies to any gauge group G with an abelian factor and any matter representation ${\bf R}$, as shown in appendix~\ref{app:factorization}. This yields a representation of the path integral that can be interpreted as a sum over Higgs vacua of terms factorized into holomorphic and anti-holomorphic contributions, corresponding to vortices and anti-vortices respectively. These formulas motivate natural conjectures for the vortex partition functions corresponding to gauge theories with gauge group G. In the absence of $U(1)$ factors in the gauge group, the factorization can be carried out formally, but the two factors may be divergent series.

%
%

\section{Higgs Branch Representation}
\label{sec:instanton}

The localization principle, under mild conditions, guarantees that the path integral does not depend either on the choice of supercharge $\cQ$ or on the choice of $V$ in the deformation term. But  different
choices can lead to different representations of the same path integral and therefore to non-trivial identities.

In section~\ref{sec:classical} we have derived a representation of the partition function as an integral  over Coulomb branch vacua.  In section~\ref{sec:factorization}, by explicitly evaluating the integral,
we have demonstrated that the partition function also has an alternative representation as a sum -- in the Higgs phase --  over vortex and anti-vortex field configurations localized at the poles.

This section aims to derive from path integral localization arguments the Higgs branch representation of the partition function.
This representation  should have a direct derivation using localization. The appropriate choice of supercharge to use to obtain this representation  is the same supercharge $\cQ$   introduced in  \rf{choice},
since it has the elegant feature of giving rise to the vortex equations at the north pole
  \begin{equation}
  \label{NPBPS}
  \begin{aligned}
    \left( D_{\hat{1}} + iD_{\hat{2}} \right) \phi&=0 & \qquad \qquad D_{\hat{i}}\left(\sigma_1+i\sigma_2\right)&= 0\\
    F_{\hat{1}\hat{2}}  + \sigma_1 + g_{\text{eff}}^{2} (\phi \bar{\phi} - \xi_{\text{eff}} \mathbbm{1} ) &= 0  & \operatorname{Re}\rmD = [\sigma_1,\sigma_2]&=0\\
    \left(  \sigma_{1} + \frac{q}{2} \right) \phi &=0 & \left(\sigma_{2}+m\right)\phi &= 0\,,
  \end{aligned}
  \end{equation}
  and anti-vortex equations at the south pole
  \begin{equation}
  \label{SPBPS}
  \begin{aligned}
    \left( D_{\hat{1}} -iD_{\hat{2}} \right) \phi&=0 & \qquad \qquad D_{\hat{i}}\left(\sigma_1+i\sigma_2\right)&= 0\\
    F_{\hat{1}\hat{2}}  + \sigma_1 - g_{\text{eff}}^{2} (\phi \bar{\phi} - \xi_{\text{eff}} \mathbbm{1} ) &= 0  & \operatorname{Re}\rmD =[\sigma_1,\sigma_2]&=0\\
    \left(  \sigma_{1} - \frac{q}{2} \right) \phi &=0 & \left(\sigma_{2}+m\right)\phi &= 0\,.
  \end{aligned}
  \end{equation}
We remark that when the effective Fayet-Iliopoulos parameters are non-vanishing, these equations admit solutions with non-vanishing $\phi$. These solutions then restrict $\sigma_{2}$ to be a diagonal matrix
with the masses of the excited chiral fields on the diagonal and the Coulomb branch configurations \rf{saddlept} parametrizing  \(\cF_{\text{Coul}}\) are lifted.
The $\cQ$-invariant field configurations admitted by \rf{NPBPS} and \rf{SPBPS} are vortex and anti-vortex configurations
at the north and south pole of the $S^2$. Since vortices and anti-vortices exist in the Higgs phase, we denote this space of supersymmetric field configurations that must be integrated over by  \(\cF_{\text{Higgs}}\).

\subsection{Localizing onto the Higgs Branch}

In this subsection we present a heuristic argument to introduce non-zero FI parameters in the localization computation, which as explained above yields to a representation of the path integral as a sum over vortex and
anti-vortex configurations. For the purpose of this argument, we take all the $R$-charges to be zero.

Recall that our choice of deformation term $\delta_{\cQ}V = \cL_{\text{v.m.}}+\cL_{\text{c.m.}}+\cL_{\text{mass}}$ does not include a FI term.
In section~\ref{sec:classical}, we performed the saddle point approximation after taking the $t \to \infty$ limit.
In this limit, the effective FI parameter vanishes $\xi_\text{eff}=0$  and the saddle point equations forbid vortices,
hence the path integral localizes to $\cF_{\text{Coul}}$.
Instead, we assume here that there is another choice of $\cQ$-exact deformation terms $\cQ V'$ leading to
a non-vanishing effective FI parameter $\xi_\text{eff}\neq0$ in the $t \to \infty$ limit\footnote{See \cite{Benini:2012ui} for a choice of $V'$.}.

The equation of motion for the $\rmD$ field arising from the deformed action $S + t \delta_{\cQ} V'$ is
\begin{equation}
    \label{irmDeom}
    ig_{\text{eff}}^{-2}\rmD + \xi_{\text{eff}} - \sum_I \phi_I\bar{\phi}_I = 0.
\end{equation}
On the space of $\cQ$-supersymmetric field configurations (see section~\ref{sec:solutions}), $\rmD$ vanishes in the bulk and we conclude that
\begin{equation}
\label{phibphieqxi}
  \sum_I \phi_I \bar{\phi}_I = \xi_{\text{eff}} \mathbbm{1}_N\,,
\end{equation}
which, together with $(a + m_I) \phi_I = 0$ imply that the Coulomb branch is lifted, localizing instead to the Higgs branch. Moreover the supersymmetry equations at the poles yield
\begin{equation}
\sigma_1 \phi_{I}\bar{\phi}_{I} = -\frac{B}{2} \phi_{I}\bar{\phi}_{I} = 0
\end{equation}
which by virtue of \rf{phibphieqxi} imply $B=\sigma_1=0$. This leads us directly to the vortex and anti-vortex equations at the north and the south poles.

The contribution of vortices and anti-vortices to the partition function of an $\cN=(2,2)$ gauge theory on $S^2$ can be obtained as follows.
Since the vortices and anti-vortices are localized at the poles, these can be studied by restricting the  $\cN=(2,2)$ gauge theory  to the local  $\bR^2$ flat space near the north and south poles of $S^2$.
Asymptotic infinity of each $\bR^2$ is identified with a small latitude  circle on $S^2$ close to the north and south pole respectively. Therefore, the contribution of   vortices and anti-vortices is captured
by the vortex/anti-vortex partition function of the gauge theory obtained by restricting our  $\cN=(2,2)$ gauge theory  at the poles.
As we will see in section~\ref{sec:ZHiggs}, integrating over vortex and anti-vortex configurations for all Higgs branch vacua exactly reproduces the partition function computed by integrating over the Coulomb branch
found in section~\ref{sec:factorization}.


\subsection{Vortex Partition Function}
\label{sec:ZHiggs}

Following the discussion in the last subsection, in the planes glued to the poles and in the presence of the FI parameter, the supersymmetry equations reduce to
\begin{equation}
\left( D_{{1}} + iD_{{2}} \right) \phi_{I}=0\,,\qquad (\sigma_2+m_{I})\phi_{I}=0 \,,\qquad F_{{1}{2}}  + \sum_{I}\phi_{I}\bar{\phi}_{I}-\xi_{\text{eff}} = 0\,,
\label{vortexnormal}
\end{equation}
in the plane attached to the north pole, and
\begin{equation}
\left( D_{{1}} -iD_{{2}} \right) \phi_{I}=0\,,\qquad (\sigma_2+m_{I})\phi_{I}=0 \,,\qquad F_{{1}{2}}  - \sum_{I}\phi_{I}\bar{\phi}_{I}+\xi_{\text{eff}} = 0\,,
\label{vortexnormal2}
\end{equation}
in the copy of $\bR^2$ attached to the south pole. These equations can be recognized as the differential equations  describing supersymmetric vortices and anti-vortices in $\cN=(2,2)$ supersymmetric gauge theories.
Therefore, in our localization computation we must integrate over the moduli space of solutions of vortices at the north pole and anti-vortices at the south pole. For simplicity, we discuss their contribution to the
partition function for $\cN=(2,2)$ SQCD with \(U(N)\) gauge group and  \(N_F\) fundamental chiral multiplets and \(\antiNF\) anti-fundamental chiral multiplets.

Since the vortices and anti-vortices exist only in the Higgs phase, let us first work out the vacuum structure
in the Higgs phase.  We first note that vortices can only exist in vacua in which the anti-fundamental fields vanish.
This follows from the known mathematical result that the vortex equations for an anti-fundamental field have no non-zero smooth solution when the background field is a connection of a bundle with positive first
Chern class $c_1=k>0$.  The vortex equations \rf{vortexnormal} and \rf{vortexnormal2} then imply that exactly \(N\) chiral multiplets take non-zero values, and diagonalizing $\sigma_2=\operatorname{diag}(a_1,\cdots,a_N)$,
one obtains that each Higgs branch of solutions to these equations is labelled by a set of distinct integers $1\leq p_1<\cdots<p_N\leq N_F$, with
\begin{equation}
a_i + m_{p_i} = 0 \qquad i=1,\ldots,N\,,
\label{fixmass}
\end{equation}
up to permutations of integers $p_i$. The contribution from vortices and anti-vortices depends on the choice of Higgs branch components. In each of these   components, the $U(N)\times S[U(N_F)\times U(\antiNF)]$
symmetry of the theory is broken to
\begin{equation}
 S[U(N)_{\text{diag}}\times U(N_F-N)] \times U(1) \times  SU(\antiNF)\,,
\end{equation}
where $U(1)$ rotates   fundamental and anti-fundamental chiral multiplets equally.

For a given Higgs branch component labeled by $\{p_i\}$, the  familiar vortex equations \rf{vortexnormal} admit a multidimensional moduli space of solutions which we denote by $\cM_{\text{vortex}}^{\{p_i\}}$.
Since the vorticity 
\begin{equation}
k=\frac{1}{2\pi}\int_{\bR^2}\operatorname{Tr}F
\end{equation}
is quantized, this moduli space splits into disconnected components $\cM_{\text{vortex}}^{\{p_i\}, k}$, each of which is a K\"ahler manifold, of dimension $2kN_F$. Taking into account the south pole anti-vortex
contributions, we find that the  solutions of the localization equations on $S^2$ span the moduli space
\begin{equation}
\cF_{\text{Higgs}}=\bigsqcup_{\{p_i\}} \ \left[ \cup_{k=0}^\infty \cM_{\text{vortex}}^{\{p_i\},k} \right] \oplus  \left[\cup_{l=0}^\infty \cM_{\text{anti-vortex}}^{\{p_i\},l}\right]\,.
\end{equation}

We now argue that the vortex partition function at the poles is captured by the partition function of the $\cN=(2,2)$ gauge theory in the $\Omega$-background, which is a supersymmetric deformation of the
$\cN=(2,2)$ gauge theory in $\bR^2$ by a  $U(1)_\varepsilon$ equivariant rotation parameter~$\varepsilon$.  Let us recall that the supercharge with which we localize an $\cN=(2,2)$ gauge theory on $S^2$ obeys
\begin{equation}
\cQ^2=J+\frac{1}{2}R\,.
\label{SUSY1}
\end{equation}
The key observation is to note that \rf{SUSY1} is precisely the supersymmetry  preserved by an $\cN=(2,2)$ gauge theory in $\bR^2$ when placed in the $\Omega$-background. The rotation generator in the
$\Omega$-background  corresponds to $J+\frac{1}{2}R$, thus giving rise to the scalar supercharge  under $U(1)_\varepsilon$  preserved by an  $\cN=(2,2)$   theory   in   the $\Omega$-background.
Therefore, the contribution to the partition function of an $\cN=(2,2)$ gauge theory on $S^2$  due to vortices and anti-vortices  localized at the   poles  is captured by the vortex/anti-vortex  partition function of
the same gauge theory placed in the $\Omega$-background originally studied by Shadchin \cite{Shadchin:2006yz} (see also \cite{Yoshida:2011au,Bonelli:2011fq,Miyake:2011yr,Fujimori:2012ab,Kim:2012uz}).

The vortex partition function in the Higgs branch component $\{p_i\}$ of an $\cN=(2,2)$ gauge theory in the $\Omega$-background is obtained by performing the functional integral of that theory around  the background
field configuration of $k$ vortices, and summing over all $k$. It admits an expansion
\begin{equation}
Z_{\text{vortex}}(\{p_i\}, M^\Omega, \widetilde{M}^\Omega, z_\Omega) = \sum_{k=0}^{\infty} z_\Omega^k Z_{k}(\{p_i\}, M^\Omega, \widetilde{M}^\Omega) \,,
\end{equation}
where  $z_\Omega=\exp(2\pi i \tau_\Omega)$ is the vortex fugacity and $Z_{k}(\{p_i\}, M^\Omega, \widetilde{M}^\Omega)$ is the  equivariant volume of the moduli space of $k$ vortices. The volume is given by
\begin{equation}
Z_{k}(\{p_i\}, M^\Omega, \widetilde{M}^\Omega) = \int_{\cM_{\text{vortex}}^{\{p_i\},k}} e^{\hat\omega}\,,
\label{volume}
\end{equation}
where $\hat\omega$ is the $U(1)_\varepsilon$  equivariant closed K\"ahler form\footnote{The form $\hat\omega$ is also equivariant under the action of the residual symmetry of the vacuum over which vortices are considered.
See \rf{symmm}.}  on $\cM_{\text{vortex}}^{\{p_i\},k}$.  Our computations of the supersymmetry transformations on $S^2$ in section~\ref{sec:supercharge} imply that the equivariant rotation parameter $\varepsilon$ for the
$\Omega$-background theory induced at the poles is given in terms of the radius of the $S^2$ by
\begin{equation}
\varepsilon=\frac{1}{r}\,.
\end{equation}
It is pleasing that the $\cN=(2,2)$ theory near the poles yields the $\Omega$-deformed theory, since the integral \rf{volume} for the $\cN=(2,2)$ theory in flat space suffers from ambiguities, such as infrared divergences.
Fortunately, a closer inspection of the $\cN=(2,2)$ gauge theory on $S^2$ near the  poles cures this problem, yielding finite, unambiguous results.
In fact, the $\Omega$-deformation was first introduced to regularize otherwise infrared divergent volume integrals such as \rf{volume}.

The vortex partition function of an $\cN=(2,2)$ gauge theory in the $\Omega$-background can be computed from the knowledge of the symplectic quotient construction of the vortex moduli space
$\cM_{\text{vortex}}^{\{p_{i}\},k}$ given in \cite{Hanany:2003hp,Eto:2005yh}. Some details of this construction are presented in appendix~\ref{vortexmatrix}. The volume \rf{volume}
is then given by the matrix integral of a  supersymmetric matrix theory action with $U(k)$ gauge group.  This matrix theory can be obtained by dimensionally reducing a certain two dimensional
$\cN=(0,2)$ $U(k)$ gauge theory to zero dimensions.  This supersymmetric matrix theory inherits the supercharge $\cQ$ of the $\cN=(2,2)$ theory in the $\Omega$-background as well as an equivariant
\begin{equation}
U(1)_\varepsilon \times S[U(N)_{\text{diag}}\times U(N_F-N)] \times U(1) \times SU(\antiNF)
\label{symmm}
\end{equation}
symmetry. The first factor $U(1)_\varepsilon$ is the rotational symmetry of the  $\Omega$-background  while  the rest is the residual symmetry of the vacuum over which vortices are studied.
The integral \rf{volume} receives contributions from isolated points in the vortex moduli space $\cM_{\text{vortex}}^{\{p_i\},k}$, corresponding to the $\cQ$-invariant configurations.
These are labeled by a partition of $k$ into $N$ non-negative integers
\begin{equation}
k = \sum_{i = 1}^N k_i\,.
\label{partition}
\end{equation}
To each such partition we associate an $N$-component vector $\vec{k}=(k_1,\ldots, k_N)$, describing how the total vortex number $k$ is distributed among the $N$ Cartan generators in $U(N)$ at this point.

For the choice of Higgs branch component of the $\cN=(2,2)$ gauge theory labelled by integers \(\{p_i\} \subseteq \{1,\ldots, N_F\}\),
the partition function of $k$-vortices admits the following contour integral
representation \cite{Shadchin:2006yz,Dimofte:2010tz} (see appendix~\ref{vortexmatrix} for details),
\begin{align}\label{contour1}
  Z_{k}(\{p_i\},M^{\Omega},\widetilde{M}^{\Omega}) = \oint_{\Gamma_{\{p_i\},k}} \prod_{I=1}^{k} \frac{d\varphi_I}{2\pi i}
  \ \cZ_{\text{vec}} (\varphi) \cdot \cZ_{\text{fund}}(M^{\Omega}, \varphi)
  \cdot \cZ_{\text{anti-fund}}(\widetilde{M}^{\Omega}, \varphi)
\end{align}
with
\begin{align}
  \cZ_{\text{vec}}(\varphi) & = \frac{1}{k! \ \vare^{k}}
  \prod_{I\neq J}^{k} \frac{\varphi_I - \varphi_J}{\varphi_I -\varphi_J - \vare}
  \\
  \cZ_{\text{fund}}(M^{\Omega},\varphi) & = \prod_{I=1}^k \prod_{s=1}^{N_F} \frac{1}{\varphi_I - M_{s}^{\Omega} }
  \\
  \cZ_{\text{anti-fund}}(\widetilde{M}^{\Omega},\varphi) & = \prod_{I=1}^k \prod_{t=1}^{\antiNF} \left(\varphi_I + \widetilde{M}_{t}^{\Omega}\right)\,.
\end{align}
For each Higgs vacuum $\{p_{i}\}$ and vorticity $\vec{k}$, the integrand in \rf{contour1} admits a pole at
\begin{align}
  \varphi_{(i,l)} = M_{p_i}^{\Omega} + (l-1) \vare \qquad l=1,2,..,k_i\,\qquad i=1,\ldots,N \, ,
\end{align}
and the contour of integration $\Gamma_{\{p_{i}\},k}$ is carefully chosen to enclose all such poles for \(\sum_{i=1}^{N} k_i = k\), and no other.  The poles of \rf{contour1} can be understood as the location of
the fixed points under the action of $\cQ$.
Each factor in \rf{contour1} reflects the contribution of the vortex collective coordinates associated to each of the $\cN=(2,2)$ multiplets: the vector multiplet and fundamental and anti-fundamental chiral multiplets.
Note here that the mass parameters in the $\Omega$-background theory can be identified with the mass parameters of the theory on $S^2$,
\begin{align}
  M_{p_i}^{\Omega} =  -i m_{p_i}
  \,, \qquad M_{s}^{\Omega} = - \vare -i m_s \ (s \not\in \{p_i\})
  \,, \qquad \widetilde{M}_{s}^{\Omega} = -i \widetilde{m}_s
  \,.
\end{align}
We observe the same shift in masses as for $\cN=2$ gauge theories on $S^4$ found in \cite{Okuda:2010ke}.
Performing the contour integral and summing over all vortex charges $\vec k$, the vortex partition function  for SQCD  takes the following form
\begin{align}
  Z_{\text{vortex}}(\{p_i\},m,\widetilde{m},z) = \smash{\sum_{k_1+\cdots+k_N=k}} z^{|\vec k|} Z_{\vec k}(\{p_i\},m,\widetilde{m})\,,
\end{align}
with
\begin{align}\label{Zvortexk}
  Z_{\vec k}(\{p_i\},m,\widetilde{m})
  = \frac{1}{\prod_i k_i!} \frac{\prod_{s=1}^{\antiNF} \prod_{i=1}^{N} ( -irm_{p_i} -ir\widetilde{m}_s )_{k_i}} {\prod_{i \neq j}(irm_{p_j} -irm_{p_i} - k_j)_{k_i}
    \prod_{s\not\in\{p\}}^{N_F} \prod_{i=1}^{N} (1 + irm_s -irm_{p_i})_{k_i}} \, .
\end{align}
This expression exactly agrees\footnote{One must analytically continue the twisted masses \(m\to\rmM\) and \(\widetilde{m}\to\widetilde{\rmM}\) to restore non-zero $R$-charges.} with the expression \rf{fpi} arising from factorization of the Coulomb branch representation of the partition function on $S^2$.
Anti-vortices localized at the south pole provide an identical contribution, expanded in terms of the anti-vortex fugacity $\bar{z}$.
The one loop determinant must be evaluated at the location of the Higgs branches, where there is a zero mode. Removing the zero mode  amounts to taking the residue of the one-loop determinant.
Summing over Higgs branch components finally leads to the Higgs branch representation of the partition function of $\cN=(2,2)$ gauge theories on $S^2$
\begin{equation}
   Z_{\text{Higgs}} (m, \tau) = \sum_{\mathclap{\substack{v_i = - m_{p_i} \\ \{p_i\} \subseteq \{1,\ldots,N_F\}}}} \, Z_{\text{cl}}(v, 0, \tau)
   \begin{aligned}[t]
     & \res_{a=v} \left[ Z_{\text{one-loop}}(a, 0, m)\right] \\
     & \!\!\!\! \times Z_{\text{vortex}} (\{p_i\}, m, (-1)^{
\changed{10}
N_F
\to
N_F+N-1
\done
}z) Z_{\text{vortex}}(\{p_i\}, m, (-1)^{
\changed{10}
\antiNF
\to
\antiNF+N-1
\done
} \bar{z})\,.
   \end{aligned}
 \end{equation}
This matches with the Coulomb branch representation  of the partition function computed earlier.


\section{Gauge Theory/Toda Correspondence}
\label{sec:Toda}

In this section we initiate the study  of  a novel  correspondence
 between two dimensional $\cN=(2,2)$ gauge theories on $S^2$ and two dimensional Liouville/Toda CFT, leaving a more complete analysis to a separate publication \cite{Gomis:2014eya}. Our correspondence has a
 well known counterpart, the AGT correspondence \cite{Alday:2009aq} (see also \cite{Wyllard:2009hg}), which relates four dimensional $\cN=2$ gauge theories and these CFTs.
 The correspondence we find  shares features with  the AGT one.  In fact, motivated by   AGT, the entry of the dictionary relating Liouville conformal blocks and vortex partition functions was already established in
 \cite{Dimofte:2010tz,Bonelli:2011fq,Bonelli:2011wx}. The partition function of   $\cN=(2,2)$ gauge theories on $S^2$ provides an elegant way of combining the vortex partition functions into modular invariant objects.
 Some of the implications of the modular properties we find in these two dimensional $\cN=(2,2)$ gauge theories are discussed in section~\ref{sec:conclusion}.

Specifically, we consider the example of  \(\cN = (2,2)\) SQED, described  by a $U(1)$ vector multiplet and   \(N_F\) of fundamental and \(N_F\) anti-fundamental chiral multiplets. We show that the partition function of
this theory is given by  a four point correlation function on the sphere for the \(A_{N_F - 1}\) Toda CFT.\footnote{See \cite{Fateev:2007ab} for an introduction to the Toda conformal field theory.}
In detail\footnote{The power of $z$ in the denominator can be removed by shifting the masses of all the chiral multiplets, which corresponds to a constant gauge transformation.}
\begin{equation}
  \label{Toda-match}
  Z_{\text{SQED}} (\rmM, \widetilde{\rmM}, \tau)
  = \frac{|1-z|^{-2b\langle {\rm \hat m}, h_1\rangle}}{|z|^{2\delta}}
  \frac{\langle V_{\alpha_2}(\infty)V_{{\rm \hat m}}(1)V_{\mu}(z,\bar{z})V_{\alpha_1}(0) \rangle}{\langle V_{\alpha_2}(\infty)V_{{\rm \hat m} - bh_{N_F}}(1)V_{\alpha_1}(0) \rangle} \, ,
\end{equation}
up to a normalization of the \(V_{\rm \hat m}\) insertion. The cross-ratio of the four-punctures is given by the vortex fugacity parameter
\begin{equation}
z = (-1)^{
\changed{10}
N_F
\to
N_F+N-1
\done
} e^{2\pi i \tau}\,,
\end{equation}
 and the masses of the chiral multiplets are encoded in the momenta \(\alpha_1\), \(\alpha_2\) and \({\rm \hat m}\), and the exponent \(\delta\).  The precise relation between parameters is given in
 \rf{param-match} and \rf{delta-match}.  Note that the three point function in the denominator is a normalization, which does not affect the dependence on \(z\).

The vertex operators\footnote{Local operators in the Toda theory take the form \(V_{\alpha} = e^{\langle\alpha, \phi\rangle}\), labelled by a momentum vector \(\alpha\) in the Cartan subalgebra of \(A_{N_F - 1}\).}
\(V_{\alpha_1}\) and \(V_{\alpha_2}\) are labelled by generic momenta, thus they each involve \(N_F - 1\) continuous parameters.  The vertex operator \(V_{{\rm \hat m}}\) is a semi-degenerate
insertion,\footnote{The theory is symmetric under the \(W_{N_F}\) algebra, an extension of the Virasoro algebra \(W_2\) involving fields with higher spins \(2, \ldots, N_F\).
To each primary operator \(V_{\alpha}\) is associated a representation of the \(W_{N_F}\) symmetry algebra.  For so called degenerate momenta, the \(W_{N_F}\) representation becomes reducible,
and must be quotiented by the space of null vectors.} labelled by a momentum \({\rm \hat m} = - \varkappa h_{N_F}\) parallel to the highest weight \(-h_{N_F}\) of the antifundamental representation of
\(A_{N_F-1}\), with one continuous parameter.
The correlator finally involves a fully degenerate insertion \(V_{\mu}\), whose momentum \(\mu = - b h_1\) is fully constrained to the highest weight \(h_1\) of the fundamental representation of \(A_{N_F-1}\).
Let us now prove~\rf{Toda-match} by expressing both sides of the equality in terms of hypergeometric series.

Restricting \rf{Coulomb-factorized} to the case of \(U(1)\) with \(\antiNF = N_F\), the partition function we are interested in is given by
\begin{equation}
  \label{U(1)-factorized}
  Z^{\antiNF = N_F}_{U(1)} = 2\pi \sum_{p=1}^{N_F} \Bigg[ z^{-i\rmM_p} \bar{z}^{-i\rmM_p}
    \frac{ \prod_{s=1}^{N_F} \gamma( -i\widetilde{\rmM}_s -i\rmM_p) } { \prod_{s=1}^{N_F} \gamma(1 + i\rmM_s -i\rmM_p) } F_p(z) F_p(\bar{z}) \Bigg] \, ,
\end{equation}
where \(z = (-1)^{
\changed{10}
N_F
\to
N_F+N-1
\done
} e^{2\pi i \tau}\), and
\begin{equation}
  \label{U(1)-block}
  F_p(z) = {}_{N_F}F_{N_F - 1} \left( \begin{smallmatrix} -i\rmM_p-i\widetilde{\rmM}_1 \, \cdots \, -i\rmM_p-i\widetilde{\rmM}_{N_F} \\ 1 + i\rmM_1 -i\rmM_p \, \widehat{\cdots} \, 1 + i\rmM_{N_F} -i\rmM_p \end{smallmatrix}
    \middle| z \right) = \sum_{k \geq 0} \frac{z^{k} \prod_{s=1}^{N_F} ( -i\widetilde{\rmM}_s -i\rmM_p )_k} { \prod_{s=1}^{N_F} (1 + i\rmM_s-i\rmM_p)_k}
\end{equation}
are hypergeometric series of type \((N_F, N_F - 1)\), skipping \(1 + i\rmM_p -i\rmM_p\) in the list of parameters of \({}_{N_F}F_{N_F-1}\).  We shall see shortly that this factorized representation of the partition function matches exactly with the \(s\)-channel expression of the Toda four point
correlator as a sum over all allowed internal momenta.  The one-loop contribution matches with the product of the three point functions in the Toda theory, while the conformal blocks are reproduced by the contribution
\(z^{-i\rmM_p}\) from the classical action together with the vortex partition functions \(F_p(z)\).

\begin{figure}
  \centering
  \includegraphics{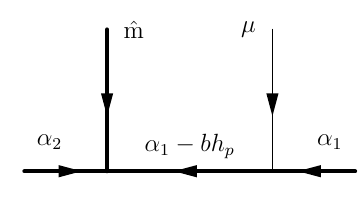}
  \caption{Toda CFT $s$-channel conformal block which reproduces the vortex partition function of SQED for a Higgs vacuum matching the choice of channel $1\leq p\leq N_F$.  The full four point correlator is equal to
  the SQED partition function on \(S^2\). The momenta \(\alpha_1\) and \(\alpha_2\) encode the masses of fundamental and anti-fundamental chiral multiplets, with \(SU(N_F)\times SU(N_F)\) flavour symmetry, while
  \({\rm \hat m}\) captures the remaining \(U(1)_{\text{diag}}\) flavour symmetry.\label{fig:toda-correlator}}
\end{figure}

As in any two dimensional conformal field theory, the four point correlator of interest can be expressed in the \(s\), \(t\), or \(u\) channels as an integral over all internal momenta of a combination of three point
functions, multiplying a holomorphic and an anti-holomorphic conformal blocks.  For our purposes, the \(s\)-channel is the most useful.  The fusion rules between the degenerate operator \(V_{\mu}\) and the generic insertion
\(V_{\alpha_1}\) only allow the internal momentum in this channel to be \(\alpha_1 - bh_p\) for some weight \(h_p\) of the fundamental representation of \(A_{N_F-1}\).  Thus, the correlation function is expressed as a
discrete sum rather than an integral over internal momenta:
\begin{equation}
  \label{4pt-factorized}
  \langle V_{\alpha_2}(\infty)V_{{\rm \hat m}}(1)V_{\mu}(z,\bar{z})V_{\alpha_1}(0) \rangle =
    \sum_{p = 1}^{N_F} C(\alpha_2, \alpha_1 - b h_p, {\rm \hat m}) C^{\alpha_1 - bh_p}_{-bh_1, \alpha_1} \cF^{(s)}_{\alpha_1 - bh_p} (z) \cF^{(s)}_{\alpha_1 - bh_p} (\bar{z}) \, .
\end{equation}
Here, \(C(\bullet, \bullet, \bullet) = \langle V_{\bullet} V_{\bullet} V_{\bullet} \rangle\) are the three point correlation functions of the theory, and \(\cF^{(s)}_{\alpha_1 - bh_p}\) is the \(s\)-channel
conformal block with internal momentum \(\alpha_1 - bh_p\) (see figure~\ref{fig:toda-correlator}).  The three point function involving the degenerate field \(\mu\) actually has poles for each allowed internal
momentum, hence we consider the residue \(C^{\alpha_1 - bh_p}_{-bh_1, \alpha_1}\) of the three point function at the given momenta.\footnote{The only non-zero two point functions are \(\langle V_{2Q - \alpha} V_{\alpha}\rangle\)
and its Weyl conjugates, where \(Q = \big(b + \frac{1}{b}\big) \rho = \big(b + \frac{1}{b}\big) \sum_{p = 1}^{N_F} \frac{N_F + 1 - 2p}{2} h_p\).  The three point function \(C^{\alpha_1 - bh_p}_{-bh_1, \alpha_1}\) appearing in \rf{4pt-factorized} is thus (the residue at) \(C(2Q - (\alpha_1 - bh_p), \alpha_1, \mu)\).}

The four point correlator of interest was shown \cite{Fateev:2007ab} -- using null-vector equations -- to obey identical holomorphic and anti-holomorphic hypergeometric differential equations of order \(N_F\),
up to a power of \(|1-z|^2\).
  Those two equations enabled them to evaluate the conformal blocks \(\cF^{(s)}\) as\footnote{We use the more symmetrical notations of appendix~B of~\cite{Gomis:2010kv}, with the change \(\alpha_2 \to 2Q - \alpha_2\)
  to make this momentum incoming rather than outgoing.}
\begin{equation}
  \label{Toda-block}
  \frac{(1-z)^{-b\varkappa/N_F}}{z^{(b^2 + 1)\frac{N_F - 1}{2} + iba_{1,p}}}
  \cF^{(s)}_{\alpha_1-bh_p}\!\left[\begin{matrix}
      {\rm \hat m} & \mu \\ \alpha_2 & \alpha_1
    \end{matrix}\right]
  =
  {}_{N_F}F_{N_F - 1}\!\left(\begin{smallmatrix}
      \left[iba_{2,s} + iba_{1,p} + b\widehat{\varkappa}/N_F\right]_{1\leq s \leq N_F}\\
      \left[1 + iba_{1,p} -iba_{1,s}\right]_{1\leq s\leq N_F, s\neq p}
    \end{smallmatrix}\middle| z \right) \, ,
\end{equation}
where \(i a_{1, s} = \langle \alpha_1 - Q, h_s\rangle\), \(i a_{2, s} = \langle \alpha_2 - Q, h_s\rangle\), and \(\widehat{\varkappa} = \varkappa - (N_F-1) b\).

The matching between the vortex partition functions and the conformal blocks occurs if and only if the \(2N_F - 1\) parameters of the hypergeometric functions are equal (up to permutation). Up to Weyl reflection, this fixes the \(2N_F - 1\)
momentum components of the Toda correlator in terms of the \(2N_F - 1\) physical masses of the gauge theory:
\begin{equation}
  \label{param-match}
  \begin{aligned}
    \alpha_1 &= Q - \frac{i}{b} \sum_{s=1}^{N_F} \rmM_s h_s \\
    \alpha_2 &= Q - \frac{i}{b} \sum_{s=1}^{N_F} \widetilde{\rmM}_s h_s \\
    \varkappa &= (N_F-1) b - \frac{i}{b} \sum_{s=1}^{N_F} \left(\rmM_s + \widetilde{\rmM}_s\right) \, .
  \end{aligned}
\end{equation}
Furthermore, the exponent \(\delta\) is fixed by comparing the powers of \(z\) appearing in \rf{Toda-block} and \rf{U(1)-block},
\begin{equation}
  \label{delta-match}
  \delta = (b^2 + 1)\frac{N_F - 1}{2} + \frac{i}{N_F}\sum_{s=1}^{N_F}\rmM_s \, .
\end{equation}

The next object to consider is the product of three point correlation functions appearing in \rf{4pt-factorized}.  We start from the explicit expression for three point functions with a semi-degenerate insertion
(equation~(1.39) in~\cite{Fateev:2007ab}),
\begin{equation}
  \begin{aligned}
    C(\alpha_1, \alpha_2, - \varkappa h_{N_F})
    &= \left[ \pi \mu \gamma(b^2) b^{2-2b^2} \right]^{\langle 2Q - \alpha_1 - \alpha_2 + \varkappa h_{N_F}, \rho/b\rangle} \\
    &\quad \cdot \frac{\Upsilon(b)^{N_F - 1} \Upsilon(\varkappa) \prod_{s<t}^{N_F} \Upsilon(\langle \alpha_1 - Q, h_t - h_s\rangle) \Upsilon(\langle \alpha_2 - Q, h_t - h_s\rangle)}{\prod_{s,t}^{N_F}
      \Upsilon\big(\frac{\varkappa}{N_F} + \langle\alpha_1 - Q, h_s\rangle + \langle\alpha_2 - Q, h_t\rangle\big)} \, ,
  \end{aligned}
\end{equation}
where the \(\Upsilon\) function is introduced in~\cite{Fateev:2007ab}.  Thanks to the relation \(\Upsilon(x + b) = \gamma(bx) b^{1-2bx} \Upsilon(x)\), one can simplify a ratio involving the first of our three point functions as
\begin{equation}
  \frac{C(\alpha_2, \alpha_1 - bh_p, -\varkappa h_{N_F})}{C(\alpha_2, \alpha_1, -(\varkappa+b) h_{N_F})} =
    \frac{\prod_{s=1}^{N_F} \gamma\big(\frac{b\widehat{\varkappa}}{N_F} + iba_{1,p} + iba_{2,s}\big) \gamma\big((b^2+1)\mathbbm{1}_{s\leq p} + ib(a_{1,s} - a_{1,p})\big)}{[-\pi\mu\gamma(b^2+1)]^{p}} \, .
\end{equation}
The second correlation function is given by (see equation~(1.51) in~\cite{Fateev:2007ab})
\begin{equation}
  C^{\alpha_1 - bh_p}_{-bh_1, \alpha_1} = [-\pi\mu\gamma(1+b^2)]^{p} \prod_{s<p} \frac{\gamma(iba_{1,s}-iba_{1,p})}{\gamma(1+b^2+iba_{1,s}-iba_{1,p}))} \, .
\end{equation}
Using the relations \rf{param-match}, the two correlation functions combine into exactly the appropriate factor  in the SQED partition function
 \rf{U(1)-factorized},
\begin{equation}
  \frac{C(\alpha_2, \alpha_1 - bh_p, -\varkappa h_{N_F})}{C(\alpha_2, \alpha_1, -(\varkappa+b) h_{N_F})}   C^{\alpha_1 - bh_p}_{-bh_1, \alpha_1}
  = \prod_{s=1}^{N_F} \frac{\gamma\big(\frac{b\hat{\varkappa}}{N_F} + iba_{1,p} + iba_{2,s}\big)}{\gamma(1 -ib(a_{1,s} - a_{1,p}))}
  = \prod_{s=1}^{N_F} \frac{\gamma(-i\rmM_p -i\widetilde{\rmM}_s)}{\gamma(1+i\rmM_s -i\rmM_p)} \, .
\end{equation}

Putting all the ingredients together, we obtain the relation \rf{Toda-match} summarizing the correspondence.  The normalization by a three point correlator in the denominator of this relation indicates that the
gauge theory partition function corresponds to the insertion of a fully degenerate momentum in a Toda three point function with two generic and one semi-degenerate vertex operator.

One noteworthy aspect of the matching is that delta-renormalizable vertex operators, whose momenta are characterized by the reality condition that \(\langle\alpha - Q, h_s\rangle \in i\mathbb{R}\) for all \(1\leq s\leq N_F\),
arise exactly when the gauge theory complex masses \(\rmM = m + \frac{i}{2} q\) are real, hence the $R$-charges are zero.  We can analytically continue the correlator to arbitrary momenta in order to capture the
partition function of gauge theories with non-zero $R$-charges.

\begin{figure}[ht]
  \hskip-2.3cm
  \includegraphics[width=20cm]{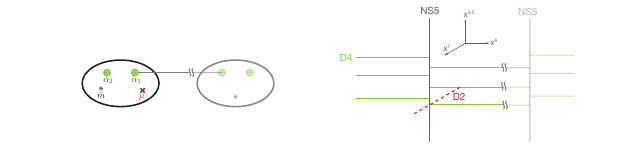}
  \caption{Decoupling limit of the AGT correspondence with a simple surface operator.\label{toda-graph}}
\end{figure}

The precise matching between the  partition function of SQED with a correlator in $A_{N_F-1}$ Toda CFT    can be given a physical explanation  using  the AGT correspondence. We  start with the punctured Riemann
surface describing four dimensional $\cN=2$ SQCD with $SU(N_F)$ gauge group and $N_F$ fundamental and $N_F$ anti-fundamental hypermultiplets. This is described by an $A_{N_F-1}$ Toda CFT correlator on the
four-punctured sphere, with two non-degenerate and two semi-degenerate punctures. We now add a degenerate puncture, which is believed to correspond to inserting a half-BPS surface operator on $S^2$ inside $S^4$
\cite{Alday:2009fs}. For the simplest degenerate field, the surface operator can be described by coupling a two dimensional $\cN=(2,2)$ gauge theory to four dimensional  SQCD (see e.g. \cite{Gukov:2006jk}).  The precise two dimensional gauge
theory can be found by realizing the simple surface operator as a D2-brane in Type IIA string theory, as summarized in  Figure~\ref{toda-graph}. For the simplest surface operator, the
corresponding   theory is two dimensional $\cN=(2,2)$ SQED  with $N_F$ flavours, which we just analyzed.  The Toda CFT correlator with the degenerate field insertion is expected to capture the four dimensional
gauge dynamics, the two dimensional gauge dynamics on the surface operator and the coupling of the four dimensional degrees of freedom to the two dimensional ones.

We can turn off the couplings of four dimensional SQCD to two dimensional SQED by sending the four dimensional gauge coupling to zero. This corresponds in the language of Toda CFT to factorizing the five point
function into the four point function that we are after times  a three point function. This factorization of the five-punctured sphere is depicted in  Figure~\ref{toda-graph}.
In this limit, only the dynamics of the two dimensional theory remain.  Moreover, in this limit,  the couplings between the four dimensional and two dimensional theories are realized as twisted mass parameters
for the chiral multiplets in the two dimensional theory.  It is therefore natural to expect that the Toda correlation function \rf{4pt-factorized} is related to the partition function  of $\cN=(2,2)$ SQED
with $N_F$ flavours, which is what we have shown explicitly in this section.

\section{Seiberg Duality}
\label{sec:Seiberg}

In this section, we apply our results to study the infrared duality of $\cN=(2,2)$ non-abelian gauge theories in two dimensions.  There are many interesting mathematical conjectures on the properties of
moduli spaces of Calabi-Yau manifolds embedded in Grassmannians, one of which is known as the R{\o}dland conjecture that a certain Calabi-Yau threefold in the Grassmannian $G(2,7)$ and the Pfaffian Calabi-Yau
in $\mathbb{CP}^6$ are in the same one-dimensional complexified K\"ahler moduli space.  In attempts to provide a physical proof of R{\o}dland's conjecture, it has been proposed~\cite{Hori:2006dk} that the
$\cN=(2,2)$ $SU(N)$ gauge theory with $N_F > N$ massless fundamental chiral multiplets without superpotential is dual to the theory with gauge group $SU(N_F-N)$,
\begin{equation}
  \label{SU-duality}
  SU(N) \; \longleftrightarrow \; SU(N_F - N) \, .
\end{equation}
Indeed the two theories, endowed with twisted masses, have the same Witten index.  Furthermore, since these theories are expected to flow in the infrared to $\cN=(2,2)$  superconformal theories with
central charges $\hat c = (N_F-N)N +1$, the two theories in this duality pair carry the same central charge.  Recently, the above duality, known as the Seiberg-like duality in two dimensions, has been
generalized to other gauge groups~\cite{Hori:2011pd}
\begin{equation}
  \label{G-dualities}
  \begin{aligned}
    O_+(N) & \; \longleftrightarrow \; SO(N_F - N +1), \qquad & N_F &\geq N \\
    SO(N) & \; \longleftrightarrow \; O_+(N_F-N +1), & N_F &\geq N \\
    O_-(N) & \; \longleftrightarrow \; O_-(N_F-N+1), & N_F &\geq N \\
    Sp(N)  & \; \longleftrightarrow \; Sp(N_F-N-1), & N_F &\geq N + 3
  \end{aligned}
\end{equation}
where $\pm$ denotes the eigenvalues of $\mathbb{Z}_2$ gauge symmetry of $O(N)$.

We prove in appendix~\ref{app:Seiberg} that the partition functions of theories with special unitary gauge groups \rf{SU-duality} are equal in two different limits, hence providing non-trivial evidence
to support the above duality for this pair of gauge groups.

In the limit of small masses and $R$-charges,  the partition function is singular, and we will express it as
\begin{equation}
  \label{ZSUN-polar-part}
  Z_{SU(N)} (\rmM) = \sum_{\substack{E\subseteq \{1,\ldots,N_F\}\\\#E=N}} \Bigg[\frac{\prod_{p\in E} \prod_{s\not\in E}^{N_F} \gamma\left(i\rmM_p-i\rmM_s\right)}{\sum\limits_{p\in E} (-i\rmM_p)}\Bigg] + O(1) \, ,
\end{equation}
where the sum ranges over sets \(E\) of \(N\) flavours, and \(O(1)\) indicates that only the singular part of $Z_{SU(N)}$ is captured by the sum.
This expression is symmetrical under the transformation
\begin{equation}
  \label{duality-dictionary}
  \begin{cases}
    E \to E' = \{1,\ldots,N_F\}\setminus E \\
    N \to N' = N_F - N \\
    \rmM_p \to \rmM'_p = - \rmM_p + \frac{1}{N_F-N} \sum_{s=1}^{N_F} \rmM_s \, .
  \end{cases}
\end{equation}
Only the \(O(1)\) term may be affected, hence we obtain the duality
\begin{equation}
  \label{Seiberg-small-mass}
  Z_{SU(N)} (\rmM) = Z_{SU(N_F-N)}(\rmM') + O(1) \, ,
\end{equation}
in the limit \(\rmM\to 0\).  Note that since \(Z_{SU(N)} \sim \rmM^{-N(N_F-N) - 1}\) at \(0\), the relation~\rf{Seiberg-small-mass} involves \(N(N_F-N)+1\) orders.  The duality was also tested explicitly
to order $O(\rmM^2)$ in the case $N=2$, $N_F=3$.

The second case which we consider in appendix~\ref{app:Seiberg} is the limit where a sum of \(N\) of the complex masses vanishes, with masses otherwise generic.  The partition function \(Z_{SU(N)}\) has a
simple pole in this limit, whose residue is shown to match the dual \(SU(N_F-N)\) theory.  This is a strong check of the Seiberg-like duality \(Z_{SU(N)} (\rmM) = Z_{SU(N_F-N)} (\rmM')\) since the masses
and $R$-charges span in this case a codimension $1$ subspace of the $N_F$-dimensional parameter space.

Using the Coulomb branch expression~\rf{semicolumb} for partition functions of \(\cN=(2,2)\) theories with arbitrary gauge groups, it should be possible to prove Seiberg-like duality for different pairs of
gauge groups, such as those given in~\rf{G-dualities}.  It should also be possible to extend the above Seiberg-like duality to theories with a homogenous superpotential of degree $d$ in the baryon operators,
with $\frac{2N_F}{(N_F-N) N_F + 1}<d \leq N_F$
\cite{Hori:2011pd}.
Due to the superpotential, the $R$-charge of each chiral multiplet is constrained to be in the range\footnote{The upper bound of this range reproduces the condition $\sum_{s=1}^{N_F} q_s<N_F - N + 1/N$ which
ensures convergence of the Coulomb branch integral~\rf{ZSUN-Coulomb}.}
\begin{equation}
  \frac{2}{N N_F} \leq q_i < \frac{N_F - N}{N_F} + \frac{1}{N N_F}\, .
\end{equation}
It would be interesting to show an agreement between the partition functions of each pair of theories for the case of $R$-charges in the above range.

\section{Discussion}
\label{sec:conclusion}

In this paper we have computed the exact partition function of two dimensional $\cN=(2,2)$ gauge theories on $S^2$. We have shown that there are two ways of representing   the partition function. It can be either
written as an integral over the Coulomb branch or as a sum over vortices and anti-vortices in the Higgs branch. By explicitly evaluating the integral representation in the Coulomb branch, we find exact agreement
with the  Higgs branch representation of the partition function. Quite pleasingly, despite that we are integrating over different field configurations,   the two results give rise to the same partition function.

 The Coulomb branch representation is found by integrating over $\cQ$-invariant field configurations that are saddle points of the deformation action. Since our deformation term   does not contain a term linear
 in $\rmD$, the intersection of the supersymmetry fixed point equations with the saddle point equations completely lifts configurations in the Higgs branch, giving rise, as supersymmetric saddle points, to the
 Coulomb branch configurations $\cF_{\text{Coul}}$, which we integrate over with a specific measure determined by the one-loop determinants.
This implies, in particular, that the   vortex and anti-vortex configurations allowed at the poles by the supersymmetry
equations are forbidden. The same result can be more straightforwardly obtained by localizing the path integral with respect to different supercharges, concretely  $Q_1$ and $Q_
2$. In this approach, the supersymmetry equations alone forbid any non-trivial configurations in the Higgs phase while precisely reproducing  the Coulomb phase field configurations $\cF_{\text{Coul}}$.

 The Higgs branch representation is instead found by  integrating over $\cQ$-invariant field configurations that are saddle points of a  deformed action that does contain a  term linear in   $\rmD$. In this case,
 the intersection of the supersymmetry equations with the equations of motion completely lifts the Coulomb branch. However, the equations now  allow for non-trivial field configurations supported in the Higgs phase,
 which we have denoted by $\cF_{\text{Higgs}}$. These field  configurations   describe  vortex and anti-vortex excitations  at the poles of the $S^2$ around   each of the Higgs branches of the theory. In this
 Higgs branch representation,
the partition function is written as a sum over Higgs branches of the product of the vortex partition function at the north pole with the anti-vortex partition function at the south pole. The deformed action that we
have considered to obtain the Higgs branch representation is the same   deformed action as before, but now the saddle point equations are analyzed at a large finite value of the parameter multiplying the deformation term.
A more  desirable and precise way to arrive at the same conclusion would be to localize the path integral with a different deformation term  $\delta_\cQ V$ that,   in the limit when the parameter multiplying it goes to
infinity, yields   a non-trivial linear term in $\rmD$. It would be interesting to explicitly construct such  a deformation term.

Conceptually, the fact that a correlation function in a supersymmetric gauge theory may admit multiple representations can be understood as follows.  When computing a supersymmetric path integral by supersymmetric
localization, several choices are available, including the choice of supercharge and of deformation term with which to localize (see section~\ref{sec:local} for details).  Under mild conditions, the localization
principle guarantees that the path integral is independent of these choices.  For different choices, however, the path integral may localize to different supersymmetric field configurations and therefore provide
alternative representations of the same correlation function.  This general picture is behind the equivalence we find between the Coulomb and Higgs branch representation of the partition function of $\cN=(2,2)$
gauge theories on $S^2$.  It would be very interesting to extend this general picture to find new dual descriptions of correlation functions in supersymmetric gauge theories,
as they can lead, at the very least, to novel identities or to a physical derivation of known ones.

The Higgs branch expression for the partition function shares features with the localization computation of the partition function and Wilson loops \cite{Pestun:2007rz}, 't Hooft loops \cite{Gomis:2011pf} and
domain walls \cite{Drukker:2010jp} in four dimensional $\cN=2$ gauge theories.  These correlation functions receive  contributions from non-perturbative field configurations localized at the north and south
poles of the corresponding sphere.  In four dimensions they are due to instantons and anti-instantons, while in two dimensions the path integral is a sum over vortices at the north pole and anti-vortices at
the south pole.  In four dimensions the contribution of instantons and anti-instantons are captured by the   instanton partition function \cite{Moore:1997dj,Nekrasov:2002qd}, while the contribution of vortices
and anti-vortices are captured by the vortex partition function  \cite{Shadchin:2006yz} (see also \cite{Yoshida:2011au,Bonelli:2011fq,Miyake:2011yr,Fujimori:2012ab,Kim:2012uz}). An important
qualitative difference, however, is that instantons and anti-instantons appear in the Coulomb phase while vortices and anti-vortices can only appear as non-trivial field configurations in the Higgs phase.
Furthermore, the four dimensional correlation functions do not have a known dual description, while in two dimensions we find that the partition function admits
a Coulomb branch representation.

Several   applications  and correspondences emerge from our results. A correspondence between the partition function of $\cN=(2,2)$ gauge theories on $S^2$ and correlation functions in Liouville/Toda CFT has
been found, extending the AGT correspondence \cite{Alday:2009aq} (see also \cite{Wyllard:2009hg}). We have explicitly presented the    $A_{N_F-1}$ Toda representation of the partition function of SQED with $N_F$
electrons and $N_F$ positron chiral multiplet fields, leaving the more complete correspondence for other theories to a separate publication \cite{Gomis:2014eya}. This correspondence can be enriched by adding defects
both in gauge theory and in Toda as in \cite{Alday:2009fs,Drukker:2009id,Drukker:2010jp} (see also \cite{Passerini:2010pr}) and it would be interesting to establish a detailed dictionary between gauge theory and
Toda CFT. In fact, we have already found the effect of inserting a supersymmetric Wilson loop in \rf{insertlooop}. When the gauge group contains $U(1)$ factors, a Wilson loop insertion
effectively shifts the FI parameter $\xi$ as well as a the topological term $\vartheta$. In the Toda CFT description, this corresponds to changing the moduli of the Riemann surface  in the holomorphic sector
and anti-holomorphic sector of the CFT differently. This can be realized by the  insertion in Toda CFT of the complex-structure--changing topological defect operator introduced in \cite{Drukker:2010jp}.

Since the correlation functions of Toda CFT are modular invariant, this correspondence implies that the gauge theories that admit a Toda CFT representation enjoy quite remarkable modularity properties in the
complexified  gauge theory parameters $\tau$ \rf{taucoup}. In particular, this implies that the results from $\xi>0$ to $\xi<0$ are related by analytic continuation, and that the partition function  in the
two regimes are the same.   In the example of SQED, the $\xi>0$ regime corresponds to the factorization of the Toda CFT correlator in the $s$-channel, and individual Higgs vacua, labelled by masses of the
fundamental chiral multiplets, match precisely with the $N_F$ channels allowed by the fusion of the degenerate insertion with the operator which encodes the fundamental masses.  The $\xi<0$ regime is described
by the $u$-channel factorization, and the sum over Higgs vacua -- which correspond to intermediate channels in Toda -- is labelled by masses of the anti-fundamental chiral multiplets.

The expansion of the partition function near $\xi=0$ corresponds to the $t$-channel factorization. In this limit, the expansion in terms of vortices and anti-vortices in SQED breaks down, and it would be interesting
to understand whether this expansion has an alternative description in terms of another two dimensional gauge theory. Studying the modular properties further may lead to a picture of dualities analogous to
\cite{Gaiotto:2009we}. Relatedly, it would also be interesting to study the  combined dynamics of   two dimensional gauge theories on $S^2$ coupled to four dimensional $\cN=2$ gauge theories on $S^4$, and their
potential interpretation as surface operators. Extending the analysis to the squashed $S^2$  is also worth pursuing.

Our findings can also be applied to the study of $\cN=(2,2)$ non-linear sigma models with K\"ahler target spaces, including Calabi-Yau manifolds. The sigma models which  describe string propagation in such
target spaces  enjoy a rich ``phase'' structure as the complefixied K\"ahler parameters are varied. This may include the appearance of different geometries in   large volume regimes as well as non-geometrical
phases. Novel tools and understanding in the study of these questions were introduced in \cite{Witten:1993yc}, where these theories were given an ultraviolet definition in terms of $\cN=(2,2)$ gauge theories.
An important insight brought by the gauge theory description was the proposal that topology changing transitions -- in particular the flop transition -- can be described by analytic continuation in  the gauge
theory couplings $\tau$. Our exact results for SQED -- which include the conifold for $N_F=2$ -- quantitatively   demonstrate that the two large volume regimes connected through a flop
transition are indeed related by analytic continuation. Furthermore,    analytic continuation in the flop transition is realized  by crossing symmetry in   our correspondence with Toda CFT. Our
formulas further demonstrate that the physics at   $\xi=0$, while corresponding to  a  singular Calabi-Yau geometry, is completely regular for  a non-vanishing topological angle $\vartheta$.

Another relevant  connection between $\cN=(2,2)$ gauge theories in the ultraviolet and non-linear sigma models in the infrared is the transmutation   of gauge  vortices
into worldsheet instantons \cite{Witten:1993yc}. Given the exact results for the gauge theory partition function found in this paper, it would be interesting to revisit this connection, which was effectively
used in \cite{Morrison:1994fr} to quantitatively study worldsheet instantons.

Finally, we have used our formulas to study Seiberg duality in two dimensions, where we have demonstrated that Seiberg dual pairs have  the same partition function in some limits.
A very rich set of dualities relating two dimensional $\cN=(2,2)$  theories  is mirror symmetry, which relates  string theory on different mirror Calabi-Yau manifolds and in different phases. It would be very
interesting to extend our results to the case of Landau-Ginzburg models and   provide a detailed picture   relating these models to their dual gauged linear sigma models. This requires extending  our analysis
by including twisted chiral multiplets and the allowing for a non-trivial K\"ahler potential.

Two dimensional $\cN=(2,2)$ non-abelian gauge theories  been recently proposed
to study non-toric Calabi-Yau manifolds, such as Calabi-Yau manifolds embedded in Grassmannians and determinantal Calabi-Yau varieties \cite{Jockers:2012zr}. Due to
the strong coupling dynamics of these gauge theories, these models have not been studied much.
Our exact results provide a new and powerful tool to investigate the strong coupling dynamics of these $\cN=(2,2)$
non-abelian gauged linear sigma models, which may hopefully lead to new insights into this large class of Calabi-Yau manifolds.
Another  direction to study further is a possible connection of our results to the physics of domain walls in three dimensional gauge theories on $S^3$, generalizing the results in
\cite{Drukker:2010jp,Hosomichi:2010vh,Hama:2010av}.
Finally, our exact results may provide  hints on a  4d/2d relation between the geometry of four-manifolds and  two dimensional gauge theories, resulting in a   novel correspondences beyond the  the
2d/4d relations of \cite{Alday:2009aq} and
3d/3d relations of  \cite{Dimofte:2011ju,Dimofte:2011py,Yamazaki:2012cp}.

\clearpage

\section*{Acknowledgements}

We would like to thank Nick Dorey, Davide Gaiotto, Kazuo Hosomichi, Seok Kim, Kimyeong Lee, Nick Manton, Takuya Okuda, Jock McOrist, Sara Pasquetti, Vasily Pestun and David Tong for helpful discussions.
Talks about this work have been given  at the ``Maths of String and Gauge Theory'' Conference in London, at the \'Ecole Normale Sup\'erieure and at the IHES in Paris, and J.G. would like to thank the organizers
for the opportunity to speak.
N.D., J.G. and B.L.F. thank the University of  Barcelona for hospitality and support under grant FPA2010-20807. J.G. is also grateful to the LPT-ENS in Paris for hospitality and support. S.L.
thanks the Perimeter Institute for hospitality where part of this work has been done.
Research at the Perimeter Institute is supported in part by the Government of Canada through NSERC and by the Province of Ontario through MRI.
J.G. also acknowledges further support from an NSERC Discovery Grant and from an ERA grant by the Province of Ontario.
The work of S.L. is supported by the Ernest Rutherford fellowship of the Science \& Technology Facilities Council ST/J003549/1.
B.L.F. has also received funding from the [European Union] Seventh Framework Programme [FP7-People-2010-IRSES] under grant agreement n$^\circ$269217.

\clearpage


\appendix
\addtocontents{toc}{\protect\setcounter{tocdepth}{1}}

\section{Notations and Conventions}
\label{app:conven}

We use the following  conventions
for   indices:
\begin{align}
 i,j,k,\dots =1,2  & \qquad \text{coordinate indices on $S^2$}
\\
 \hat i, \hat j, \hat k, \dots =\hat{1},\hat{2} & \qquad \text{tangent space indices}
\\
\alpha,\beta,\gamma,\dots = 1,2 &  \qquad \text{Dirac spinor indices}
\\
m,n,p=1,2,3  & \qquad \text{indices for $SU(2)$ generators}
\end{align}

\subsection{\texorpdfstring{$S^{2}$}{S\texttwosuperior} Conventions}

We work in polar coordinates $(x^{1},x^{2})=(\theta,\varphi)$ where the metric on $S^{2}$ can be written as
\begin{equation}
\rmd s^2=r^2\left(\rmd\theta^2+\sin^2\theta \rmd\varphi^2\right)\,.
\end{equation}
The canonical choice of orientation is
\begin{equation}
\varepsilon_{12}=\sqrt{h}\,\varepsilon_{\hat{1}\hat{2}} = r^{2}\sin\theta\,,
\end{equation}
with the corresponding volume-form
\begin{equation}
\rmd^{2}x\sqrt{h} = r^2\sin\theta \rmd\theta\wedge\rmd\varphi\,.
\end{equation}
The simplest choice of zweibein is
\begin{equation}
e^{\hat{1}}=r\rmd\theta\qquad\text{and}\qquad e^{\hat{2}}= r\sin\theta \rmd\varphi \,,
\end{equation}
with the spin connection given by
\begin{equation}
\label{spinconnection}
\omega^{\hat{i}\hat{j}}=-\varepsilon^{\hat{i}\hat{j}}\cos\theta \rmd\varphi \,.
\end{equation}
By $D_{i}$ we denote the gauge-covariant derivative
\begin{equation}
D_{i} = \nabla_{i} -iA_{i}\, ,
\end{equation}
where $\nabla_i$ is the usual covariant derivative and $A_{i}$ is the gauge field. The corresponding curvature is given by
\begin{equation}
F_{ij}=\varepsilon_{ij}F_{\hat{1}\hat{2}} = \nabla_{i}A_{j} - \nabla_{j}A_{i} -i[A_{i},A_{j}]\, .
\end{equation}

\subsection{Spinors and the Clifford Algebra}
Our conventions for spinors are the same as in \cite{Wess:1992cp} and are listed below. Let $\tau_{m}$ denote the standard Pauli matrices given by
\begin{equation}
  \tau_{1}=\begin{pmatrix} 0 & 1 \\ 1 & 0  \end{pmatrix} , \quad
  \tau_{2}=\begin{pmatrix} 0 & -i\\ i & 0  \end{pmatrix} , \quad
  \tau_{3}=\begin{pmatrix} 1 & 0 \\ 0 & -1 \end{pmatrix} .
\end{equation}

We take our spinors to be anti-commuting Dirac spinors $\epsilon_{\alpha}$. These spinors are acted on by the $\gamma$-matrices defined by
\begin{equation}
(\gamma_{\hat{m}})_{\alpha}{}^{\beta} :\quad \gamma_{\hat{m}}=\tau_{\hat{m}}\,.
\end{equation}
Evidently, the matrices $\gamma^{\hat{i}}$ satisfy the two dimensional Clifford algebra
\begin{equation}
\left\{\gamma^{\hat{i}},\gamma^{\hat{j}}\right\}=2\delta^{\hat{i}\hat{j}}\,,
\end{equation}
and $\gamma^{\hat{3}}=-i\gamma^{\hat{1}}\gamma^{\hat{2}}$ is the two dimensional chirality matrix.\footnote{In terms of the $\sigma$ and $\bar{\sigma}$ matrices introduced in \cite{Wess:1992cp},
the $\gamma$-matrices are given by $\gamma^{m}{}_{\alpha}{}^{\beta} = \frac{i}{2}\epsilon^{mnp}\sigma_{n\,\alpha\dot{\alpha}}\bar{\sigma}_{p}{}^{\dot{\alpha}\beta}$.}

The spinor indices are raised and lowered by the (anti-symmetric) charge conjugation matrix as
\begin{equation}
  \epsilon^\alpha=C^{\alpha\beta}\epsilon_\beta\qquad \text{and}\qquad \epsilon_\alpha= C_{\alpha\beta}\epsilon^\beta\,,
\end{equation}
with the consistency condition
\begin{equation}
  C_{\alpha \gamma}C^{\gamma\beta}=\delta^{\beta}_{\alpha}\,.
\end{equation}
More explicitly, we take $C^{12}=C_{21}=1$ and $C^{21}=C_{12}=-1$.

We adopt the Northwest-Southeast convention for the implicit contraction of the spinor indices, i.e.\@ for two spinors $\epsilon$ and $\lambda$ we define
\begin{equation}
\epsilon\lambda \equiv \epsilon^{\alpha}\lambda_{\alpha}=\lambda\epsilon \qquad \text{and} \qquad
\epsilon\gamma^{\hat{m}}\lambda \equiv \epsilon^{\alpha}(\gamma^{\hat{m}})_{\alpha}^{\phantom{\alpha}\beta}\lambda_{\beta}= - \lambda\gamma^{\hat{m}}\epsilon\,.
\end{equation}
Note that the $\gamma$-matrices with both spinor indices lowered
\begin{equation}
(\gamma^{\hat{m}})_{\alpha\beta} \equiv C_{\beta\delta}\gamma^{\hat{m}}{}_{\alpha}{}^{\delta}\,,
\end{equation}
are symmetric and are numerically equal to $(-\tau_3,-i,\tau_1)$ for $\hat{m}=(1,2,3)$ respectively.

\subsection{Fierz Identities}
Let $\bar{\epsilon},\lambda$ and $\epsilon$ be anticommuting spinors. The following Fierz identities   are used extensively in our calculations
\begin{align}
\label{Fierz1}
 (\bar{\epsilon}\lambda)\epsilon + (\lambda\epsilon)\bar{\epsilon} + (\bar{\epsilon}\epsilon)\lambda &= 0\,,
\\
\label{Fierz2}
 (\bar{\epsilon}\gamma_{\hat{m}}\lambda)\gamma^{\hat{m}}\epsilon+(\bar{\epsilon}\lambda)\epsilon+2(\bar{\epsilon}\epsilon)\lambda &=0\,.
\end{align}

\section{Supersymmetry Transformations on \texorpdfstring{$S^2$}{S\texttwosuperior}}
\label{app:sca}

The $\cN=(2,2)$ superconformal algebra in the $S^{2}$ basis is spanned by the bosonic generators
\begin{equation}
J_{m}, K_{m}, R, \cA\,,
\end{equation}
and the supercharges
\begin{equation}
Q_{\alpha}, S_{\alpha}, \bar{Q}_{\alpha}, \bar{S}_{\alpha}\,.
\end{equation}
$J_m$ generate the $SU(2)$ isometries of $S^2$ while $K_m$ generate the conformal symmetries of $S^2$.
$R$ and  $\cA$ are each a $U(1)$ $R$-symmetry generator, the first being non-chiral and the latter being chiral.

The  $\cN=(2,2)$ superconformal algebra is given by
\begin{equation}
\label{Neq2sca}
\begin{aligned}
  \{ S_{\alpha}, Q_{\beta} \} &= \gamma^{m}_{\alpha \beta} J_{m} - \frac{1}{2} C_ {\alpha \beta}R \quad &
  [J_{m},S^{\alpha}] &= -\frac{1}{2} \gamma_{m}{}^{\alpha \beta}S_{\beta}\quad &
  [R,S_{\alpha}] &= + S_{\alpha}
\\ \{ \bar{S}_{\alpha}, \bar{Q}_{\beta} \} &= -\gamma^{m}_{\alpha \beta} J_{m} - \frac{1}{2} C_{\alpha \beta}R&
  [J_{m},Q^{\alpha}] &= -\frac{1}{2} \gamma_{m}{}^{\alpha \beta}Q_{\beta} &
  [R,Q_{\alpha}] &= - Q_{\alpha}
\\
 \{ Q_{\alpha}, \bar{Q}_{\beta} \} &= \gamma^{m}{}_{\alpha \beta} K_{m} + \frac{1}{2} C_{\alpha \beta} \cA  &
  [J_{m},\bar{Q}^{\alpha}] &= -\frac{1}{2} \gamma_{m}{}^{\alpha \beta}\bar{Q}_{\beta} &
  [R,\bar{Q}_{\alpha}] &= + \bar{Q}_{\alpha}
\\
      \{ S_{\alpha}, \bar{S}_{\beta} \} &= \gamma^{m}{}_{\alpha \beta} K_{m} - \frac{1}{2} C_{\alpha \beta} \cA \quad  &
  [J_{m},\bar{S}^{\alpha}] &= -\frac{1}{2} \gamma_{m}{}^{\alpha \beta}\bar{S}_{\beta} &
  [R,\bar{S}_{\alpha}] &= - \bar{S}_{\alpha}
\\
  [J_{m},J_{n}] &= i \epsilon_{mnp} J^{p}\quad  &
  [K_{m},S^{\alpha}] &= -\frac{1}{2} \gamma_{m}{}^{\alpha \beta}\bar{Q}_{\beta} &
  [\cA, S_{\alpha}] &= \bar{Q}_{\alpha}
\\
  [K_{m},K_{n}] &= -i \epsilon_{mnp} J^{p} &
  [K_{m},Q^{\alpha}] &= -\frac{1}{2} \gamma_{m}{}^{\alpha \beta}\bar{S}_{\beta} &
  [\cA, Q_{\alpha}] &= - \bar{S}_{\alpha}
\\
 [J_{m},K_{n}] &= i \epsilon_{mnp} K^{p}	 &
  [K_{m},\bar{Q}^{\alpha}] &= -\frac{1}{2} \gamma_{m}{}^{\alpha \beta}S_{\beta} &
  [\cA, \bar{Q}_{\alpha}] &= - S_{\alpha}
\\
  & &
  [K_{m},\bar{S}^{\alpha}] &= -\frac{1}{2} \gamma_{m}{}^{\alpha \beta}Q_{\beta} &
  [\cA, \bar{S}_{\alpha}] &= Q_{\alpha}\,.
\end{aligned}
\end{equation}
This algebra admits a $\mathbb{Z}_2$ automorphism, under which
\begin{equation}
\begin{aligned}
J_m, R, Q_\alpha,S_\alpha&\rightarrow J_m, R, Q_\alpha,S_\alpha\\
K_m, \cA, \bar Q_\alpha,\bar S_\alpha&\rightarrow -K_m,-\cA,- \bar Q_\alpha,-\bar S_\alpha\,.
\end{aligned}
\end{equation}
 Therefore, $\{J_{m},R,S,Q\}$  generate a subalgebra. It is the  $SU(2|1)$ algebra in  \rf{SU(2|1)}, which describes the $\cN=(2,2)$ supersymmetry algebra on $S^2$.

\subsection{Realization of \texorpdfstring{$SU(2|1)$}{SU(2|1)} on the Fields}

A simple way to obtain the $SU(2|1)$ supersymmetry transformations is to first construct the  $\cN=(2,2)$ superconformal  transformations and then restrict   to those of the $SU(2|1)$
subalgebra. This logic applies  in any dimension and gives a first principles construction of the supersymmetry transformations that does not require guesswork, at least as long as the space admits a conformal Killing spinor.

The superconformal transformations are easily obtained from the Poincar\'e supersymmetry transformations in flat space by demanding that once the flat metric is replaced by a curved metric,
that the supersymmetry transformations are covariant under Weyl transformations.  In this process, the constant supersymmetry parameters of flat space are replaced by  conformal Killing spinors, which obey
\begin{equation}
\nabla_i \epsilon=\gamma_i\tilde \epsilon\qquad \nabla_i \bar\epsilon=\gamma_i \hskip+2pt\tilde\bar\hskip-2pt\epsilon\,.
\end{equation}
Using that the fields and conformal Killing spinors transform with definite   weight under a Weyl transformation
\begin{equation}
g_{ij}\rightarrow e^{2\Omega(x)}g_{ij}
\label{Weyl}
\end{equation}
we obtain the required superconformal transformations by imposing Weyl covariance. The terms that need to be modified in the vector and chiral multiplet    flat space supersymmetry transformations
(which can be obtained by dimensionally reducing the four dimensional $\cN=1$ supersymmetry transformations in  \cite{Wess:1992cp} to two dimensions) to make them Weyl covariant
are\footnote{The coefficients of the extra terms are fixed by demanding that the combination   transforms \emph{covariantly} under Weyl transformations and, in general, depend on the Weyl weight of the
fields as well as the dimension of space.}
\begin{equation}
\begin{aligned}
\bar{\epsilon}\slashed{D}\lambda \quad &\longrightarrow\quad  \bar{\epsilon}\slashed{D}\lambda-\lambda\slashed{\nabla}\bar{\epsilon}
\\
\epsilon\slashed{D}\bar{\lambda} \quad &\longrightarrow\quad  \epsilon\slashed{D}\bar{\lambda}-\bar{\lambda}\slashed{\nabla}\epsilon
\\
\slashed{D}\sigma_{1,2} \epsilon   \quad &\longrightarrow\quad  \slashed{D}\sigma_{1,2} \epsilon + \sigma_{1,2} \slashed{\nabla} \epsilon
\\
\slashed{D}\sigma_{1,2} \bar{\epsilon}   \quad &\longrightarrow\quad  \slashed{D}\sigma_{1,2} \bar{\epsilon} + \sigma_{1,2} \slashed{\nabla} \bar{\epsilon}
\\
\slashed{D}\phi \epsilon \quad &\longrightarrow\quad  \slashed{D}\phi \epsilon+\frac{q}{2}\phi\slashed{\nabla} \epsilon
\\
\slashed{D}\bar\phi \bar \epsilon \quad &\longrightarrow\quad \slashed{D}\bar\phi \bar \epsilon + \frac{q}{2}\phi\slashed{\nabla} \bar\epsilon
\\
\slashed{D} \psi \epsilon   \quad &\longrightarrow\quad \slashed{D} \psi \epsilon  - \frac{q}{2} \psi \slashed{\nabla} \epsilon
\\
\slashed{D} \bar\psi \bar \epsilon   \quad &\longrightarrow\quad \slashed{D} \bar\psi \bar\epsilon  - \frac{q}{2} \bar\psi \slashed{\nabla} \bar\epsilon\,,
\end{aligned}
\end{equation}
where we have used   the following Weyl weights $w$
\medskip
\begin{center}
  \begin{tabular}{*{16}{>{\(}c<{\)}}}
    \toprule
    \multicolumn{2}{c}{SUSY} &
    \multicolumn{6}{c}{vector multiplet} &
    \multicolumn{6}{c}{chiral multiplet}
    \\
    \cmidrule(lr{1mm}){1-2}\cmidrule(lr{1mm}){3-8}\cmidrule(lr{1mm}){9-14}
    \epsilon & \bar{\epsilon}  &
    A_\mu & \sigma_1 & \sigma_2 & \lambda & \bar{\lambda} & D &
    \phi & \psi & F & \bar{\phi} & \bar{\psi} & \bar{F}
    \\
    -\frac{1}{2} & -\frac{1}{2}  &
    0 & 1 & 1 & \frac{3}{2} & \frac{3}{2} & 2 &
    \frac{q}{2}& \frac{q+1}{2} &  \frac{q+2}{2} & \frac{q}{2} & \frac{q+1}{2} &  \frac{q+2}{2} \\
    \bottomrule
  \end{tabular}
\end{center}
\medskip

\noindent
where $w$ is the charge $\varphi\rightarrow e^{-w \Omega(x)} \varphi$ under the  Weyl transformation \rf{Weyl}.

In this way, we obtain the two dimensional $\cN=(2,2)$ superconformal transformations  for  the vector multiplet
\begin{align}
\begin{split}
  {\delta} A_{i} &=
    -\frac{i}{2}\left(\bar{\epsilon}\gamma_{i}\lambda+\epsilon\gamma_{i}\bar{\lambda}\right),
\\
  {\delta} \sigma_{1}& =
    \frac{1}{2}\left(\bar{\epsilon}\lambda-\epsilon\bar{\lambda}\right),
\\
   {\delta} \sigma_{2} &=
    -\frac{i}{2}\left(\bar{\epsilon}\gamma_{\hat 3}\lambda+\epsilon\gamma_{\hat 3}\bar{\lambda}\right),
\\
   {\delta} \rmD &=
    -\frac{i}{2}\bar{\epsilon}\left(\slashed{D}\lambda+\left[\sigma_{1},\lambda\right]-i\left[\sigma_{2},\gamma^{\hat 3}\lambda\right]\right)
\\
   & \quad +\frac{i}{2}\lambda\slashed{\nabla}\bar{\epsilon} + \frac{i}{2}\epsilon\left(\slashed{D}\bar{\lambda}-\left[\sigma_{1},\bar{\lambda}\right]-i\left[\sigma_{2},\gamma^{\hat 3}\bar{\lambda}\right]\right)
    -\frac{i}{2}\bar{\lambda}\slashed{\nabla}\epsilon ,
\\
   {\delta} \lambda &=
    \left(i\gamma^{\hat 3}F_{\hat{1}\hat{2}}-\gamma^{\hat 3}\slashed{D}\sigma_{2}+i\slashed{D}\sigma_{1}-\gamma^{\hat 3}[\sigma_{1},\sigma_{2}]-\rmD\right)\epsilon
    +i\sigma_{1}\slashed{\nabla}\epsilon-\sigma_{2}\gamma^{\hat 3}\slashed{\nabla}\epsilon ,
\\
   {\delta} \bar{\lambda} &=
    \left(i\gamma^{\hat 3}F_{\hat{1}\hat{2}}-\gamma^{\hat 3}\slashed{D}\sigma_{2}-i\slashed{D}\sigma_{1}+\gamma^{\hat 3}[\sigma_{1},\sigma_{2}]+\rmD\right)\bar{\epsilon}
    -i\sigma_{1}\slashed{\nabla}\bar{\epsilon}-\sigma_{2}\gamma^{\hat 3}\slashed{\nabla}\bar{\epsilon}\,,
\end{split}
\label{superconff}
\\ \intertext{and chiral multiplet}
\begin{split}
  {\delta} \phi&= \bar{\epsilon}\psi
\\
  {\delta} \bar{\phi}&= \epsilon\bar{\psi}
\\
  {\delta} \psi&=
    i\left(\slashed{D}\phi+\sigma_{1}\phi-i\sigma_{2}\phi\gamma^{\hat 3}+\frac{q}{2}\phi\slashed{\nabla}\right)\epsilon+\bar{\epsilon}F
\\
   {\delta} \bar{\psi}&=
    i\left(\slashed{D}\bar{\phi}+\bar{\phi}\sigma_{1}+i\bar{\phi}\sigma_{2}\gamma^{\hat 3}+\frac{q}{2}\bar{\phi}\slashed{\nabla}\right)\bar{\epsilon}+\epsilon \bar{F}
\\
   {\delta} F&=
    -i\left(D_{i}\psi\gamma^{i}+\sigma_{1}\psi-i\sigma_{2}\psi\gamma^{\hat 3}+\lambda\phi+\frac{q}{2}\psi\slashed{\nabla}\right)\epsilon
\\
   {\delta} \bar{F}&=
    -i\left(D_{i}\bar{\psi}\gamma^{i}+\bar{\psi}\sigma_{1}+i\bar{\psi}\sigma_{2}\gamma^{\hat 3}-\bar{\phi}\bar{\lambda}+\frac{q}{2}\bar{\psi} \slashed{\nabla}\right)\bar{\epsilon}\,.
  \end{split}
  \label{superconfcm}
\end{align}

 The  spinors $\epsilon$ and $\bar{\epsilon}$ serve as the parameters of the   superconformal transformations, such that each
independent conformal Killing spinor is associated with one of the supercharges in the superconformal algebra. On $S^{2}$, we can take the  conformal Killing spinors to satisfy
\begin{equation}
\nabla_{i}\epsilon_{s}=\frac{s}{2r}\gamma_{i}\gamma^{\hat 3}\epsilon_{s}
\qquad \text{and} \qquad
\nabla_{i}\bar{\epsilon}_{\bar{s}}=\frac{\bar{s}}{2r}\gamma_{i}\gamma^{\hat 3}\bar \epsilon_{\bar{s}}
\end{equation}
with $s,\bar{s}=\pm$. There are four independent solutions to these equations
\begin{align}
\epsilon_{s} & =
  \exp\left(-\frac{is\theta}{2}\gamma_{\hat{2}}\right)\exp\left(\frac{i\varphi}{2}\gamma^{\hat{3}}\right)\epsilon_{\circ}^{s}\,,
\\
\bar{\epsilon}_{\bar{s}} &=
  \exp\left(-\frac{i\bar{s}\theta}{2}\gamma_{\hat{2}}\right)\exp\left(\frac{i\varphi}{2}\gamma^{\hat{3}}\right)\bar{\epsilon}_{\circ}^{\bar{s}}\,.
\end{align}
parametrized by four independent constant spinors $\epsilon_{\circ}^{\pm}$ and $\bar{\epsilon}_{\circ}^{\pm}$.
A general superconformal transformation is then generated by a linear combination of the supercharges parametrized as follows
\begin{equation}
\begin{aligned}
  \delta_{\epsilon_{+}} = \epsilon_{\circ}^{+}\tilde{\gamma}_{+} Q,
  \qquad \delta_{\epsilon_{-}} = \epsilon_{\circ}^{-}\tilde{\gamma}_{-} \bar{S},
  \qquad \bar{\delta}_{\bar{\epsilon}_{+}} = \bar{\epsilon}_{\circ}^{+} \tilde{\gamma}_{+} \bar{Q},
  \qquad \bar{\delta}_{\bar{\epsilon}_{-}} = -\bar{\epsilon}_{\circ}^{-} \tilde{\gamma}_{-} S
\end{aligned}
\end{equation}
where $\tilde{\gamma}_{\pm}$ satisfy
\begin{equation}
\begin{aligned}
  \tilde{\gamma}_{\pm} &= \frac{1}{\sqrt{2}} \left( \mathbbm{1} \pm i \gamma^{3} \right) = \pm i \gamma^{3} \tilde{\gamma}_{\mp}
    \\
  \tilde{\gamma}_{+}^{2} &= - \tilde{\gamma}_{-}^{2} = i \gamma^{3}, \quad \tilde{\gamma}_{+} \tilde{\gamma}_{-} = \mathbbm{1}      \,.
\end{aligned}
\end{equation}
Using the conformal Killing spinor equations above, the superconformal algebra is realized on the vector multiplet fields as
\begin{equation}
\label{sca-fields}
\begin{aligned}
\left[\delta_{\epsilon},\delta_{\bar{\epsilon}}\right] \lambda &=
    -\cL_{\xi}\lambda+i\left[\Lambda,\lambda\right]+ i\frac{s-\bar{s}}{2}\alpha\lambda
    +i\frac{s+\bar{s}}{2}\Theta\gamma^{\hat 3}\lambda-3i\frac{s+\bar{s}}{2}\alpha\lambda \,,
\\
\left[\delta_{\epsilon},\delta_{\bar{\epsilon}}\right] \bar{\lambda} &=
    -\cL_{\xi}\bar{\lambda}+i\left[\Lambda,\bar{\lambda}\right] -i\frac{s-\bar{s}}{2}\alpha\bar{\lambda}
    -i\frac{s+\bar{s}}{2}\Theta\gamma^{\hat 3}\bar{\lambda}-3i\frac{s+\bar{s}}{2} \alpha \bar{\lambda} \,,
\\
\left[\delta_{\epsilon},\delta_{\bar{\epsilon}}\right] A_{i} &=
    -(\cL_{\xi} A)_{i} + D_{i}\Lambda \,,
\\
\left[\delta_{\epsilon},\delta_{\bar{\epsilon}}\right] \sigma_{1} &=
    -\cL_{\xi} \sigma_{1}+i\left[\Lambda,\sigma_{1}\right] - (s+\bar{s})\Theta \sigma_{2} -i (s+\bar{s}) \alpha \sigma_{1} \,,
\\
\left[\delta_{\epsilon},\delta_{\bar{\epsilon}}\right] \sigma_{2} &=
    -\cL_{\xi} \sigma_{2}+i\left[\Lambda,\sigma_{2}\right] + (s+\bar{s})\Theta \sigma_{1} -i (s+\bar{s}) \alpha \sigma_{2} \,,
\\
\left[\delta_{\epsilon},\delta_{\bar{\epsilon}}\right] \rmD &=
    -\cL_{\xi} \rmD+i\left[\Lambda,\rmD\right] - 2i (s+\bar{s}) \alpha \rmD \,,
\end{aligned}
\end{equation}
and $[\delta_{\epsilon},\delta_{\epsilon}]=[\delta_{\bar{\epsilon}},\delta_{\bar{\epsilon}}]=0$ on all the fields. Therefore $\left[\delta_{\epsilon},\delta_{\bar{\epsilon}}\right]$ generates a space-time transformation
as well as a gauge transformation, an $R$ and $\cA$ $R$-symmetry transformation and a Weyl transformation. The parameters of these transformations are given by
\begin{equation}
\label{ddbparameters}
\begin{aligned}
\xi^{i} &= -i\bar{\epsilon}\gamma^{i}\epsilon \,,
\\
\Lambda &=(\bar{\epsilon}\epsilon)\sigma_{1}-i(\bar{\epsilon}\gamma^{\hat 3}\epsilon)\sigma_{2} +\xi^iA_i\,,
\\
\Theta &=\frac{1}{2r}\bar{\epsilon}\epsilon \,,
\\
\alpha &= -\frac{1}{2r}\bar{\epsilon}\gamma^{\hat 3}\epsilon\,,
\end{aligned}
\end{equation}
where we have omitted the subscript $s$ and $\bar{s}$ on the spinors. Note that the spacetime transformation is realized by the Lie derivative on bosonic fields and by the Lie-Lorentz derivative \rf{LLd} on the fermions.
More explicitly, the Lie-Lorentz derivative along the vector field $\xi$ is given by
\begin{equation}
\cL_{\xi} = \xi^{j}\nabla_{j}-i\frac{s-\bar{s}}{2} \Theta\gamma^{\hat 3}\,.
\end{equation}
The superconformal algebra is realized on the chiral multiplet fields as
\begin{equation}
\label{sca-cfields}
\begin{aligned}
\left[\delta_{\epsilon},\delta_{\bar{\epsilon}}\right] \psi &=
  -\cL_{\xi}\psi + i\Lambda \psi + i \frac{s-\bar{s}}{2} (1-q)\alpha\psi -i\frac{s+\bar{s}}{2}\Theta\gamma^{\hat 3}\psi
  -i\frac{s+\bar{s}}{2}(q+1)\alpha \psi \,,
\\
\left[\delta_{\epsilon},\delta_{\bar{\epsilon}}\right] \bar{\psi} &=
  -\cL_{\xi}\bar{\psi} -i\bar{\psi}\Lambda + i \frac{s-\bar{s}}{2} (q-1)\alpha\bar{\psi} +i\frac{s+\bar{s}}{2}\Theta\gamma^{\hat 3}\bar{\psi}
  -i\frac{s+\bar{s}}{2}(q+1)\alpha \bar{\psi} \,,
\\
\left[\delta_{\epsilon},\delta_{\bar{\epsilon}}\right] \phi &=
  -\cL_{\xi}\phi + i\Lambda \phi -i \frac{s-\bar{s}}{2} q\alpha\phi
  -i\frac{s+\bar{s}}{2}q\alpha \phi \,,
\\
\left[\delta_{\epsilon},\delta_{\bar{\epsilon}}\right] \bar{\phi} &=
  -\cL_{\xi}\bar{\phi} -i\bar{\phi}\Lambda + i \frac{s-\bar{s}}{2} q\alpha\bar{\phi}
  -i\frac{s+\bar{s}}{2}q\alpha \bar{\phi} \,,
\\
\left[\delta_{\epsilon},\delta_{\bar{\epsilon}}\right] F &=
  -\cL_{\xi}F + i\Lambda F + i \frac{s-\bar{s}}{2} (2-q)\alpha F
  -i\frac{s+\bar{s}}{2}(q+2)\alpha F \,,
\\
\left[\delta_{\epsilon},\delta_{\bar{\epsilon}}\right] \bar{F} &=
  -\cL_{\xi}\bar{F} -i\bar{F}\Lambda + i \frac{s-\bar{s}}{2} (q-2)\alpha \bar{F}
  -i\frac{s+\bar{s}}{2}(q+2)\alpha \bar{F} \,,
\end{aligned}
\end{equation}
where the parameters of the transformations are the same as those for the vector multiplet fields \rf{ddbparameters}.

To obtain the $SU(2|1)$ supersymmetry transformations, we restrict the superconformal transformations \rf{superconff} and \rf{superconfcm} we have constructed to those
associated with $Q_\alpha$ and $S_\alpha$,  which are parametrized by $\epsilon_{+}$ and $\bar{\epsilon}_{-}$. The corresponding  realization of the algebra on the  fields
is given by \rf{sca-fields} and \rf{sca-cfields} with $s=1$ and $\bar{s}=-1$.

In the main text, we find it convenient to perform the field redefinition $\rmD\rightarrow \rmD+\sigma_2/r$, after which we obtain the supersymmetry transformations presented in section~\ref{sec:gauge}.

\section{Supersymmetric Configurations}
\label{app:susyv}

In this appendix we present the derivation of the choice of SUSY parameters and the corresponding supersymmetric configurations.

\subsection{Choice of Supercharge}

The conformal Killing spinor equations on $S^{2}$ are
\begin{align}
\label{epsilonplus}
\nabla_{i}\epsilon&=+\frac{1}{2r}\gamma_{i}\gamma^{\hat 3}\epsilon\,,
\\
\label{bepsilonminus}
\nabla_{i}\bar{\epsilon}&=-\frac{1}{2r}\gamma_{i}\gamma^{\hat 3}\bar{\epsilon}\,,
\end{align}
with the general solutions of the form
\begin{align}
\epsilon & =
  \exp\left(-\frac{i\theta}{2}\gamma_{\hat{2}}\right)\exp\left(\frac{i\varphi}{2}\gamma^{\hat{3}}\right)\epsilon_{\circ}\,,
\\
\bar{\epsilon} &=
  \exp\left(+\frac{i\theta}{2}\gamma_{\hat{2}}\right)\exp\left(\frac{i\varphi}{2}\gamma^{\hat{3}}\right)\bar{\epsilon}_{\circ}\,.
\end{align}
Here, the hatted $\gamma$ indices denote the tangent space (flat) indices\footnote{See appendix~\ref{app:conven}.}. The corresponding bilinear \(\xi^i = -i\bar{\epsilon}\gamma^{i} \epsilon\) is given by
\begin{align}
\xi^1 & = -\cos\varphi\left(i\bar{\epsilon}_{\circ}\hat{\gamma}^{1}\epsilon_{\circ}\right)-\sin\varphi\left(i\bar{\epsilon}_{\circ}\hat{\gamma}^{2}\epsilon_{\circ}\right)\,,\\
\xi^2 & = -\bar{\epsilon}_{\circ}\epsilon_{\circ}+\cot\theta\sin\varphi\left(i\bar{\epsilon}_{\circ}\hat{\gamma}^{1}\epsilon_{\circ}\right)-
\cot\theta\cos\varphi\left(i\bar{\epsilon}_{\circ}\hat{\gamma}^{2}\epsilon_{\circ}\right)\,.
\end{align}
We wish to find spinors such that \(\xi^1\) vanishes while \(\xi^2\) is a non-zero constant. The vanishing on \(\xi^{1}\) for all angles \(\varphi\) requires
\(\bar{\epsilon}_\circ \gamma^1 \epsilon_\circ = \bar{\epsilon}_\circ \gamma^2 \epsilon_\circ = 0\). This can be achieved by choosing \(\epsilon_\circ\) and \(\bar{\epsilon}_\circ\) to be chiral spinors with opposite
chirality. We choose the constant spinors such that
\begin{align}
\gamma^{\hat 3}\epsilon_{\circ}&=+\epsilon_{\circ}\,,
\\
\gamma^{\hat 3}\bar{\epsilon}_{\circ}&=-\bar{\epsilon}_{\circ}\,,
\end{align}
and the conformal Killing spinors reduce to
\begin{align}\label{eq:ourCKS}
\epsilon &= \exp\left(-\frac{i\theta}{2}\gamma_{\hat{2}}+\frac{i\varphi}{2}\right)\epsilon_{\circ}\,,
\\
\bar{\epsilon} &= \exp\left(+\frac{i\theta}{2}\gamma_{\hat{2}}-\frac{i\varphi}{2}\right)\bar{\epsilon}_{\circ}\,.
\end{align}
The spinor bilinears constructed out of these spinors take the form
\begin{align}
\bar{\epsilon}\epsilon &= \bar{\epsilon}_{\circ}\epsilon_{\circ} \cos\theta\,,
\\
\xi &= -\frac{1}{r} \bar{\epsilon}_{\circ}\epsilon_{\circ}\frac{\partial}{\partial\varphi}\,,
\\
\alpha&=-\frac{1}{2r}\bar{\epsilon}_{\circ}\epsilon_{\circ}\,.
\end{align}

\subsection{Supersymmetric Saddle Point Equations}

Since after localization, only supersymmetric configurations can contribute, we write $\cQ f = 0 $ for all fermionic fields, with $\cQ$ parametrized by the particular choice of \(\epsilon\) and \(\bar{\epsilon}\) we just derived.
Let us fix the relative normalization of $\epsilon_\circ$ and $\bar{\epsilon}_{\circ}$ such that
\begin{equation}
\bar{\epsilon}_{\circ} = -i\gamma^{\hat{2}}\epsilon_{\circ}
\end{equation}
We thus obtain the explicit expressions
\begin{align}
  \epsilon
  & = e^{i\varphi/2}
    \left( \cos\frac{\theta}{2}
    -i \sin\frac{\theta}{2} \gamma^{\hat{2}} \right) \epsilon_{\circ}
  &
  \bar{\epsilon}
  & = e^{-i\varphi/2}
    \left( \sin\frac{\theta}{2}
    -i \cos\frac{\theta}{2} \gamma^{\hat{2}} \right) \epsilon_{\circ}
  \\
  \gamma^{\hat{1}} \epsilon
  & = e^{i\varphi/2}
    \left( \sin\frac{\theta}{2}
    -i \cos\frac{\theta}{2} \gamma^{\hat{2}} \right) \epsilon_{\circ}
  &
  \gamma^{\hat{1}} \bar{\epsilon}
  & = e^{-i\varphi/2}
    \left( \cos\frac{\theta}{2}
    -i \sin\frac{\theta}{2} \gamma^{\hat{2}} \right) \epsilon_{\circ}
  \\
  \gamma^{\hat{2}} \epsilon
  & = e^{i\varphi/2}
    \left( -i\sin\frac{\theta}{2}
    + \cos\frac{\theta}{2} \gamma^{\hat{2}} \right) \epsilon_{\circ}
  &
  \gamma^{\hat{2}} \bar{\epsilon}
  & = e^{-i\varphi/2}
    \left(-i \cos\frac{\theta}{2}
    + \sin\frac{\theta}{2} \gamma^{\hat{2}} \right) \epsilon_{\circ}
  \\
  \gamma^{\hat{3}} \epsilon
  & = e^{i\varphi/2}
    \left( \cos\frac{\theta}{2}
    + i \sin\frac{\theta}{2} \gamma^{\hat{2}} \right) \epsilon_{\circ}
  &
  \gamma^{\hat{3}} \bar{\epsilon}
  & = e^{-i\varphi/2}
    \left(  \sin\frac{\theta}{2}
    + i \cos\frac{\theta}{2} \gamma^{\hat{2}} \right) \epsilon_{\circ}
\end{align}
Thanks to those expressions for various gamma matrices acting on our conformal Killing spinors, \(\delta\lambda = 0\) and \(\delta \bar{\lambda} = 0\) may be written as
\begin{align}
  \begin{split}
    0 = \delta \lambda & = \left[
      \sin\frac{\theta}{2} \left(iV_{\hat{1}} + V_{\hat{2}} \right)
      + i\cos\frac{\theta}{2} \left( V_{\hat{3}} + i\rmD \right)
    \right] e^{i\frac{\varphi}{2}} \epsilon_0 \\
    & \quad + \left[
      \cos\frac{\theta}{2} \left( V_{\hat{1}} + iV_{\hat{2}} \right)
      - \sin\frac{\theta}{2} \left( V_{\hat{3}} -i \rmD \right)
    \right] e^{i\frac{\varphi}{2}}\gamma^{\hat{2}} \epsilon_0
  \end{split}
  \\
  \begin{split}
    0 = \delta \bar{\lambda} & = \left[
      \cos\frac{\theta}{2} \left( i\bar{V}_{\hat{1}} + \bar{V}_{\hat{2}} \right)
      + i\sin\frac{\theta}{2} \left( \bar{V}_{\hat{3}} -i \rmD \right)
    \right] e^{-i\frac{\varphi}{2}} \epsilon_0 \\
    & \quad + \left[
      \sin\frac{\theta}{2} \left( \bar{V}_{\hat{1}} + i \bar{V}_{\hat{2}} \right)
      - \cos\frac{\theta}{2} \left( \bar{V}_{\hat{3}} + i\rmD \right)
    \right] e^{-i\frac{\varphi}{2}} \gamma^{\hat{2}} \epsilon_0 .
  \end{split}
\end{align}
while $\delta\psi = 0$ and $\delta\bar{\psi} = 0$ yields
\begin{align}
\begin{split}
  0= \delta \psi & =
    i \left[ \sin\frac{\theta}{2} \left( D_{-}\phi -ie^{-i\varphi} F \right)  + \cos\frac{\theta}{2} \left(\sigma_{1} -i\sigma_{2} + \frac{q}{2r}\right)\phi \right] e^{i\frac{\varphi}{2}}\epsilon_{\circ}
    \\
    &\quad + \left[ \cos\frac{\theta}{2} \left( D_{+} \phi -ie^{-i\varphi} F \right) + \sin\frac{\theta}{2} \left(\sigma_{1} + i\sigma_{2} - \frac{q}{2r}\right)\phi \right]
      e^{i\frac{\varphi}{2}}\gamma^{\hat{2}}\epsilon_{\circ} ,
\end{split}
\\
\begin{split}
  0 = \delta \bar{\psi} & =
    i \left[ \cos\frac{\theta}{2} \left( D_{-}\bar{\phi} -ie^{i\varphi} \bar{F}\right) + \sin\frac{\theta}{2}\,\bar{\phi} \left(\sigma_{1} + i\sigma_{2} + \frac{q}{2r}\right) \right] e^{-i\frac{\varphi}{2}}\epsilon_{\circ}
    \\
    &\quad + \left[ \sin\frac{\theta}{2} \left( D_{+}\bar{\phi} -ie^{i\varphi} \bar{F}\right) + \cos\frac{\theta}{2}\,\bar{\phi} \left(\sigma_{1} -i\sigma_{2} + \frac{q}{2r}\right) \right]
      e^{-i\frac{\varphi}{2}}\gamma^{\hat{2}}\epsilon_{\circ} \, .
\end{split}
\end{align}
Here $D_{\pm}= D_{\hat{1}} \pm i D_{\hat{2}}$ and for future reference, we define $\sigma_{\pm}=\sigma_{1}\pm i\sigma_{2}$.
Since $\epsilon_{\circ}$ and $\gamma^{\hat{2}}\epsilon_{\circ}$ are linearly independent, each square bracket must vanish separately. Using the reality conditions
\begin{equation}
\begin{aligned}
A_{i}^{\dagger}&= A_{i}& \hspace{70pt} \bar{\phi}^{\dagger}&=\phi
\\
\sigma_{\pm}^{\dagger}&=\sigma_{\mp} & \bar{F}^{\dagger}&=F
\end{aligned}
\end{equation}
we can write the equations as
\begin{align}
\begin{split}
  \sin\frac{\theta}{2} D_{\pm}\sigma_{+} + \cos\frac{\theta}{2}\left(F_{\hat{1}\hat{2}}+\frac{\sigma_{1}}{r} + i\rmD \mp i\left[ \sigma_{1}, \sigma_{2}\right] \right) &=0
  \\
  \cos\frac{\theta}{2} D_{\pm}\sigma_{-} - \sin\frac{\theta}{2}\left(F_{\hat{1}\hat{2}}+\frac{\sigma_{1}}{r} -i\rmD \pm i\left[ \sigma_{1}, \sigma_{2}\right] \right) &= 0 \,
\end{split}
\\[6pt]
\begin{split}
  \sin\frac{\theta}{2} \left( D_{-}\phi \pm ie^{-i\varphi} F \right) + \cos\frac{\theta}{2} \left( \sigma_{\mp}+\frac{q}{2r} \right)\phi  &= 0
  \\
  \cos\frac{\theta}{2} \left( D_{+}\phi \pm ie^{-i\varphi} F \right) + \sin\frac{\theta}{2} \left( \sigma_{\pm}-\frac{q}{2r} \right)\phi  &= 0 \, .
\end{split}
\end{align}
Taking linear combinations of each set of these equations and using the reality conditions, we obtain the desired SUSY equations
\begin{align}
\label{seq1}
\begin{aligned}
  D_{\hat{2}} \sigma_1 = D_{\hat{2}} \sigma_2 = D_{\hat{1}} \sigma_2 & = 0 &
  \operatorname{Re} \rmD = [ \sigma_1 , \sigma_2 ] & = 0 \\
  D_{\hat{1}} \sigma_1 - \operatorname{Im} \rmD \sin \theta & = 0 & \qquad
   F_{\hat{1}\hat{2}}  + \frac{\sigma_1}{r} - \operatorname{Im}\rmD \cos\theta & = 0 \,,
\end{aligned}
\\[3pt]
\label{seq2}
\begin{aligned}
  \cos\frac{\theta}{2}D_{+}\phi + \sin\frac{\theta}{2} \left( \sigma_{1}-\frac{q}{2r} \right)\phi &=0 & \quad \sigma_{2}\phi &=0
  \\
  \sin\frac{\theta}{2}D_{-}\phi + \cos\frac{\theta}{2} \left( \sigma_{1}+\frac{q}{2r} \right)\phi &=0 & F &=0 \, .
\end{aligned}
\end{align}

\subsection{\texorpdfstring{$\cQ$}{Q}-Supersymmetric Field Configurations}

To compute the path integral using localization on supersymmetric configurations, we need to find the space of solutions of equations \rf{seq1} and \rf{seq2}.

Let us first analyze the vector multiplet field equations.

  For concreteness, we choose the coordinate patch $0<\theta<\pi$, where we can gauge away the $\rmd\theta$-component of the gauge field\footnote{Every 1-form $w=w_{\theta}\rmd\theta$ on $S^{2}$ is, up to $\rmd\varphi$
  terms, closed and therefore exact -- since the $H^{1}(S^{2})=0$.}. The general solution to \rf{seq1} takes the form
  \begin{equation}
  A= r\sigma_{1} \cos\theta\, \rmd\varphi, \qquad \sigma_{1}=\sigma_{1}(\theta), \qquad \sigma_{2}=\sigma_{2}(\varphi)\,.
  \end{equation}
  Imposing the chiral multiplet supersymmetry equations \rf{seq2} and plugging in the above form for the vector multiplet fields we obtain
  \begin{equation}
  \begin{aligned}
    \left(\sin\theta \ \partial_{\theta} + \frac{q}{2} \cos\theta + \sigma_{1} \right) \phi &=0
    \\
    \left(\partial_{\varphi} + i\frac{q}{2}\right)\phi &= 0
  \end{aligned}
  \hspace{50pt}
  \begin{aligned}
    F &= 0
    \\[6pt]
    (\sigma_{2} + m ) \phi &= 0
  \end{aligned}
  \end{equation}
  where we have also included the mass term which, as explained in section~\ref{sec:gauge} is just a shift in $\sigma_{2}$ by a diagonal matrix valued in the flavor symmetry group.
  For generic values of $R$-charges $q$, the only solution of the above equations which is periodic in $\varphi$ is
  \begin{equation}
  \phi=0\,.
  \end{equation}
  Consequently, in the absence of \emph{effective} Fayet-Iliopoulos parameters,
  the reality conditions necessary for having a convergent path integral constrain the vector multiplet auxiliary field to vanish, i.e.\@
  \begin{equation}
    \Im \rmD= - g^{2} \phi\bar{\phi} = 0\,.
  \end{equation}
  The vanishing of the auxiliary field in turn forces $\sigma_{1}$ to be a constant and the general solution to the supersymmetry equations \rf{seq1} and \rf{seq2} takes the form
  \begin{equation}
  \begin{aligned}
    A &=\frac{B}{2}(\kappa-\cos\theta) \rmd \varphi   \qquad &     \sigma_1&=-\frac{B}{2r}
    \\
    \sigma_2&=a  &   \rmD&=0
    \\[4pt]
    \phi=\bar{\phi}^{\dagger}&=0   &  F=\bar{F}^{\dagger}&=0
  \end{aligned}
  \end{equation}
  where $\delta A = \frac{\kappa B}{2}\rmd\varphi$ is the appropriate gauge transformation to extend the solution to the coordinate patches including the north pole (with $\kappa=1$) or the south pole (where $\kappa=-1$).
  We conclude that for general $R$-charge assignments, $\cF_0$ -- the space  of smooth solutions to the supersymmetry fixed point equations -- is parametrized by two constant matrices, $a$ and $B$,
  where $B$ is further constrained by the first Chern class quantization to take integer values.

  We note in passing that for special values of the $R$-charges, there exist non-trivial solutions to the chiral multiplet supersymmetry equations which take the form
  \begin{equation}
  \phi = e^{\frac{i}{2}(\kappa B -q)\varphi} \frac{  (\sin\frac{\theta}{2})^{ \frac{B-q}{2} }  }{  (\cos\frac{\theta}{2})^{ \frac{B+q}{2} }  } \phi_{\circ},\qquad\text{subject to}\qquad (a+m)\phi_{\circ} = 0\,.
  \end{equation}

\section{One-Loop Determinants}
\label{app:one-loop}

Here we present the computation of the one-loop determinants in the localization computation of  the partition function. Our starting point is the quadratic part of the vector and chiral multiplet actions
\rf{vm_action} and   \rf{chiralmult} in the background \rf{saddlept} with the addition of the gauge fixing ghosts $\bar{c},c$ and the Lagrange multiplier ${b}$. The various terms are
\begin{align}
\nonumber
S^{\text{v.m.}}_{\text{b}} &=\int \rmd^{2}x \sqrt{h} \Tr\Bigg\{
A^{i} \left(\rmM^{2}+\frac{1}{r^{2}} \right) A_{i} + \frac{i}{2r^{2}} \varepsilon_{ij}A^{i}\big[B,A^{j}\big] + \frac{2}{r} \sigma_{1}\varepsilon^{ij}D_{i}A_{j}
\\ & \phantom{ =\ \int \rmd^{2}x \sqrt{h}\ \Tr\ }
+\sigma_{1}\left(\rmM^{2}+\frac{1}{r^{2}}\right)\sigma_{1} +\sigma_{2}\rmM^{2}\sigma_{2} +\rmD^{2} - \cG^{2}
\Bigg\}\,,\label{quaddr}
\\
S^{\text{v.m.}}_{\text{f}} &=\int \rmd^{2}x \sqrt{h} \Tr\Bigg\{
\bar{\lambda}\left( i\slashed{D}-\frac{i}{2r} \big[B, \cdot\ \big] + \gamma^{\hat 3}\big[a ,\cdot\ \big] \right) \lambda
\Bigg\}\,,
\\
S_{\text{ghost}} &=\int \rmd^{2}x \sqrt{h} \Tr\Bigg\{\bar{c}\rmM^{2}c-b\cG(A_{i},\sigma_{1},\sigma_{2})\Bigg\}\,,
\\
S^{\text{c.m.}}_{\text{b}} &=\int \rmd^{2}x\ \sqrt{h}\quad \Bigg\{
\bar{\phi}\left(\rmM^{2}+i\frac{q-1}{r} a -\frac{q^{2}-2q}{4r^{2}}\right)\phi+\bar{F}F
\Bigg\}\,,
\\
S^{\text{c.m.}}_{\text{f}} &=\int \rmd^{2}x\ \sqrt{h}\quad \Bigg\{
\bar{\psi} \left(-i\slashed{D}-\frac{i}{2r}B-\Big(a+\frac{iq}{2r}\Big)\gamma^{\hat 3} \right) \psi
\Bigg\}\,,
\end{align}
where $\cG$ is the gauge fixing condition corresponding to the choice of gauge
\begin{equation}
\cG(A_{i},\sigma_{1},\sigma_{2}) = D_{i}A^{i}+\frac{i}{2r}\big[B,\sigma_{1}\big]-i\big[ a,\sigma_{2} \big] =0,
\label{gaugefix}
\end{equation}
and $\rmM^{2}$ is given by
\begin{equation}
\rmM^{2} = -D_{i}^{2}+\frac{1}{4r^{2}}  B^{2}+a^{2}\,,
\end{equation}
where $a$ and $B$ act in the appropriate representations. We note that \rf{gaugefix} is the background gauge field choice $D_M A^M=0$ in   four dimensions dimensionally reduced to two dimensions.
This choice simplifies computations considerably.

The integral over $b$ imposes the background field gauge \rf{gaugefix} while integrating out the auxiliary fields $\rmD$ and $F$ yields a trivial factor. We now analyze the rest.

\subsection{Dirac Operator in Monopole Background}

Before  computing the one-loop determinant contribution of fermionic fields, let us first derive the spectrum of the Dirac operator in the background \rf{saddlept}. Since the index of the Dirac operator,
acting in the representation $\bf R$ of the gauge algebra, is given by
\begin{equation}
\operatorname{ind}(\slashed D)=\frac{1}{2\pi}\int_{S^2}\Tr F=\Tr B \, ,
\end{equation}
we anticipate $|\Tr B|$  zero-modes. Excluding these modes, we may diagonalize the Dirac operator using spinor monopole harmonics. For each weight $w$ of the representation $\bf R$ and each mode $(J,m)$
such that $J>|B_{w}|/2$ and $-J\leq m\leq J$ we have
\begin{equation}
\label{tracelessDirac}
(i\slashed{D})_{J,m}=
\left(\begin{array}{cc}
       \lambda_{J,m} & 0 \\
	0 & -\lambda_{J,m}
      \end{array}
\right)
\end{equation}
since $i\slashed{D}$ is traceless. The spectrum of $i\slashed{D}$ can easily be derived from the spectrum of $-\slashed{D}^{2}$ when expressed in terms of the scalar Laplacian
\begin{equation}
\label{dslash2}
(i\slashed{D})^{2}=
\left(\begin{array}{cc}
 -(D^{-}_{i})^{2}+\frac{1-B_{w}}{2r^{2}}  &  0
  \\
  0  & -(D^{+}_{i})^{2}+\frac{1+B_{w}}{2r^{2}}
\end{array}\right)\,.
\end{equation}
Here $(D_{i}^{\pm})^2\equiv(\partial_{i}-i\frac{B_{w}\pm 1}{2}\omega_{i})^2$ denotes the scalar Laplacian in the monopole background with monopole charge $\frac{B_{w}\pm 1}{2}$. The connection $\omega_{i}$
is expressed in terms of the spin connection \rf{spinconnection} as  $\omega_i= \omega^{\hat 1\hat 2}_i$. In the rest of this subsection, we drop the subscript in $B_{w}$ to avoid cluttering the notation.

The eigen-value of the scalar Laplacian in the $(J,m)$ mode is given by
\begin{equation}
\label{d2jm}
-(D^{\pm}_{i})^{2}_{J,m}= \frac{J(J+1)}{r^2}-\frac{\left(B\pm1\right)^2}{4r^2}\,,
\end{equation}
where $J$ runs from $\frac{|B\pm1|}{2}$ to $\infty$ in integer steps and the multiplicity in each mode is $2J+1$. Using this expression for the eigenvalues and the relation between the eigenvalues of the
scalar Laplacian, which can be easily read off from \rf{tracelessDirac} and \rf{dslash2}, we conclude that the spectrum of the Dirac operator consists of
\begin{align}
  0,&  \qquad  & \text{with multiplicity $|B|$, and} \\
 +\sqrt{ \frac{(J+\frac{1}{2})^2-(\frac{B}{2})^{2}}{r^2} },&\qquad  J={\frac{|B|+1}{2}, \dots} \qquad  & \text{with multiplicity $2J+1$,} \\
 -\sqrt{ \frac{(J+\frac{1}{2})^2-(\frac{B}{2})^{2}}{r^2} },&\qquad  J={\frac{|B|+1}{2}, \dots} \qquad  & \text{with multiplicity $2J+1$}
\end{align}
for $J=\frac{|B|+1}{2},\ldots$
We also note   that the fermonic zero-modes are  spinors of a definite chirality, which depends on the sign of $B$.

\subsection{Chiral Multiplet Determinant}

Using the spectrum of the Dirac operator we just derived, we can easily compute the fermionic determinant of the chiral multiplet. First, note that $\gamma^{\hat 3}$ anticommutes with $\slashed{D}$, hence,
a shift in $\slashed{D}$ by $\gamma^{\hat 3}$
results in a shift in the square of the eigenvalues. Therefore, we have
\begin{align}
\nonumber
\det \Delta^{\text{c.m.}}_{\text{f}}  &= \det\bigg[ -i\slashed{D}-\frac{iB}{2r}-\left(a+\frac{iq}{2r}\right)\gamma^{\hat 3} \bigg]
\\
\nonumber
&=
\prod_{w} \Biggl[ (-i)^{B_w} \left(\frac{q+ |B_{w}| }{2r}  -i a_{w}\right)^{|B_w|}
\nonumber \\ & \hspace*{0.5cm} \times
\prod_{J=\frac{|B_{w}|+1}{2}}^{\infty}  \left[ -\left(\frac{B_{w}}{2r}\right)^2- \left(  \frac{(J+\frac{1}{2})^2-(\frac{ B_{w} }{2})^{2}}{r^2}+\left(a_{w}+\frac{iq}{2r}\right)^{2} \right) \right]^{2J+1} \Biggr] \\
&=
\prod_{w} \Biggl[ (-i)^{B_w} \prod_{J=0}^{\infty} \Bigg[
  \left(\frac{J}{r}+\frac{|B_{w}|+q}{2r}-i a_{w} \right)^{2J+|B_{w}|}
  \nonumber \\ & \hspace*{3cm}\times  (-i)^{2|B_w|}
  \left(\frac{J+1}{r}+\frac{|B_{w}|-q}{2r}+i a_{w} \right)^{2J+|B_{w}|+2}
\Bigg]\Biggr]\,.\nonumber
\\
&=
\prod_{w} \Biggl[ (-1)^{(B_w + |B_{w}|)/2} \prod_{J=0}^{\infty} \Bigg[
  \left(\frac{J}{r}+\frac{|B_{w}|+q}{2r}-i a_{w} \right)^{2J+|B_{w}|}
  \nonumber \\ & \hspace*{4.5cm}\times 
  \left(\frac{J+1}{r}+\frac{|B_{w}|-q}{2r}+i a_{w} \right)^{2J+|B_{w}|+2}
\Bigg]\Biggr]\,.\nonumber
\end{align}
Here we have used the notation $x_w\equiv x\cdot w$, where $w$ are the weights of the representation $\bf R$ under which the chiral multiplet transforms. In the last line zeta-function regularization is used to regularize the sign factor in the infinite product 
\begin{equation}
\left[\prod_{J=0}^{\infty} (-i)^{2|B_{w}|}\right]_{\text{reg}} = (-i)^{(1+\zeta(0))2|B_{w}|} = (-i)^{|B_{w}|}.
\end{equation}

The bosonic determinant may be written as
\begin{align}
\nonumber
(\det\Delta^{\text{c.m.}}_{\text{b}})^{\frac{1}{2}} &=
  \prod_{w}\prod_{J=\frac{|B_{w}|}{2}}^{\infty}
    \Bigg[
      \left(\frac{J+\frac{1}{2}}{r}\right)^{2} + \left(a_{w}+i\frac{q-1}{2r}\right)^2
    \Bigg]^{2J+1}
\\
&=
\prod_{w} \prod_{J=0}^{\infty} \left[
  \left(\frac{J}{r}+\frac{|B_{w}|+q}{2r}-i a_{w} \right) \cdot
  \left(\frac{J+1}{r}+\frac{|B_{w}|-q}{2r}+i a_{w} \right)
\right]^{2J+|B_{w}|+1}.
\end{align}
Putting the two together we have the one-loop contribution from the chiral multiplet fields:
\begin{equation}
Z^{\text{c.m.}}_{\text{one-loop}}(a,B,m) = \prod_{w\in {\bf R}} \left[ (-1)^{(B_w + |B_{w}|)/2} \prod_{J=0}^{\infty} \left[
  \frac{ J+1+\frac{|B_{w}|-q}{2}+ir a_{w} }{ J+\frac{|B_{w}|+q}{2}-ir a_{w} }\right] \right]
\end{equation}
These infinite products can be regularized using Euler's gamma function
\begin{equation}
\frac{1}{\Gamma(z)}=\left[ \prod_{J=0}^\infty (z+J)\right]_{\text{reg}}
\end{equation}
to yield, in the presence of a twisted mass $m$  introduced by shifting $a\rightarrow a+m$
\begin{align}
  Z_{\text{one-loop}}^{\text{c.m.}}(a,B,m)
  & =\prod_{w\in {\bf R}} \left[ (-1)^{(B_w + |B_{w}|)/2}\frac{\Gamma\left(\frac{q}{2} -ir(a_w +m) + \frac{|B_w|}{2}\right)} {\Gamma\left(1 - \frac{q}{2} + ir(a_w+m) + \frac{|B_w|}{2}\right)} \right] \,.
\end{align}
The chiral multiplet determinant has a pole when  $a+m$ has a zero and $q$ is a non-positive integer. More precisely, there is a pole whenever
  $|B|\leq -q$  with $B-q$   even when acting on $\phi$. These poles are due to the zero modes found in
  \rf{zeromodes}, which exist precisely under these conditions. In evaluating the determinant for these tuned values of $q$,  the zero modes must be excluded, thus yielding a  finite  result.

 \subsection{Vector Multiplet Determinant}

The fermion contribution  to the vector multiplet one-loop determinant is the same as that of a chiral multiplet in the adjoint representation with   $R$-charge $q=0$. It is given by
\begin{align}
\nonumber
\det &\Delta^{\text{v.m.}}_{\text{f}}  =
\prod_{\alpha\in \Delta} (-1)^{(B_{\alpha}+|B_{\alpha}|)/2} \prod_{J=0}^{\infty} \Bigg[
  \left(\frac{J}{r}+\frac{|B_{\alpha}|}{2r}-i a_{\alpha} \right)^{2J+|B_{\alpha}|}\left(\frac{J+1}{r}+\frac{|B_{\alpha}|}{2r}+i a_{\alpha} \right)^{2J+|B_{\alpha}|+2}
\Bigg]
\\
&=
\prod_{\alpha\in \Delta_{+}} (-1)^{B_{\alpha}} \prod_{J=0}^{\infty} \Bigg\{
  \left[ \left(\frac{J}{r}+\frac{|B_{\alpha}|}{2r}\right)^{2} + a_{\alpha}^{2} \right]^{2J+|B_{\alpha}|}
  \left[ \left(\frac{J+1}{r}+\frac{|B_{\alpha}|}{2r}\right)^{2} + a_{\alpha}^{2} \right]^{2J+|B_{\alpha}|+2}
\Bigg\}\,.
\end{align}
where  $\alpha\in\Delta_{+}$ are  the positive roots of the Lie algebra of $G$.

In order to compute the contribution from the bosonic fields, we need to write down the mode expansion of the fields. For the scalars fields $\sigma_{1}$ and $\sigma_{2}$, we may use the expansion in the
standard scalar monopole harmonics
\begin{equation}
\sigma_{s}^{\alpha}=\sum_{J=\frac{|B_\alpha|}{2}}^{\infty}\sum_{m=-J}^{J}
\frac{1}{r}\sigma_{s,J,m}^{\alpha}Y^{\frac{|B.\alpha|}{2}}_{J,m}
\end{equation}
where we have introduced a factor of $\frac{1}{r}$ for normalization and $s=1,2$.
As for the gauge field, the mode expansion is much more subtle. A basis of monopole vector spherical harmonics is given in \cite{Weinberg:1993sg}. Expanding the gauge field in this basis we find
\begin{align}
  A^{\alpha}_{i}=\sum_{\lambda=\pm}\sum_{J=J^{\lambda}_{0}}^{\infty}\sum_{m=-J}^{J} A^{\alpha,\lambda}_{J,m} \left(C^{\lambda, \frac{B_{\alpha} }{2}}_{J,m} \right)_{i}\,,
\end{align}
where $J_{0}^{\pm}=\frac{|B_{\alpha} |}{2}\mp1$ for $\frac{|B_{\alpha}| }{2}\ge 1$ and $J_{0}^{\pm}=\frac{|B_{\alpha} |+1}{2}\mp \frac{1}{2} $ otherwise.
The reality condition on the gauge field then implies $A_{-\alpha}=A_{\alpha}^{*}$ and   for scalars   $\sigma_{s,-\alpha}=\sigma_{s,\alpha}^{*}$. The explicit form of
$\left(C^{\lambda, \frac{B_{\alpha}}{2}}_{J,m} \right)_{i}$
is not necessary for our computation and will be omitted here. All we need are some basic properties of the basis elements which are
\begin{align}
  \delta^{\lambda^{\prime}}_{\lambda}\delta^{J^{\prime}}_{J}\delta^{m^{\prime}}_{m}
  &=\int \rmd^{2}x \sqrt{h} \left(C^{\lambda^{\prime}, \frac{B_{\alpha} }{2}}_{J^{\prime},m^{\prime}} \right)^{*}_{i} \left(C^{\lambda, \frac{B_{\alpha}}{2}}_{J,m} \right)^{i}\,,
\\
  -D_{j}^{2} \left(C^{\lambda, \frac{B_{\alpha} }{2}}_{J,m} \right)^{i} &= \frac{1}{r^{2}}\left[ J(J+1)-\left(\frac{|B_{\alpha} |}{2}-\lambda\right)^{2} \right] \left(C^{\lambda, \frac{B_{\alpha} }{2}}_{J,m} \right)^{i}\,,
\\
  D_{i}\left(C^{\lambda, \frac{B_{\alpha} }{2}}_{J,m} \right)^{i} &= -\frac{1}{\sqrt{2}r^{2}}\sqrt{J(J+1)-\frac{|B_{\alpha}|}{2}\left(\frac{|B_{\alpha}|}{2}-\lambda\right)} Y^{\frac{|B_{\alpha}|}{2}}_{J,m}\,,
\\
  i\varepsilon_{ij} \left(C^{\lambda, \frac{B_{\alpha}}{2}}_{J,m} \right)^{j} &= - \lambda \left(C^{\lambda, \frac{B_{\alpha}}{2}}_{J,m} \right)_{i}\,.
\end{align}
Using the above expansion for the gauge field and the scalars and performing the integral over $S^2$,
the bosonic part of the vector multiplet action in \rf{quaddr}  can be written as
\begin{align}
\nonumber
S_{\text{b}}^{\text{v.m.}} &\simeq
 \sum_{\lambda=\pm}\sum_{J=J^{\lambda}_{0}}^{\infty}\sum_{m=-J}^{J}
  A^{-\alpha,\lambda}_{J,m} \left[
  \frac{J(J+1)}{r^{2}}+ a_{\alpha}^{2} + \lambda\frac{B_{\alpha}}{2r^{2}}\right]
  A^{\alpha,\lambda}_{J,m}
\\
\nonumber
& \quad
  -\sum_{\lambda=\pm}\sum_{J=\frac{|B_{\alpha}|}{2}}^{\infty}\sum_{m=-J}^{J} \sigma_{1,J,m}^{-\alpha}
  i\lambda\sqrt{2}\ \frac{\sqrt{J(J+1)-\frac{|B_{\alpha}|}{2}\left(\frac{|B_{\alpha}|}{2}-\lambda\right)} }{r^{2}}
  A^{\alpha,\lambda}_{J,m}
\\
\nonumber
& \quad
  +\sum_{s=1,2}\sum_{J=\frac{|B_{\alpha}|}{2}}^{\infty}\sum_{m=-J}^{J} \sigma_{s,J,m}^{-\alpha}
  \left[\frac{J(J+1)}{r^{2}}+ a_{\alpha}^{2}+\frac{2-s}{r^{2}} \right]\sigma_{s,J,m}^{\alpha}\,,
\end{align}
where there is an implicit summation over all roots $\alpha\in \Delta$.

In order to compute the determinant, it is best to break it down into three factors. The first one isolates the $J=\frac{| B_{\alpha} |}{2}-1$ contribution, which is only non-trivial
when $\frac{| B_{\alpha} |}{2}-1$ is non-negative. In this case we have
\begin{equation}
\det(\Delta_{\text{b},1}^{\text{v.m.}})=
 \prod_{\alpha\in\Delta,|B_{\alpha}|\ge 2}\left[ \left( \frac{B_{\alpha}}{2r} \right)^{2} + a_{\alpha}^{2} \right]^{| B_{\alpha} |-1}\,.
\end{equation}
The second factor is
\begin{equation}
\det(\Delta_{\text{b},2}^{\text{v.m.}})=\frac{ \det(\rmM^{2}) }{\prod_{\alpha\in\Delta} \left[ \left( \frac{B_{\alpha}}{2r} \right)^{2} + a_{\alpha}^{2} \right]^{| B_{\alpha} |+1} }
\end{equation}
where the numerator is just the contribution of $\sigma_{2}$ and the denominator is a factor that we have included to shift the lowest mode of $A^{-}$ (which has $J=|B_\alpha|/2+1$). With this shift,
the rest of the determinant is given by
\begin{align*}
& \det(\Delta_{{\text b},3}^{\text{v.m.}})
\\[-6pt]
& \; = \prod_{\alpha}\prod_{ J=\frac{| B_{\alpha} |}{2} }^{\infty}
\left|
\begin{array}{ccc}
  \frac{J(J+1)-\frac{| B_{\alpha} |}{2} }{ r^{2} }+ a_{\alpha}^{2}  &
    0 &
      \mathllap{-}\frac{i}{ r^{2} } \sqrt{ \frac{ J(J+1)- \frac{| B_{\alpha} |}{2} \big[ \frac{| B_{\alpha} |}{2} + 1 \big] }{ 2 } }
  \\
  0 &
    \frac{J(J+1)+\frac{| B_{\alpha} |}{2} }{ r^{2} }+ a_{\alpha}^{2} &
      \frac{i}{ r^{2} } \sqrt{ \frac{ J(J+1)- \frac{| B_{\alpha} |}{2} \big[ \frac{| B_{\alpha} |}{2} - 1 \big] }{ 2 } }
  \\
  \frac{i}{ r^{2} } \sqrt{ \frac{ J(J+1)- \frac{| B_{\alpha} |}{2} \big[ \frac{| B_{\alpha} |}{2} + 1 \big] }{ 2 } } &
    -\frac{i}{ r^{2} } \sqrt{ \frac{ J(J+1)- \frac{| B_{\alpha} |}{2} \big[ \frac{| B_{\alpha} |}{2} - 1 \big] }{ 2 } } &
      \frac{ J(J+1)+1 }{ r^{2} }+a_{\alpha}^{2}
\end{array}
\right|^{2J+1}
\\[6pt]
& \; =
  \prod_{\alpha\in\Delta}\prod_{J=\frac{|B_{\alpha}|}{2}}^{\infty}\left[
  \left( \frac{J(J+1)}{r^{2}} + a_{\alpha}^{2} \right) \left( \frac{J^{2}}{r^{2}}+ a_{\alpha}^{2} \right)
  \left( \left(\frac{J+1}{r}\right)^{2}+a_{\alpha}^{2} \right) \right]^{2J+1}
\\
& \; = \det(\rmM^{2}) \prod_{\alpha\in\Delta} \prod_{J=0}^{\infty}\left[
  \left( \left(\frac{J}{r}+\frac{|B_{\alpha}|}{2r} \right)^{2}+ a_{\alpha}^{2} \right)
  \left( \left(\frac{J+1}{r}+\frac{|B_{\alpha}|}{2r}\right)^{2}+a_{\alpha}^{2} \right) \right]^{2J+|B_{\alpha}|+1}\,,
\end{align*}
where
\begin{equation}
 \det(\rmM^{2})=\prod_{\alpha\in\Delta}\prod_{J=\frac{|B_\alpha|}{2}}^\infty \left[\frac{J(J+1)}{r^2}+a_\alpha^2\right]^{2J+1}\,.
\end{equation}
Note the shift in the lowest mode of $A^{-}$ at the top left component in the matrix. As we mentioned earlier, this a factor that we multiply and divide by hand to avoid isolating the $J=\frac{|B_{\alpha}|}{2}$ mode.
Note also that in this case the off-diagonal terms $(1,3)$ and $(3,1)$ vanish. Including the contribution from the ghosts -- which is $\det(\rmM^{2})$ -- the one-loop partition function of the vector-multiplet becomes
\begin{align}
\nonumber
\frac{ \det(\Delta^{\text{v.m.}}_{\text{b}})^{\frac{1}{2}} }{\det(\rmM^{2})}
&=
  \frac{ \prod_{\alpha\in\Delta_{+}} \prod_{J=0}^{\infty}\left[
    \left( \left(\frac{J}{r}+\frac{|B_{\alpha}|}{2r} \right)^{2}+ a_{\alpha}^{2} \right)
    \left( \left(\frac{J+1}{r}+\frac{|B_{\alpha}|}{2r}\right)^{2}+a_{\alpha}^{2} \right) \right]^{2J+|B_{\alpha}|+1} }{
  \prod_{\alpha\in\Delta_{+}} \left[ \left( \frac{B_{\alpha}}{2r} \right)^{2} + a_{\alpha}^{2} \right]^{|B_{\alpha}|+1}
    \prod_{\alpha\in\Delta_{+},|B_{\alpha}|\ge 2}
    \left[ \left( \frac{B_{\alpha}}{2r} \right)^{2} + a_{\alpha}^{2} \right]^{-|B_{\alpha}|+1} }
\\
\nonumber
\\
&=
  \left| \det(\Delta^{\text{v.m.}}_{\text{f}}) \right| \prod_{\alpha\in\Delta_{+}}
    \left[\frac{1}{ \left( \frac{B_{\alpha}}{2r} \right)^{2} + a_{\alpha}^{2}} \right]^{|B_{\alpha}|}
  \prod_{\alpha\in\Delta_{+},|B_{\alpha}|\ge 2}
    \left[\frac{1}{ \left( \frac{B_{\alpha}}{2r} \right)^{2} + a_{\alpha}^{2} }\right]^{1-|B_{\alpha}|}\,.
\end{align}
Therefore, we find that
\begin{equation}
  Z^{\text{v.m.}}_{\text{one-loop}}(a,B)
  =\prod_{\substack{\alpha\in\Delta^{+}\\ B_{\alpha}\neq 0}} \left[ (-1)^{B_{\alpha}}  \left[\left(\frac{B_{\alpha}}{2r}\right)^{2} + a_{\alpha}^2\right]\right]\,.
\end{equation}

\section{One-Loop Running of FI Parameter}
\label{FIrunning}

Consider a two dimensional $\cN=(2,2)$ gauge theory with  a $U(1)$ gauge group factor in the presence of an FI parameter $\xi$. When the sum of the $U(1)$ charges of the chiral multiplets
$Q=\sum_{i}Q_{i}$ is non-vanishing, the  FI parameter gets renormalized according to
\begin{equation}
\xi(\mu)=\xi+{\frac{1}{2\pi}}\sum_{j} Q_j\ln\left(\frac{\mu}{M_{\text{UV}}} \right)\,.
\label{runningb}
\end{equation}
In our localization computation, some care has been taken to regularize the theory in a $\cQ$-invariant way. We accomplish this by introducing an ``expectator'' chiral multiplet of charge $-Q$, mass $M$,
and $R$-charge $q=0$. In this enriched theory the FI parameter does not run. However, we recover the original theory by decoupling the expectator chiral multiplet by taking its mass $M$ to be large.
We now demonstrate by analyzing the one-loop determinant of the expectator chiral multiplet that this yields the running of the FI parameter with $M_{\text{UV}}=M$ and $\mu=1/r$.

The relevant one-loop determinant of the expectator chiral multiplet is
\begin{equation}
\label{lnzm}
  \ln Z_{\text{one-loop}}^{\text{c.m.}}(a,B,M) = \ln \left[ \frac{\Gamma\left(\frac{QB+q}{2}+irQa-irM\right)}{\Gamma\left(1+\frac{QB-q}{2}-irQa+irM\right)}\right] + O(1)\,.
\end{equation}
The asymptotic expansion of $\Gamma(z)$ with large imaginary argument is given by
\begin{equation}
  \ln \Gamma(z) = \left(z-\frac{1}{2}\right)\ln z - z + O(1)
\end{equation}
where the   terms of order $1$ depend on the sign of $\operatorname{Im} z$ but are irrelevant for renormalization of $\xi$. Using this asymptotic form for large mass $M$ in \rf{lnzm} yields\begin{align}
\nonumber
 \ln Z_{\text{one-loop}}^{\text{c.m.}}(a,B,M)&\mathrel{\mathop{\simeq}\limits_{rM\gg1}}
    2irM \left( 1- \ln rM \right) + (q-1) \ln rM + 2irQa \ln rM
\\
  &\qquad = 2irM \left( 1- \ln rM\right) + (q-1) \ln rM + 4\pi i ra \frac{1}{2\pi}Q\ln\left(\frac{M}{\varepsilon}\right)\,,
\end{align}
where $\varepsilon=\frac{1}{r}$. Note that the first two terms do not have any physical effect since they just rescale the partition function by an $a$-independent factor.  The last term, however,
combines with the on-shell classical piece of the action
\begin{equation}
 \ln Z_{0} \simeq -4\pi i ra \xi
\end{equation}
to account for the running of the FI parameter
\begin{equation}
\ln Z_{0}\cdot Z_{\text{one-loop}}^{\text{c.m.}}(a,B,M) \simeq -4\pi i ra \xi_{\text{ren}}\,,
\end{equation}
with
\begin{equation}
\xi_{\text{ren}} = \xi + \frac{1}{2\pi} \sum_{i}Q_{i}\ln\left(\frac{\varepsilon}{M}\right)\,.
\end{equation}

\section{Factorization for any \texorpdfstring{$\cN=(2,2)$}{N=(2,2)} Gauge Theory}
\label{app:factorization}

We repeat in this appendix in full generality the proof of
section~\ref{sec:factorization} that the partition function can be
written as a finite sum of terms, each of which is a product of a
holomorphic and an antiholomorphic functions of the complex
parameter~$\tau$ associated to each $U(1)$ gauge factor.  We start
from~\rf{semicolumb} with arbitrary gauge group $G$ and matter
representation $R$, which we recall in a more compact form below
as~\rf{ZCoulomb-compact}.  The vector multiplet one-loop determinant
in the original expression can be recast in terms of the one-loop
determinant of an adjoint chiral multiplet with $i\rmM = -1$ (in this
appendix we take $r=1$),
\begin{equation}
  \begin{aligned}
    & \prod_{\alpha\in\Delta^+} (-1)^{(\alpha\cdot B)} \left[\left(\alpha\cdot a\right)^2 +
        \left(\frac{\alpha\cdot B}{2}\right)^{2}\right]
    \\
    &
    = \prod_{\alpha\in\Delta^+}
    (-1)^{\alpha\cdot B}
    \frac{\Gamma(1 -i \alpha \cdot a + |\alpha\cdot B|/2)}
         {\Gamma(  -i \alpha \cdot a + |\alpha\cdot B|/2)}
    \frac{\Gamma(1 + i \alpha \cdot a + |\alpha\cdot B|/2)}
         {\Gamma(    i \alpha \cdot a + |\alpha\cdot B|/2)}
    \\ &
    =
    \prod_{\alpha\in\Delta}
    (-1)^{(\alpha\cdot B)^+}
    \frac {\Gamma(1 -i\alpha\cdot a + |\alpha\cdot B|/2)}
          {\Gamma(    i\alpha\cdot a + |\alpha\cdot B|/2)} \, .
  \end{aligned}
\end{equation}
The classical factor is
\begin{equation}
  \prod_{\substack{\text{abelian}\\\text{factors}}}
  e^{-4\pi i \xi \Tr a+i\vartheta \Tr B}
  =
  e^{2\pi i \mathtt{t}\cdot(ia + B/2)} e^{- 2\pi i \bar{\mathtt{t}}\cdot(ia - B/2)}
\end{equation}
where the (non-integer) weight $\mathtt{t}$ depends holomorphically on
the complexified parameters $\tau = \vartheta / (2\pi) + i\xi$ for each
abelian factor in~$G$:
\begin{equation}
  \mathtt{t}
  = \sum_{\substack{\text{abelian}\\\text{factors}}}
  \left(\frac{\vartheta}{2\pi} + i\xi\right) \Tr
  \, .
\end{equation}
Next we show that in the factor corresponding to one weight $w_I$ of the
representation of a chiral multiplet~$I$, the sign can be absorbed by
modifying the arguments of Gamma functions,
\begin{equation}
  (-1)^{(w_I\cdot B)^+}
  \frac {\Gamma\left(-i\rmM_I -iw_I\cdot a + |w_I\cdot B|/2\right)}
  {\Gamma\left(1 +i\rmM_I + iw_I\cdot a + |w_I\cdot B|/2\right)}
  = \frac {\Gamma\left(-i\rmM_I -iw_I\cdot a - w_I\cdot B/2\right)}
  {\Gamma\left(1 +i\rmM_I + iw_I\cdot a - w_I\cdot B/2\right)} \, .
\end{equation}
When $w_I \cdot B$ is negative, this identity is trivial, while for
positive (integer) $w_I \cdot B$ it results from Euler's identity $\Gamma(x)
\Gamma(1 - x) = \pi / [\sin \pi x]$ and anti-periodicity of the sine
function,
\begin{equation}
    (-1)^{w_I\cdot B} \pi \big / \bigl[ \sin \pi \left(-i\rmM_I -iw_I\cdot a + w_I\cdot B/2\right)\bigr]
    = \pi \big/ \bigl[\sin \pi \left(-i\rmM_I -iw_I\cdot a - w_I\cdot B/2\right)\bigr]
    \, .
\end{equation}
From this we deduce
\begin{equation}\label{ZCoulomb-compact}
  \begin{aligned}
    Z_{\text{Coulomb}}(\rmM, \mathtt{t}, \bar{\mathtt{t}}) = & \frac{1}{|\cW(G)|}
    \sum_{B} \int_{\tfrak} \rmd a\,  e^{2\pi i \mathtt{t}\cdot(ia + B/2)} e^{-2\pi i\bar{\mathtt{t}}\cdot(ia - B/2)}
    \\ & \quad
    \times \prod_{I,w_I} \frac {\Gamma(-i \rmM_I - w_I \cdot (ia+B/2))} {\Gamma(1 + i \rmM_I + w_I\cdot (ia-B/2))} \, ,
  \end{aligned}
\end{equation}
with a sum ranging over all GNO-quantized $B$ (including gauge
equivalent values), an integral ranging over the Cartan subalgebra
$\tfrak$, and a product over weights of the representation $R$ in which
the chiral multiplets transform, as well as weights of an additional
adjoint representation for the vector multiplet determinant.

Just as we did in section~\ref{sec:factorization} for the case of SQCD,
we close each of the integration contours in a direction that depends on
the matter content and the sign of $\xi$ for each abelian gauge factor.
Each factor in the integrand of $Z$ has poles whenever the numerator
Gamma function has a non-positive integer argument while the denominator
one does not, namely when
\begin{equation}\label{one-pole-coordinate}
  iw_I \cdot a = -i \rmM_I + |w_I \cdot B|/2 + n
\end{equation}
for some non-negative integer $n$.  Evaluating the $N =
\operatorname{rank}(\mathfrak{g})$ integrals in~\rf{ZCoulomb-compact}
yields a sum over common poles obeying~\rf{one-pole-coordinate} for
$N$ different choices of a flavor $I$ and a weight $w_I$, such that the
chosen $w_I$ span weight space\footnote{If the chosen $w_I$ did not span
  weight space, the conditions \rf{pole-coordinates} would not
  constrain $a$ to a given element in the Cartan subalgebra.}.
Explicitly,
\begin{equation}\label{pole-coordinates}
  i w_j \cdot a = -i \rmM_{p_j} + n_j + |w_j \cdot B| / 2\, ,
  \qquad \text{for all $1\leq j\leq N$} \, .
\end{equation}
Note that the contours do not enclose all such combined poles.  The
combinations of flavors $p_j$ and weights $w_j$ over which we sum thus
obey further constraints, such as restricting $p_j$ to (anti)fundamental
flavors in the case of SQCD.  Those constraints are complicated to
obtain in general, hence preventing this analysis from providing a fully
explicit factorized expression of the partition function.  However, they
do not affect any of the analysis proving that factorization does indeed
occur.

We introduce the dual basis to $w_j$, given by elements $\lambda_j$ of
the Cartan subalgebra such that $w_j\cdot \lambda_k = \delta_{jk}$.  For
every weight $w$ that appears in the Coulomb branch expression, all $w
\cdot \lambda_j$ are rational, and
\begin{equation}
  w = \sum_{j=1}^{N} (w\cdot \lambda_j) w_j \, .
\end{equation}
The partition function is expressed in terms of
\begin{equation}
  w \cdot (ia\pm B/2)
  = \sum_{j=1}^{N} (w\cdot \lambda_j) (-i\rmM_{p_j} + n_j + (w_j \cdot B)^\pm) \, ,
\end{equation}
where we use the notation $(x)^\pm = (|x| \pm x) / 2$.  Contrarily to
the SQCD case where all $w\cdot \lambda_j$ are $0$ or $\pm 1$, the
integers $n_j$ and $(w_j \cdot B)^\pm$ may not lead to integer shifts of
$w\cdot (ia\pm B/2)$ hence of the Gamma function arguments.  This was a
key ingredient in section~\ref{sec:factorization} to extract the
Pochhammer symbols in terms of which the partition function factorizes.
We recover integer shifts by splitting the sums over $n_j$~and~$w_j\cdot B$
depending on residues modulo the lowest common denominator $\mu_j$ of
all $w\cdot \lambda_j$.  Namely, for each $1\leq j\leq N$ we use
Euclidean division to write
\begin{equation}
  n_j + (w_j \cdot B)^\pm = k_j^\pm \mu_j + d_j^\pm \, ,
\end{equation}
with a quotient $0\leq k_j^\pm$ and a remainder $0 \leq d_j^\pm <
\mu_j$.  Clearly, each choice of integers $k_j^\pm$ and $d_j^\pm$ in
those ranges corresponds to integers $n_j$ and a vector $B$ in the
Cartan subalgebra, determined by
\begin{align}
  n_j & = \min(k_j^+ \mu_j + d_j^+, k_j^- \mu_j + d_j^-) \\
  w_j \cdot B & = k_j^+\mu_j + d_j^+ - k_j^- \mu_j - d_j^- \, .
\end{align}
However, the element $B$ thus constructed may not obey GNO quantization,
which requires that for every weight $w$,
\begin{equation}\label{GNO-quantization}
  w \cdot B
  = \sum_{j=1}^{N} (w\cdot\lambda_j) (w_j\cdot B)
  = \sum_{j=1}^{N} (w\cdot\lambda_j) (k_j^+\mu_j+d_j^+-k_j^-\mu_j-d_j^-)
\end{equation}
is an integer.  Since all $\mu_j (w\cdot \lambda_j)$ are integers,
\rf{GNO-quantization} reduces to a condition on $d_j^\pm$, only,
with no restriction on $k_j^\pm \geq 0$.

Hence, the sums over $n$ and $B$ split into a sum over (allowed
combinations of) degeneracy parameters $d_j^\pm$, and a sum over vortex
parameters $k_j^\pm$.  We have thus expressed the partition function as
\begin{equation}\label{ZCoulomb-pre-factorized}
  \begin{aligned}
    & Z(m,\mathtt{t},\bar{\mathtt{t}}) =
    \sum_{\{(p_j,w_j)\}} \sum_{\{d_j^\pm\}} \sum_{k_j^\pm \geq 0}
    \operatorname{res} \Bigg[
    e^{2\pi i\sum_{j=1}^{N} (\mathtt{t}\cdot\lambda_j) (-i\rmM_{p_j} + d_j^+ + k_j^+\mu_j)}
    e^{-2\pi i\sum_{j=1}^{N} (\bar{\mathtt{t}}\cdot\lambda_j) (-i\rmM_{p_j} + d_j^- + k_j^- \mu_j)}
    \\ & \quad
    \times \prod_{I,w_I} \frac {\Gamma\bigl(-i \rmM_I - \sum_{j=1}^{N} (w_I\cdot \lambda_j) (-i\rmM_{p_j} + d_j^+ + k_j^+ \mu_j)\bigr)} {\Gamma\bigl(1 + i \rmM_I + \sum_{j=1}^{N} (w_I\cdot \lambda_j) (-i\rmM_{p_j} + d_j^- + k_j^- \mu_j)\bigr)} \Bigg] \, ,
  \end{aligned}
\end{equation}
up to constant factors, and replacing the $N$~singular Gamma functions
by their residue at that pole.  The vorticities $k_j^\pm$ introduce
integer shifts in the arguments of Gamma functions, indeed, by
construction of $\mu_j$, all $\mu_j (w\cdot\lambda_j)$ are integers.
This enables us to extract from the summand the factors that only
depends on the choice of flavors, weights, and degeneracy parameters,
$p_j$, $w_j$, and~$d_j^\pm$,
\begin{align}
  Z_{\text{cl}} & = e^{2\pi i \sum_{j=1}^{N} (\mathtt{t}\cdot\lambda_j) (-i\rmM_{p_j} + d_j^+)} e^{-2\pi i\sum_{j=1}^{N} (\bar{\mathtt{t}}\cdot\lambda_j) (-i\rmM_{p_j} + d_j^-)}
  \\
  \operatorname{res} Z_{\text{one-loop}} & = \operatorname{res} \prod_{I,w_I} \gamma\biggl(-i \rmM_I - \sum_{j=1}^{N} (w_I\cdot\lambda_j) (-i\rmM_{p_j} + d_j^+)\biggr) \, ,
\end{align}
where, once more, gamma functions should be replaced by their residue
when appropriate.  After removing these $k_j^\pm$-independent factors,
we are left with
\begin{align}
  &
  \begin{aligned}
    & \sum_{k_j^\pm \geq 0} \Bigg[
    e^{2\pi i \sum_{j=1}^{N} \mu_j (\mathtt{t}\cdot\lambda_j) k_j^+}
    e^{-2\pi i \sum_{j=1}^{N} \mu_j (\bar{\mathtt{t}}\cdot\lambda_j) k_j^-}
    \\ & \quad
    \times \prod_{I,w_I} \frac {\bigl(-i \rmM_I - \sum_{j=1}^{N} (w_I\cdot\lambda_j) (-i\rmM_{p_j} + d_j^+)\bigr)_{-\sum_{j=1}^{N} \mu_j (w_I\cdot\lambda_j) k_j^+}} {\bigl(1 + i \rmM_I + \sum_{j=1}^{N} (w_I\cdot\lambda_j) (-i\rmM_{p_j} + d_j^-)\bigr)_{\sum_{j=1}^{N} \mu_j (w_I\cdot\lambda_j) k_j^-}} \Bigg]
  \end{aligned}
  \\
  &
  \begin{aligned}
    & = \sum_{k_j^- \geq 0}
    \frac{e^{-2\pi i\sum_{j=1}^{N} \mu_j (\bar{\mathtt{t}}\cdot\lambda_j) k_j^-}}
      {
        \prod_{I, w_I}
        \bigl(1 + i \rmM_I + \sum_{j=1}^{N} (w_I\cdot\lambda_j)
        (-i\rmM_{p_j} + d_j^-)\bigr)_{\sum_{j=1}^{N} \mu_j (w_I\cdot\lambda_j) k_j^-}
      }
    \\
    & \quad \times \sum_{k_j^+ \geq 0}
    \frac
      {
        e^{2\pi i\sum_{j=1}^{N} \mu_j (\mathtt{t}\cdot\lambda_j) k_j^+}
        \:
        \prod_{I, w_I} (-1)^{\sum_{j=1}^{N} \mu_j (w_I\cdot\lambda_j) k_j^+}
      }
      {
        \prod_{I, w_I}
        \bigl(1 + i \rmM_I + \sum_{j=1}^{N} (w_I\cdot\lambda_j)
        (-i\rmM_{p_j} + d_j^+)\bigr)_{\sum_{j=1}^{N} \mu_j (w_I\cdot\lambda_j) k_j^+}
      }
    \, .
  \end{aligned}
\end{align}
The partition function reduces to a finite sum of factorized terms,
\begin{equation}
  Z(\mathtt{t}, \bar{\mathtt{t}}, \rmM) = \sum_{\{(p_j,w_j)\}} \sum_{\{d_j^\pm\}}
  Z_{\text{cl}}(\mathtt{t},\bar{\mathtt{t}},\rmM)
  \operatorname{res} Z_{\text{one-loop}} (\rmM)
  Z_{\text{vortex}}(\mathtt{t}, \rmM)
  Z_{\text{anti-vortex}}(\bar{\mathtt{t}}, \rmM) \, ,
\end{equation}
where each of the factors additionally depends on the choice of
vacuum $\{p_j,w_j,d_j^\pm\}$.  This extends the result of
section~\ref{sec:factorization} to a general gauge group~$G$ and a
general chiral multiplet representation~$R$ of~$G$.

\section{Vortex Partition Function}
\label{vortexmatrix}

We describe in this appendix the procedure used to evaluate the contribution from vortex (and anti-vortex) configurations.  For simplicity, we only consider the case of SQCD, the two dimensional
$\cN=(2,2)$ $U(N)$ supersymmetric gauge theory with $N_F \geq N$ fundamental chiral multiplets of masses \((M_1,\ldots,M_{N_F})\) and $\antiNF\leq N_F$ anti-fundamental chiral multiplets of masses
\((\widetilde{M}_1,\ldots,\widetilde{M}_{N_F})\).  The flavour group is \(U(1)_{\text{anti-diag}} \times SU(N_F) \times SU(\antiNF)\), hence $\sum_{s=1}^{N_F} M_s = \sum_{s=1}^{\antiNF} \widetilde{M}_s$.

As we show in section~\ref{sec:instanton}, the presence of vortex/anti-vortex solutions requires the scalar field \(\sigma_2\) to take specific values, labelled by a choice of \(N\) masses \(M_{p_1}, \ldots, M_{p_N}\).
For such a choice of Higgs vacuum, the moduli space of solutions to the vortex equations~\rf{NPBPS} splits into discrete components \(\cM_{\text{vortex}}^{\{p_i\},k}\), where the vorticity \(k\) is defined by
\begin{equation}
k = \frac{1}{2\pi} \int_{\bR^2} \Tr F \, .
\end{equation}
The equivariant volume of the moduli space \(\cM_{\text{vortex}}\) can be expressed as a finite dimensional integral \cite{Shadchin:2006yz}.  We denote by \(\hat{\mathbf{M}}\) the diagonal \(N\times N\)
matrix with eigenvalues \(M_{p_i}\), by \(\check{\mathbf{M}}\) the diagonal matrix whose eigenvalues are masses of the other \(N_F - N\) (non-excited) fundamental chiral multiplets, and by \(\widetilde{\mathbf{M}}\)
the matrix of anti-fundamental masses.

\subsection{Vortex Matrix Model}
\label{sec:matrix-model}

The moduli space $\cM_{\text{vortex}}^{\{p_i\},k}$ of configurations with $k$ vortices admits an ADHM-like construction, which can be understood as the supersymmetric vacua of a certain gauged matrix
model preserving two supercharges \cite{Yoshida:2011au,Hanany:2003hp,Kim:2012uz}.
The relevant representations of the supersymmetry algebra can be obtained from the dimensional reduction of $N=(2,0)$ supersymmetry in two dimensions.
This gauged matrix model involves one $U(k)$ vector multiplet $\Phi=(\varphi, \l, \bar{\l} , D)$, and is coupled to one adjoint chiral multiplet $\cX=(X,\chi)$,
$N$ fundamental chiral multiplets $\cI = (I , \mu)$, $N_F - N$ anti-fundamental chiral multiplet $\cJ=(J, \n)$ and $\antiNF$ fundamental fermi multiplets $\Xi=(\xi,G)$.
The matrix model preserves three global symmetry groups $U(1)_R$, $U(1)_J$ and $U(1)_A$, which can be identified as the $R$-symmetry group, the rotational symmetry group $J$ and the axial $R$-symmetry group
of the given two dimensional theory, respectively.
As mentioned before, $U(1)_A$ may suffers from an axial anomaly.
Under these three $U(1)$ symmetry groups, the supercharges $Q$ and $\bar Q$ have charges $(-1,+1,-1)$ and $(+1,-1,-1)$.
For later convenience, we summarize global and gauge charges of the matrix model variables in the table below.
\begin{center}
\setlength{\tabcolsep}{4pt}
\begin{tabular}{c<{\ }rr<{\quad}rr>{\quad}rr<{\quad}r>{\quad}rrr}
 \toprule
 & $X$ & $\chi$ & $I$ & $\mu$ & $J$ & $\n$ & $\xi$ & $\bar{\varphi}$ & $\l$ & $\bar \l$ \\
 \midrule
 $U(1)_R$ & $0$ & $-1$ & $0$ & $-1$ & $0$ & $-1$ & $-1$ & $0$ & $-1$ & $+1$ \\
 $U(1)_{2J}$ & $-2$ & $-1$ & $0$ & $+1$ & $0$ & $+1$ & $+1$ & $0$ & $+1$ & $-1$ \\
 $U(1)_{A}$ & $0$ & $-1$ & $0$ & $-1$ & $0$ & $-1$ & $+1$ & $+2$ & $+1$ & $+1$ \\
 \midrule
 $U(1)_\varepsilon$ & $-2$ & $-2$ & $0$ & $0$ & $0$ & $0$ & $0$ & $0$ & $0$ & $0$ \\
 \midrule
 $U(k)$ & \multicolumn{2}{c}{\text{\textbf{adj}}} & \multicolumn{2}{c}{{$\mathbf{k}$}} &
 \multicolumn{2}{c}{$\bar{\mathbf{k}}$} & $\mathbf{k}$ & \multicolumn{3}{c}{\qquad\text{\textbf{adj}}} \\
 \bottomrule
\end{tabular}
\end{center}
Here the $U(1)_\varepsilon$ symmetry group can be identified as a twisted rotational symmetry group $J+R/2$ of the two dimensional theory.
Note that the complex scalar field $X$ represents the position of the $k$ vortices while $I$ and $J$ represent orientation modes.
The supersymmetric vacuum equation with a positive FI parameter $r \sim 1/g^2 >0$ is given by
\begin{equation}
  \label{matrix-model-0}
  \begin{aligned}
    [X, X^\dagger] + I I^\dagger - J^\dagger J & = r \mathbbm{1}_k
    \\
    \varphi I - I \hat{\mathbf{M}} & = 0 &
    [\varphi, \bar \varphi] & = 0
    \\
    J \varphi - \check{\mathbf{M}} J & = 0 &
    [\varphi , X] & = 0 \, ,
  \end{aligned}
\end{equation}
where $X$, $I$ and $J$ denote $k\times k$, $k \times N$ and $(N_F - N) \times k$ matrices.  The choice of Higgs vacuum in the original two dimensional gauge theory is encoded in the matrices \(\hat{\mathbf{M}}\) and
\(\check{\mathbf{M}}\).
The solutions of \rf{matrix-model-0} describe the moduli space $\cM_{\text{vortex}}^{\{p_i\},k}$ of $k$ vortices, and the volume of the moduli space can be identified as the partition function of this matrix model.

\subsection{Vortex Partition Function}

Since the matrix model  describing moduli space of vortices in \(\mathbb{R}^2\) has an infinite volume, it must be modified by turning on a chemical potential associated to the twisted rotational
symmetry group $U(1)_\varepsilon$.  The chemical potential $\varepsilon$ can be understood as the Omega deformation parameter in the given two dimensional theory, which is the inverse radius of the sphere $S^2$.

In the context of the matrix model, the chemical potential can be introduced by weakly gauging $U(1)_\varepsilon$, hence modifying \rf{matrix-model-0} to the deformed supersymmetry vacuum equation
\begin{equation}
  \label{matrix-model-epsilon}
  \begin{aligned}
    [X, X^\dagger] + I I^\dagger - J^\dagger J & = r \mathbbm{1}_k
    \\
    \varphi I - I \hat{\mathbf{M}} & = 0 &
    [\varphi, \bar \varphi] & = 0
    \\
    J \varphi - \check{\mathbf{M}} J & = 0 &
    [\varphi , X] & = \varepsilon X \, ,
  \end{aligned}
\end{equation}
and adding a new (deformed) fermion equation
\begin{equation}
  \varphi \xi + \xi \widetilde{\mathbf{M}} = 0 \,.
\end{equation}
Due to the chemical potential $\varepsilon$, the space of vacua is reduced to isolated points,  fixed points of supersymmetry.

We explain how to characterize such fixed points.
Suppose without loss of generality that $\varepsilon$ is positive definite.
One can show from the deformed supersymmetry vacuum equations that $J = 0$ and the $N$ chiral multiplets $I$ are each an eigenvector of the operator $\varphi$.  More specifically, denoting by $|\alpha \rangle$ an
eigenvector of the operator $\varphi$ with eigenvalue $\alpha$,
\begin{align}
  I = | M_{p_1}  \rangle \oplus \cdots \oplus | M_{p_N}  \rangle \, .
\end{align}
Then, the vector space of dimension $k$ on which $\varphi$ acts can be spanned by generators constructed by successive actions of $X$ on $|M_{p_i} \rangle$
\begin{align}
  \label{fixedpoint}
  | M_{p_i} + l \varepsilon  \rangle \stackrel{\text{def}}{\propto} X^{l} |M_{p_i} \rangle \qquad
  (l=0,1,..,k_i-1)\,,
\end{align}
with $\sum_{i=1}^N k_i = k$.
As a consequence, the fixed points are characterized by $N$ one-dimensional Young diagrams.  The number of boxes $k_i$ of the $i$-th 1-d Young diagram determines the vorticity of the $i$-th $U(1)$ factor in the
Cartan subalgebra of $U(N)$.
The matrix components of $X$ are then determined using the first relation of \rf{matrix-model-epsilon}.

The partition function of the matrix model can be reduced to a Gaussian integral around such fixed points.  The results are nicely expressed as the following contour-integral expression \cite{Shadchin:2006yz,Dimofte:2010tz}
\begin{align}\label{contourF}
  Z_{\vec{k}}(\{p_i\},M,\widetilde{M})
  = \oint_{\Gamma_{\{p_i\},k}} \prod_{I=1}^{k} \frac{d\varphi_I}{2\pi i}
  \,\cZ_{\text{vec}} (\varphi) \cdot \cZ_{\text{fund}}(M, \varphi)
  \cdot \cZ_{\text{anti-fund}}(\widetilde{M}, \varphi)
\end{align}
with
\begin{align}
  \cZ_{\text{vec}}(\varphi) & = \frac{1}{k! \,\varepsilon^{k}}
  \prod_{I\neq J}^{k} \frac{\varphi_I - \varphi_J}{\varphi_I -\varphi_J - \varepsilon}
  \\
  \cZ_{\text{fund}}(M, \varphi) & = \prod_{I=1}^k \prod_{s=1}^{N_F} \frac{1}{\varphi_I - M_{s} }
  \\
  \cZ_{\text{anti-fund}}(\widetilde{M}, \varphi) & = \prod_{I=1}^k \prod_{s=1}^{\antiNF} \left(\varphi_I + \widetilde{M}_s\right)\,,
\end{align}
where the contour $\Gamma_{\{p_i\},k}$ is chosen such that it encircles poles at
\begin{align}
  \varphi_I = \varphi_{(i,l)} = M_{p_i} + (l-1) \varepsilon \qquad (l=1,2,..,k_i)\,,
\end{align}
which can be understood as the fixed points \rf{fixedpoint}.
The vortex partition function of the two dimensional gauge theory in a specific choice of Higgs branch component $\{p_i\}$ thus takes the form
\begin{align}
  Z_{\text{vortex}}(\{p_i\},M,\widetilde{M},z)
  = \sum_{k_1+\cdots+k_N=k} z^{|\vec k|} Z_{\vec{k}}(\{p_i\}, M, \widetilde{M}) \,.
\end{align}

The residues of \rf{contourF} can be expressed as Pochhammer raising factorials \((x)_n = x(x+1) \cdots (x+n-1)\) and the full vortex partition function of SQCD in the Higgs vacuum labelled by \(\{p_i\}\) is
\begin{equation}
  Z_{\text{vortex}}^{\text{SQCD}} 
  = \sum_{\vec{k}} \frac{z^{|\vec{k}|}}{\vec{k}!} \,
    \frac{\prod_{i=1}^{N} \prod_{s=1}^{\antiNF} \big(\frac{1}{\varepsilon}(M_{p_i}+\widetilde{M}_s)\big)_{k_i}} {\prod_{i\neq j}^{N} \big(\frac{1}{\varepsilon}(M_{p_i}-M_{p_j})-k_j\big)_{k_j}
      \prod_{i=1}^{N} \prod_{s\not\in\{p_j\}}^{N_F}\big(\frac{1}{\varepsilon}(M_{p_i}-M_s)\big)_{k_i}} \, ,
\end{equation}
where \(\vec{k}! = k_1!\cdots k_N!\).

\section{\texorpdfstring{$SU(N)$}{SU(N)} Partition Function in Various Limits}
\label{app:Seiberg}

We prove first that the partition function on \(S^2\) of the \(\cN = (2,2)\) \(SU(N)\) gauge theory with \(N_F\) fundamental chiral multiplets obeys~\rf{ZSUN-polar-part}.
The Coulomb branch representation of this partition function is
\begin{equation}
  \label{ZSUN-Coulomb}
  \begin{aligned}
    Z_{SU(N)} (\rmM) &= \frac{1}{N!} \sum_{B_1 + \cdots + B_N = 0} \int \rmd a_1 \cdots \rmd a_{N-1} \Bigg[ \prod_{i<j}^{N} \left((a_i - a_j)^2 + \left(\frac{B_i - B_j}{2}\right)^2\right) \\
    & \qquad\qquad \prod_{s = 1}^{N_F} \prod_{i = 1}^{N} \frac{(-1)^{(B_i + |B_i|)/2} \Gamma(-ia_i -i\rmM_s + |B_i| / 2)}{\Gamma(1 + ia_i + i\rmM_s + |B_i|/2)}\Bigg]_{a_1 + \cdots + a_N = 0} \, .
  \end{aligned}
\end{equation}
The integral can be computed in the same way as the \(U(N)\) partition function evaluated in section~\ref{sec:factorization}. It turns out that closing the contour for $a$ towards $i\infty$ or $-i\infty$ gives the same series representation of the partition function,
\begin{equation}
  \label{ZSUN-series}
  \begin{aligned}
    & Z_{SU(N)}(\rmM) = \frac{(2\pi)^{N-1}}{N!} \sum_{B_1+\cdots+B_N=0} \sum_{n_1,\ldots,n_{N-1}\geq 0} \sum_{p_1,\ldots,p_{N-1}=1}^{N_F} \Bigg\{ (-1)^{N_F \sum_i |B_i|/2}
    \\ & \quad \cdot \prod_{j=1}^{N-1} \Bigg[\frac{(-1)^{n_j}}{n_j!(n_j+|B_j|)!} \prod_{s\neq p_j}^{N_F} \frac{\Gamma(-i\rmM_{s}+i\rmM_{p_j}-n_j)}{\Gamma(1+i\rmM_{s}-i\rmM_{p_j}+n_j+|B_j|)}\Bigg]
    \\ & \quad \cdot \prod_{s=1}^{N_F} \Bigg[\frac{\Gamma\big(-i\rmM_{s}+\frac{1}{2}|B_N|+\sum_{j=1}^{N-1}\big(-i\rmM_{p_j}+n_j+\frac{1}{2}|B_j|\big)\big)}{\Gamma\big(1+i\rmM_{s}+\frac{1}{2}|B_N|-
      \sum_{j=1}^{N-1}\big(-i\rmM_{p_j}+n_j+\frac{1}{2}|B_j|\big)\big)}\Bigg]
    \\ & \quad \cdot \prod_{i<j<N} \Bigg[\bigg(\frac{B_i-B_j}{2}\bigg)^2 - \bigg(-i\rmM_{p_i}+n_i+\frac{|B_i|}{2}+i\rmM_{p_j}-n_j-\frac{|B_j|}{2}\bigg)^2\Bigg]
    \\ & \quad \cdot \prod_{i<N} \Bigg[\bigg(\frac{B_i-B_N}{2}\bigg)^2 - \bigg(-i\rmM_{p_i}+n_i+\frac{|B_i|}{2}+\sum_{j=1}^{N-1}\bigg(-i\rmM_{p_j}+n_j+\frac{|B_j|}{2}\bigg)\bigg)^2\Bigg] \Bigg\} \, .
  \end{aligned}
\end{equation}

The argument of every Gamma function appearing in \rf{ZSUN-series} is an integer shifted by a term of order $\rmM$.  We can thus expand each as a series in powers of $\rmM$,
\begin{align}
  \Gamma(1 + n + i\rmM) &= n! (1 + O(\rmM)) \\
  \Gamma(  - n + i\rmM) &= \frac{1}{i\rmM} \frac{(-1)^n}{n!} (1 + O(\rmM))
\end{align}
for any integer $n\geq 0$.  Note that singularities arise from Gamma functions with non-positive (integer) arguments while positive arguments lead to a regular behavior.

Expanding all Gamma functions as power series in \(\rmM\), every term in \rf{ZSUN-series} has the form
\begin{equation}
  \label{Bneq0-series}
  \sum_{p_1=1}^{N_F} \cdots \sum_{p_{N-1}=1}^{N_F} \left[ \left( \prod_{j=1}^{N-1} \prod_{s\neq p_j}^{N_F} \frac{1}{\rmM_{p_j}-\rmM_s} \right) (\text{series in powers of \(\rmM\)})\right] \, ,
\end{equation}
except the term corresponding to \(n_1 = \cdots = n_{N-1} = |B_1| = \cdots = |B_N| = 0\), which has an additional singular factor \(1/\big(\rmM_s + \sum_{j=1}^{N-1} \rmM_{p_j}\big)\).  Since
\begin{equation}
  \sum_{p=1}^{N_F} \left[ \rmM_p^l \prod_{s\neq p}^{N_F} \frac{1}{\rmM_p-\rmM_s} \right] = \!\!\!\!\sum_{\substack{l_1,\ldots,l_{N_F}\geq 0\\ l_1+\cdots+l_{N_F}=l-N_F+1}} \!\!\!\! \rmM_1^{l_1}\cdots \rmM_{N_F}^{l_{N_F}}
\end{equation}
is a polynomial for any integer \(l\geq 0\) (zero if \(l\leq N_F-2\)), the expression \rf{Bneq0-series} is in fact a power series, hence is \(O(1)\) at \(\rmM\to 0\).  The same argument applied to the \(n = B = 0\)
term implies that all \(\rmM_p - \rmM_s\) factors vanish from the denominator in this case as well, leaving only denominators of the form \(\rmM_s + \sum_{j=1}^{N-1} \rmM_{p_j}\).

We have just shown that to order \(O(1)\), only the term with \(n = B = 0\) is relevant.  We can thus rewrite \rf{ZSUN-series} as
\begin{equation}
  \label{Bn0-series}
  \begin{aligned}
    Z_{SU(N)}(\rmM) & = \frac{(2\pi)^{N-1}}{N!} \sum_{p_1,\ldots,p_{N-1}=1}^{N_F} \Bigg\{ \prod_{i<j<N} \left(\rmM_{p_i}-\rmM_{p_j}\right)^2 \prod_{j=1}^{N-1}\prod_{s\neq p_j}^{N_F}
      \frac{\Gamma(i\rmM_{p_j}-i\rmM_{s})}{\Gamma(1-i\rmM_{p_j}+i\rmM_{s})}
    \\ & \quad \cdot \prod_{i<N} \bigg(\rmM_{p_i}+\sum_{j=1}^{N-1}\rmM_{p_j}\bigg)^2
      \prod_{s=1}^{N_F}\frac{\Gamma\big(-i\rmM_{s}-i\sum_{j=1}^{N-1}\rmM_{p_j}\big)}{\Gamma\big(1+i\rmM_{s}+i\sum_{j=1}^{N-1}\rmM_{p_j}\big)} \Bigg\} + O(1) \, .
  \end{aligned}
\end{equation}
Terms where \(p_i = p_j\) for \(1\leq i\neq j \leq N\) vanish because of the first product.  We then use the relation \(\Gamma(x) = \frac{1}{x} \Gamma(1 + x)\) to separate the singularities from some Gamma functions.
This cancels some factors coming from the vector multiplet determinant, yielding
\begin{equation}
  \label{N-1-sums}
  \! \begin{aligned}
    Z_{SU(N)}(\rmM) & = \frac{(2\pi)^{N-1}}{N!} \!\! \sum_{1\leq p_1\neq\ldots\neq p_{N-1}\leq N_F} \! \Bigg\{ \frac{\prod_{i<N} \big[i\rmM_{p_i}+\sum_{j=1}^{N-1}i\rmM_{p_j}\big]}{\prod_{s\not\in\{p_j\}}^{N_F}
      \Big[ \big(-i\rmM_{s}-i\sum_{j=1}^{N-1}\rmM_{p_j}\big) \prod_{j=1}^{N-1} (i\rmM_{p_j}-i\rmM_{s}) \Big]}
    \\ & \quad \cdot \prod_{j=1}^{N-1}\prod_{s\neq p_j}^{N_F} g(i\rmM_{p_j}-i\rmM_{s}) \prod_{s=1}^{N_F}g\left(-i\rmM_{s}-i\sum_{j=1}^{N-1}\rmM_{p_j}\right) \Bigg\} + O(1) \, ,
  \end{aligned}
\end{equation}
where
\begin{equation}
  g(i\rmM) = \frac{\Gamma(1+i\rmM)}{\Gamma(1-i\rmM)} = \exp\bigg(-2\gamma (i \rmM) - 2\sum_{j=1}^{\infty} \frac{\zeta(2j+1)}{(2j+1)} (i\rmM)^{2j+1}\bigg) = 1 + O(\rmM) \,.
\end{equation}

Noting that the sums in \rf{N-1-sums} only involve \((N-1)\) flavours \(p_i\), we use the relation
\begin{equation}
  \prod_{s\not\in\{p_i\}}^{N_F} \frac{1}{-i\rmM_s-\sum_{j=1}^{N-1} i\rmM_{p_j}}
  = \sum_{t\not\in\{p_i\}}^{N_F} \bigg[ \frac{1}{-i\rmM_{t} - \sum_{j=1}^{N-1} i\rmM_{p_j}} \prod_{s\not\in\{p_i,t\}}^{N_F} \frac{1}{i\rmM_t -i\rmM_s} \bigg]
\end{equation}
to obtain a sum over all \(N\)-element subsets \(E=\{p_1,\ldots,p_{N-1},p_N = t\}\subseteq\{1,\ldots,N_F\}\):
\begin{equation}
  \label{Nsubset-series}
  \begin{aligned}
    & Z_{SU(N)}(\rmM) = (2\pi)^{N-1} \sum_{\substack{E\subseteq\{1,\ldots,N_F\}\\\#E=N}} \Bigg[ \frac{\prod_{p\in E} \prod_{s\not\in E}^{N_F} \gamma(i\rmM_{p}-i\rmM_{s})}{ -i\sum_{p\in E} \rmM_{p}}
    \\ & \quad \cdot \frac{1}{N} \sum_{t\in E} \Bigg\{ \prod_{p\in E\setminus\{t\}} \frac{i\rmM_p-i\rmM_t+\sum_{s\in E} i\rmM_s}{i\rmM_{p} -i\rmM_{t}}
      \prod_{s=1}^{N_F} \frac{g(i\rmM_{s}-i\rmM_{t})}{g\big(i\rmM_{s}-i\rmM_{t}+i\sum_{p\in E}\rmM_{p}\big)} \Bigg\} \Bigg] + O(1) \, ,
  \end{aligned}
\end{equation}
The ratios of \(g\) have the form
\begin{equation}
  \label{Gamma-ratio}
  \frac{g(x)}{g\big(x+i\sum_{p\in E}\rmM_{p}\big)} = 1 + \left[\sum_{p\in E} i\rmM_p \right] (\text{series in powers of \(\rmM\)}) \, .
\end{equation}
All terms beyond the zeroth order have a factor of \(\sum_{p\in E} \rmM_p\), cancelling the corresponding pole in \rf{Nsubset-series}.  Those terms thus only have poles of the form \(1/(i\rmM_p -i\rmM_t)\),
and we have proven earlier that those poles cannot remain.  Thus, replacing the ratios \rf{Gamma-ratio} by \(1\) only changes the partition function by terms which are regular at \(\rmM\to 0\), in other words,
\(O(1)\) terms.  The sum over \(t\) in \rf{Nsubset-series} is then
\begin{equation}
  \frac{1}{N} \sum_{t\in E} \prod_{p\in E\setminus\{t\}} \frac{i\rmM_p-i\rmM_t+\sum_{s\in E} i\rmM_s}{i\rmM_{p} -i\rmM_{t}}
    = \frac{1}{N \sum_{s\in E} i\rmM_s} \oint \frac{\rmd z}{2\pi i} \prod_{p\in E} \frac{z + i\rmM_p+\sum_{s\in E} i\rmM_s}{z + i\rmM_{p}} = 1 \, .
\end{equation}
Inserting this result back into \rf{Nsubset-series} gives \rf{ZSUN-polar-part}, up to the normalization factor \((2\pi)^{N-1}\).

This concludes the proof of \rf{ZSUN-polar-part}, which in turn implies that the partition function of the \(SU(N)\) and \(SU(N_F-N)\) theories, with a particular matching of the mass parameters,
are equal at order \(O(1)\) in the limit of small masses and $R$-charge.

For any given value of \(N\) and \(N_F\), the Gamma functions appearing in~\rf{ZSUN-series} can be expanded in power series using
\begin{equation}
  \frac{(-1)^n\Gamma(x - n)}{\Gamma(1 - x + n + |B|)}
  = \frac{1}{x}  \frac{\exp\left[-2\gamma x - 2\sum_{j=1}^{\infty} \frac{\zeta(2j+1)}{(2j+1)} x^{2j+1}\right]}{\prod_{j=1}^{n} [j - x] \prod_{j=1}^{n + |B|} [ j - x]} \, .
\end{equation}
The partition function of the \(SU(2)\) gauge theory with \(N_F = 3\) fundamental chiral multiplets was computed in this manner up to order \(O(\rmM^2)\) and is, as expected, equal to the partition function of
the theory of three free chiral multiplets, with masses given by~\rf{duality-dictionary}.  The signs coming from the chiral multiplet one-loop determinants~\rf{chiralloop} are crucial: the matching would
otherwise fail with a difference of order \(1\).

The study of the \(\rmM \to 0\) limit which was just performed highlights the value of considering limits where the partition function has a pole.

The Gamma functions appearing in the series expression \rf{ZSUN-series} of \(Z_{SU(N)}(\rmM)\) have poles at
\begin{align}
  \label{MM-poles}
  i\rmM_{t}-i\rmM_{u} &= k\geq 0 \\
  \label{sumM-poles}
  \sum_{j=1}^{N} i\rmM_{s_j} & = k\geq 0 \, ,
\end{align}
where \(k\) is an integer, and \(t\), \(u\) and \(s_j\) are flavour indices.  We ignore in this paper the poles \rf{MM-poles}: in fact, those poles cancel amongst the various terms in the full partition function,
which is thus regular at \(i\rmM_{t} -i\rmM_{u} = k\).  We concentrate on the poles \rf{sumM-poles}, labelled by a choice of \(N\) chiral multiplets \(E \subseteq\{1,\ldots,N_F\}\), \(\#E = N\) and a total
vorticity \(k\).  For definiteness, we choose \(E = \{1,\ldots,N\}\), that is, \(s_j = j\).

Since the \(\cN=(2,2)\) \(SU(N)\) gauge theories with \(N_F > N\) massless fundamental chiral multiplets flow to an infrared fixed point, the $R$-charge of each chiral multiplet should be non-negative.
However, note that \(\sum_{s=1}^{N} i\rmM_{s} = k\geq 0\) implies that the sum of the $R$-charges of the \(N\) chiral multiplets is \(-2k\leq 0\).  Hence only the poles with \(k = 0\) are in the physical parameter space.
The poles corresponding to a non-zero vorticity \(k > 0\) can however be reached by analytically continuing with respect to the complex masses \(\rmM\).

The terms in \rf{ZSUN-series} which are singular when \(\sum_{s=1}^{N} i\rmM_{s} = k\) are precisely those for which \(1\leq p_1\neq\ldots\neq p_{N-1}\leq N\), and for which the integer \(n_N\) defined by
\begin{equation}
  \sum_{i=1}^{N} \left(n_i + \frac{|B_i|}{2}\right) = k
\end{equation}
is non-negative.  As we will see, the number of such terms is finite.  Defining \(p_N\) by \(\{p_1,\ldots,p_N\} = \{1,\ldots,N\}\), the residue of the partition function at this pole takes a symmetric form, where the Gamma
functions involving sums of masses are recast in terms of the mass \(i\rmM_{p_N}\).  Following the same procedure as in section~\ref{sec:factorization}, we introduce the coordinates
\(k_i^{\pm} = n_i + \frac{1}{2}|B_i| \pm \frac{1}{2} B_i \geq 0\) which factorize the summand into two identical contributions, leading to the expression
\begin{equation}
  \label{ZSUN-k-residue}
  \res_{i\rmM_1+\cdots+i\rmM_N = k} Z_{SU(N)}(\rmM) = (2\pi)^{N-1} (-1)^{N_F k} \left[\prod_{p=1}^{N} \prod_{s=N+1}^{N_F} \gamma(-i\rmM_{s}+i\rmM_{p})\right] [Z_k(\{1,\ldots,N\}, \rmM)]^2
\end{equation}
with
\begin{equation}
  Z_k(\{1,\ldots,N\}, \rmM) = \hskip-12pt \sum_{k_1+\cdots+k_N=k} \left[\prod_{p=1}^{N}\frac{1}{k_p!} \prod_{p\neq s}^{N} \frac{1}{(i\rmM_p -i\rmM_s - k_p)_{k_s}} \prod_{p=1}^{N} \prod_{s=N+1}^{N_F}
    \frac{1}{(1+i\rmM_s-i\rmM_p)_{k_p}} \right]\, .
\end{equation}
The Seiberg-like duality \rf{Seiberg-small-mass} between \(SU(N)\) and \(SU(N_F-N)\) gauge theories on \(S^2\) is satisfied in this limit if
\begin{equation}
  \label{Seiberg-any-k}
  Z_k(\{1,\ldots,N\}, \rmM) = \pm Z_k(\{N+1,\ldots,N_F\}, \rmM') \,,
\end{equation}
with the mass matching \rf{duality-dictionary}.

The case of a zero total vorticity $k$ is elementary: since \((x)_0 = 1\),
\begin{equation}
  Z_0(\{1,\ldots,N\}, \rmM) = 1 \, .
\end{equation}
The duality is shown in the case of a total vorticity $k = 1$ using a one-dimensional contour integral: the sum over partitions \((k_1,\ldots,k_N)\) of \(k = 1\) simply ranges over \(N\) terms,
interpreted as the residues at \(N\) poles of a complex function with \(N_F\) poles on the Riemann sphere.  The resulting contour integral is thus equal to the sum over residues of the \(N_F - N\)
remaining poles, and this reproduces the desired dual object.
\begin{equation}
  \begin{aligned}
    Z_1(\{1,\ldots,N\}, \rmM)
    &= \sum_{p=1}^{N} \frac{1}{\prod_{s\neq p}^{N} (i\rmM_s -i\rmM_p) \prod_{s=N+1}^{N_F} (1+i\rmM_s-i\rmM_p)}
    \\
    &= \oint \frac{\rmd z}{2\pi i} \frac{1}{\prod_{s=1}^{N} (i\rmM_s + z) \prod_{s=N+1}^{N_F} (1+i\rmM_s+z)}
    \\
    &= \sum_{p=N+1}^{N_F} \frac{(-1)^{N_F}}{\prod_{s=1}^{N} (1+i\rmM'_s -i\rmM'_p) \prod_{N+1\leq s\neq p\leq N_F} (i\rmM'_s-i\rmM'_p)}
    \\
    &= Z_1(\{N+1,\ldots,N_F\}, \rmM') \, .
  \end{aligned}
\end{equation}

Since the objects in consideration are rather explicit finite sums, it should be possible to prove \rf{Seiberg-any-k} for arbitrary \(k\geq 0\).  If one can additionally show that the two partition
functions have the same asymptotic behavior at infinity, then the exact Seiberg-like duality between \(Z_{SU(N)}(\rmM)\) and \(Z_{SU(N_F-N)}(\rmM')\) follows, for arbitrary masses and $R$-charges, by
noting that their difference is a bounded entire function, hence is a constant.

A different approach to studying the \(SU(N)\) partition function is to note that integrating over \(\xi\) and \(\vartheta\) in~\rf{ZUN-Coulomb} constrains \(B\) and \(a\) to be traceless, hence
reproducing the corresponding \(SU(N)\) partition function in the Coulomb branch representation,
\begin{equation}
  Z_{SU(N)}(\rmM) = \int \rmd \xi \rmd \vartheta Z_{U(N)}(\rmM, \tau) \, .
\end{equation}
Integrating instead the Higgs branch representation of \(Z_{U(N)}(\rmM, \tau)\) singles out the Fourier components where the exponents of \(z\) and \(\bar{z}\) in \rf{Coulomb-factorized} vanish.
The \(\vartheta\) constraint imposes that the total vorticity at the north pole be the same as that at the south pole.  The \(\xi\) constraint relates the masses to the vorticity \(k\) precisely
as \(\sum_{s\in E} i\rmM_s = k\) for a set \(E\) of \(N\) flavours.  The residues~\rf{ZSUN-k-residue} of \(Z_{SU(N)}(\rmM)\) at the corresponding poles are reproduced by the coefficient of
\((z\bar{z})^{-\sum_{s\in E}i\rmM_s + k}\) in the Higgs branch representation of the \(U(N)\) partition function, which explains the appearance of the \(k\)-vortex partition functions in the
factorization of \(Z_{SU(N)}(\rmM)\).

\clearpage

\bibliography{refs}

\providecommand{\href}[2]{#2}\begingroup\raggedright\begin{thebibliography}{10}

\bibitem{Witten:1988ze}
E.~Witten, ``Topological quantum field theory,''
\href{http://dx.doi.org/10.1007/BF01223371}{{\em Commun. Math. Phys.}
  {\bfseries 117} (1988) 353}.

\bibitem{Witten:1991zz}
E.~Witten, ``Mirror manifolds and topological field theory,''
\href{http://arxiv.org/abs/hep-th/9112056}{{\ttfamily arXiv:hep-th/9112056
  [hep-th]}}.

\bibitem{Pestun:2007rz}
V.~Pestun, ``Localization of gauge theory on a four-sphere and supersymmetric
  {W}ilson loops,'' \href{http://dx.doi.org/10.1007/s00220-012-1485-0}{{\em
  Commun. Math. Phys.} {\bfseries 313} (2012) 71--129},
\href{http://arxiv.org/abs/0712.2824}{{\ttfamily arXiv:0712.2824 [hep-th]}}.

\bibitem{Shadchin:2006yz}
S.~Shadchin, ``On {F}-term contribution to effective action,''
  \href{http://dx.doi.org/10.1088/1126-6708/2007/08/052}{{\em JHEP} {\bfseries
  08} (2007) 052},
\href{http://arxiv.org/abs/hep-th/0611278}{{\ttfamily arXiv:hep-th/0611278
  [hep-th]}}.

\bibitem{Pasquetti:2011fj}
S.~Pasquetti, ``Factorisation of {${\cal N}=2$} theories on the squashed
  3-sphere,'' \href{http://dx.doi.org/10.1007/JHEP04(2012)120}{{\em JHEP}
  {\bfseries 04} (2012) 120},
\href{http://arxiv.org/abs/1111.6905}{{\ttfamily arXiv:1111.6905 [hep-th]}}.

\bibitem{Krattenthaler:2011da}
C.~Krattenthaler, V.~P. Spiridonov, and G.~S. Vartanov, ``Superconformal
  indices of three-dimensional theories related by mirror symmetry,''
  \href{http://dx.doi.org/10.1007/JHEP06(2011)008}{{\em JHEP} {\bfseries 06}
  (2011) 008},
\href{http://arxiv.org/abs/1103.4075}{{\ttfamily arXiv:1103.4075 [hep-th]}}.

\bibitem{Hama:2011ea}
N.~Hama, K.~Hosomichi, and S.~Lee, ``{SUSY} gauge theories on squashed
  three-spheres,'' \href{http://dx.doi.org/10.1007/JHEP05(2011)014}{{\em JHEP}
  {\bfseries 05} (2011) 014},
\href{http://arxiv.org/abs/1102.4716}{{\ttfamily arXiv:1102.4716 [hep-th]}}.

\bibitem{Kim:2009wb}
S.~Kim, ``The complete superconformal index for {${\cal N}=6$}
  {C}hern--{S}imons theory,''
  \href{http://dx.doi.org/10.1016/j.nuclphysb.2012.07.015,
  10.1016/j.nuclphysb.2009.06.025}{{\em Nucl. Phys.} {\bfseries B821} (2009)
  241--284}, \href{http://arxiv.org/abs/0903.4172}{{\ttfamily arXiv:0903.4172
  [hep-th]}}.
[Erratum: Nucl. Phys.B864,884(2012)].

\bibitem{Kapustin:2009kz}
A.~Kapustin, B.~Willett, and I.~Yaakov, ``Exact results for {W}ilson loops in
  superconformal {C}hern--{S}imons theories with matter,''
  \href{http://dx.doi.org/10.1007/JHEP03(2010)089}{{\em JHEP} {\bfseries 03}
  (2010) 089},
\href{http://arxiv.org/abs/0909.4559}{{\ttfamily arXiv:0909.4559 [hep-th]}}.

\bibitem{Jafferis:2010un}
D.~L. Jafferis, ``The exact superconformal {R}-symmetry extremizes {Z},''
  \href{http://dx.doi.org/10.1007/JHEP05(2012)159}{{\em JHEP} {\bfseries 05}
  (2012) 159},
\href{http://arxiv.org/abs/1012.3210}{{\ttfamily arXiv:1012.3210 [hep-th]}}.

\bibitem{Hama:2010av}
N.~Hama, K.~Hosomichi, and S.~Lee, ``Notes on {SUSY} gauge theories on
  three-sphere,'' \href{http://dx.doi.org/10.1007/JHEP03(2011)127}{{\em JHEP}
  {\bfseries 03} (2011) 127},
\href{http://arxiv.org/abs/1012.3512}{{\ttfamily arXiv:1012.3512 [hep-th]}}.

\bibitem{Imamura:2011su}
Y.~Imamura and S.~Yokoyama, ``Index for three dimensional superconformal field
  theories with general {R}-charge assignments,''
  \href{http://dx.doi.org/10.1007/JHEP04(2011)007}{{\em JHEP} {\bfseries 04}
  (2011) 007},
\href{http://arxiv.org/abs/1101.0557}{{\ttfamily arXiv:1101.0557 [hep-th]}}.

\bibitem{Alday:2009aq}
L.~F. Alday, D.~Gaiotto, and Y.~Tachikawa, ``{L}iouville correlation functions
  from four-dimensional gauge theories,''
  \href{http://dx.doi.org/10.1007/s11005-010-0369-5}{{\em Lett. Math. Phys.}
  {\bfseries 91} (2010) 167--197},
\href{http://arxiv.org/abs/0906.3219}{{\ttfamily arXiv:0906.3219 [hep-th]}}.

\bibitem{Alday:2009fs}
L.~F. Alday, D.~Gaiotto, S.~Gukov, Y.~Tachikawa, and H.~Verlinde, ``Loop and
  surface operators in {${\cal N}=2$} gauge theory and {L}iouville modular
  geometry,'' \href{http://dx.doi.org/10.1007/JHEP01(2010)113}{{\em JHEP}
  {\bfseries 01} (2010) 113},
\href{http://arxiv.org/abs/0909.0945}{{\ttfamily arXiv:0909.0945 [hep-th]}}.

\bibitem{Witten:1993yc}
E.~Witten, ``Phases of {${\cal N}=2$} theories in two-dimensions,''
  \href{http://dx.doi.org/10.1016/0550-3213(93)90033-L}{{\em Nucl. Phys.}
  {\bfseries B403} (1993) 159--222},
\href{http://arxiv.org/abs/hep-th/9301042}{{\ttfamily arXiv:hep-th/9301042
  [hep-th]}}.

\bibitem{Aspinwall:1993yb}
P.~S. Aspinwall, B.~R. Greene, and D.~R. Morrison, ``Multiple mirror manifolds
  and topology change in string theory,''
  \href{http://dx.doi.org/10.1016/0370-2693(93)91428-P}{{\em Phys. Lett.}
  {\bfseries B303} (1993) 249--259},
\href{http://arxiv.org/abs/hep-th/9301043}{{\ttfamily arXiv:hep-th/9301043
  [hep-th]}}.

\bibitem{Seiberg:1994pq}
N.~Seiberg, ``Electric - magnetic duality in supersymmetric non{A}belian gauge
  theories,'' \href{http://dx.doi.org/10.1016/0550-3213(94)00023-8}{{\em Nucl.
  Phys.} {\bfseries B435} (1995) 129--146},
\href{http://arxiv.org/abs/hep-th/9411149}{{\ttfamily arXiv:hep-th/9411149
  [hep-th]}}.

\bibitem{Hori:2006dk}
K.~Hori and D.~Tong, ``Aspects of non-{A}belian gauge dynamics in
  two-dimensional {${\cal N}=(2,2)$} theories,''
  \href{http://dx.doi.org/10.1088/1126-6708/2007/05/079}{{\em JHEP} {\bfseries
  05} (2007) 079},
\href{http://arxiv.org/abs/hep-th/0609032}{{\ttfamily arXiv:hep-th/0609032
  [hep-th]}}.

\bibitem{Benini:2012ui}
F.~Benini and S.~Cremonesi, ``Partition functions of ${\mathcal{n}=(2,2)}$
  gauge theories on {S$^{2}$} and vortices,''
  \href{http://dx.doi.org/10.1007/s00220-014-2112-z}{{\em Commun. Math. Phys.}
  {\bfseries 334} no.~3, (2015) 1483--1527},
\href{http://arxiv.org/abs/1206.2356}{{\ttfamily arXiv:1206.2356 [hep-th]}}.

\bibitem{Festuccia:2011ws}
G.~Festuccia and N.~Seiberg, ``Rigid supersymmetric theories in curved
  superspace,'' \href{http://dx.doi.org/10.1007/JHEP06(2011)114}{{\em JHEP}
  {\bfseries 06} (2011) 114},
\href{http://arxiv.org/abs/1105.0689}{{\ttfamily arXiv:1105.0689 [hep-th]}}.

\bibitem{Witten:1991mk}
E.~Witten, ``The {N} matrix model and gauged {WZW} models,''
\href{http://dx.doi.org/10.1016/0550-3213(92)90235-4}{{\em Nucl. Phys.}
  {\bfseries B371} (1992) 191--245}.

\bibitem{Gomis:2011pf}
J.~Gomis, T.~Okuda, and V.~Pestun, ``Exact results for 't {H}ooft loops in
  gauge theories on {$S^4$},''
  \href{http://dx.doi.org/10.1007/JHEP05(2012)141}{{\em JHEP} {\bfseries 05}
  (2012) 141},
\href{http://arxiv.org/abs/1105.2568}{{\ttfamily arXiv:1105.2568 [hep-th]}}.

\bibitem{Hori:2013ika}
K.~Hori and M.~Romo, ``Exact results in two-dimensional (2,2) supersymmetric
  gauge theories with boundary,''
\href{http://arxiv.org/abs/1308.2438}{{\ttfamily arXiv:1308.2438 [hep-th]}}.

\bibitem{Honda:2013uca}
D.~Honda and T.~Okuda, ``Exact results for boundaries and domain walls in 2d
  supersymmetric theories,''
\href{http://arxiv.org/abs/1308.2217}{{\ttfamily arXiv:1308.2217 [hep-th]}}.

\bibitem{pasquetti}
{S. Pasquetti}. Private communication.

\bibitem{Gomis:2014eya}
J.~Gomis and B.~Le~Floch, ``{M2}-brane surface operators and gauge theory
  dualities in {T}oda,''
\href{http://arxiv.org/abs/1407.1852}{{\ttfamily arXiv:1407.1852 [hep-th]}}.

\bibitem{Yoshida:2011au}
Y.~Yoshida, ``Localization of vortex partition functions in
  {$\mathcal{N}=(2,2)$} super {Y}ang--{M}ills theory,''
\href{http://arxiv.org/abs/1101.0872}{{\ttfamily arXiv:1101.0872 [hep-th]}}.

\bibitem{Bonelli:2011fq}
G.~Bonelli, A.~Tanzini, and J.~Zhao, ``Vertices, vortices and interacting
  surface operators,'' \href{http://dx.doi.org/10.1007/JHEP06(2012)178}{{\em
  JHEP} {\bfseries 06} (2012) 178},
\href{http://arxiv.org/abs/1102.0184}{{\ttfamily arXiv:1102.0184 [hep-th]}}.

\bibitem{Miyake:2011yr}
A.~Miyake, K.~Ohta, and N.~Sakai, ``Volume of moduli space of vortex equations
  and localization,'' \href{http://dx.doi.org/10.1143/PTP.126.637}{{\em Prog.
  Theor. Phys.} {\bfseries 126} (2011) 637--680},
\href{http://arxiv.org/abs/1105.2087}{{\ttfamily arXiv:1105.2087 [hep-th]}}.

\bibitem{Fujimori:2012ab}
T.~Fujimori, T.~Kimura, M.~Nitta, and K.~Ohashi, ``Vortex counting from field
  theory,'' \href{http://dx.doi.org/10.1007/JHEP06(2012)028}{{\em JHEP}
  {\bfseries 06} (2012) 028},
\href{http://arxiv.org/abs/1204.1968}{{\ttfamily arXiv:1204.1968 [hep-th]}}.

\bibitem{Kim:2012uz}
H.-C. Kim, J.~Kim, S.~Kim, and K.~Lee, ``Vortices and 3 dimensional
  dualities,''
\href{http://arxiv.org/abs/1204.3895}{{\ttfamily arXiv:1204.3895 [hep-th]}}.

\bibitem{Hanany:2003hp}
A.~Hanany and D.~Tong, ``Vortices, instantons and branes,''
  \href{http://dx.doi.org/10.1088/1126-6708/2003/07/037}{{\em JHEP} {\bfseries
  07} (2003) 037},
\href{http://arxiv.org/abs/hep-th/0306150}{{\ttfamily arXiv:hep-th/0306150
  [hep-th]}}.

\bibitem{Eto:2005yh}
M.~Eto, Y.~Isozumi, M.~Nitta, K.~Ohashi, and N.~Sakai, ``Moduli space of
  non-{A}belian vortices,''
  \href{http://dx.doi.org/10.1103/PhysRevLett.96.161601}{{\em Phys. Rev. Lett.}
  {\bfseries 96} (2006) 161601},
\href{http://arxiv.org/abs/hep-th/0511088}{{\ttfamily arXiv:hep-th/0511088
  [hep-th]}}.

\bibitem{Dimofte:2010tz}
T.~Dimofte, S.~Gukov, and L.~Hollands, ``Vortex counting and {L}agrangian
  3-manifolds,'' \href{http://dx.doi.org/10.1007/s11005-011-0531-8}{{\em Lett.
  Math. Phys.} {\bfseries 98} (2011) 225--287},
\href{http://arxiv.org/abs/1006.0977}{{\ttfamily arXiv:1006.0977 [hep-th]}}.

\bibitem{Okuda:2010ke}
T.~Okuda and V.~Pestun, ``On the instantons and the hypermultiplet mass of
  {${\cal N}=2*$} super {Y}ang--{M}ills on {$S^{4}$},''
  \href{http://dx.doi.org/10.1007/JHEP03(2012)017}{{\em JHEP} {\bfseries 03}
  (2012) 017},
\href{http://arxiv.org/abs/1004.1222}{{\ttfamily arXiv:1004.1222 [hep-th]}}.

\bibitem{Wyllard:2009hg}
N.~Wyllard, ``{A(N-1)} conformal toda field theory correlation functions from
  conformal {${\cal N}=2$} {$SU(N)$} quiver gauge theories,''
  \href{http://dx.doi.org/10.1088/1126-6708/2009/11/002}{{\em JHEP} {\bfseries
  11} (2009) 002},
\href{http://arxiv.org/abs/0907.2189}{{\ttfamily arXiv:0907.2189 [hep-th]}}.

\bibitem{Bonelli:2011wx}
G.~Bonelli, A.~Tanzini, and J.~Zhao, ``The {L}iouville side of the vortex,''
  \href{http://dx.doi.org/10.1007/JHEP09(2011)096}{{\em JHEP} {\bfseries 09}
  (2011) 096},
\href{http://arxiv.org/abs/1107.2787}{{\ttfamily arXiv:1107.2787 [hep-th]}}.

\bibitem{Fateev:2007ab}
V.~A. Fateev and A.~V. Litvinov, ``Correlation functions in conformal {T}oda
  field theory. {I}.,''
  \href{http://dx.doi.org/10.1088/1126-6708/2007/11/002}{{\em JHEP} {\bfseries
  11} (2007) 002},
\href{http://arxiv.org/abs/0709.3806}{{\ttfamily arXiv:0709.3806 [hep-th]}}.

\bibitem{Gomis:2010kv}
J.~Gomis and B.~Le~Floch, ``'t {H}ooft operators in gauge theory from {T}oda
  {CFT},'' \href{http://dx.doi.org/10.1007/JHEP11(2011)114}{{\em JHEP}
  {\bfseries 11} (2011) 114},
\href{http://arxiv.org/abs/1008.4139}{{\ttfamily arXiv:1008.4139 [hep-th]}}.

\bibitem{Gukov:2006jk}
S.~Gukov and E.~Witten, ``Gauge theory, ramification, and the geometric
  {L}anglands program,''
\href{http://arxiv.org/abs/hep-th/0612073}{{\ttfamily arXiv:hep-th/0612073
  [hep-th]}}.

\bibitem{Hori:2011pd}
K.~Hori, ``Duality in two-dimensional {$(2,2)$} supersymmetric non-abelian
  gauge theories,'' \href{http://dx.doi.org/10.1007/JHEP10(2013)121}{{\em JHEP}
  {\bfseries 10} (2013) 121},
\href{http://arxiv.org/abs/1104.2853}{{\ttfamily arXiv:1104.2853 [hep-th]}}.

\bibitem{Drukker:2010jp}
N.~Drukker, D.~Gaiotto, and J.~Gomis, ``The virtue of defects in 4d gauge
  theories and 2d {CFT}s,''
  \href{http://dx.doi.org/10.1007/JHEP06(2011)025}{{\em JHEP} {\bfseries 06}
  (2011) 025},
\href{http://arxiv.org/abs/1003.1112}{{\ttfamily arXiv:1003.1112 [hep-th]}}.

\bibitem{Moore:1997dj}
G.~W. Moore, N.~Nekrasov, and S.~Shatashvili, ``Integrating over {H}iggs
  branches,'' \href{http://dx.doi.org/10.1007/PL00005525}{{\em Commun. Math.
  Phys.} {\bfseries 209} (2000) 97--121},
\href{http://arxiv.org/abs/hep-th/9712241}{{\ttfamily arXiv:hep-th/9712241
  [hep-th]}}.

\bibitem{Nekrasov:2002qd}
N.~A. Nekrasov, ``{S}eiberg--{W}itten prepotential from instanton counting,''
  \href{http://dx.doi.org/10.4310/ATMP.2003.v7.n5.a4}{{\em Adv. Theor. Math.
  Phys.} {\bfseries 7} no.~5, (2003) 831--864},
\href{http://arxiv.org/abs/hep-th/0206161}{{\ttfamily arXiv:hep-th/0206161
  [hep-th]}}.

\bibitem{Drukker:2009id}
N.~Drukker, J.~Gomis, T.~Okuda, and J.~Teschner, ``Gauge theory loop operators
  and {L}iouville theory,''
  \href{http://dx.doi.org/10.1007/JHEP02(2010)057}{{\em JHEP} {\bfseries 02}
  (2010) 057},
\href{http://arxiv.org/abs/0909.1105}{{\ttfamily arXiv:0909.1105 [hep-th]}}.

\bibitem{Passerini:2010pr}
F.~Passerini, ``Gauge theory {W}ilson loops and conformal {T}oda field
  theory,'' \href{http://dx.doi.org/10.1007/JHEP03(2010)125}{{\em JHEP}
  {\bfseries 03} (2010) 125},
\href{http://arxiv.org/abs/1003.1151}{{\ttfamily arXiv:1003.1151 [hep-th]}}.

\bibitem{Gaiotto:2009we}
D.~Gaiotto, ``{${\cal N}=2$} dualities,''
  \href{http://dx.doi.org/10.1007/JHEP08(2012)034}{{\em JHEP} {\bfseries 08}
  (2012) 034},
\href{http://arxiv.org/abs/0904.2715}{{\ttfamily arXiv:0904.2715 [hep-th]}}.

\bibitem{Morrison:1994fr}
D.~R. Morrison and M.~R. Plesser, ``Summing the instantons: Quantum cohomology
  and mirror symmetry in toric varieties,''
  \href{http://dx.doi.org/10.1016/0550-3213(95)00061-V}{{\em Nucl. Phys.}
  {\bfseries B440} (1995) 279--354},
\href{http://arxiv.org/abs/hep-th/9412236}{{\ttfamily arXiv:hep-th/9412236
  [hep-th]}}.

\bibitem{Jockers:2012zr}
H.~Jockers, V.~Kumar, J.~M. Lapan, D.~R. Morrison, and M.~Romo, ``Nonabelian 2d
  gauge theories for determinantal {C}alabi--{Y}au varieties,''
  \href{http://dx.doi.org/10.1007/JHEP11(2012)166}{{\em JHEP} {\bfseries 11}
  (2012) 166},
\href{http://arxiv.org/abs/1205.3192}{{\ttfamily arXiv:1205.3192 [hep-th]}}.

\bibitem{Hosomichi:2010vh}
K.~Hosomichi, S.~Lee, and J.~Park, ``{AGT} on the {S}-duality wall,''
  \href{http://dx.doi.org/10.1007/JHEP12(2010)079}{{\em JHEP} {\bfseries 12}
  (2010) 079},
\href{http://arxiv.org/abs/1009.0340}{{\ttfamily arXiv:1009.0340 [hep-th]}}.

\bibitem{Dimofte:2011ju}
T.~Dimofte, D.~Gaiotto, and S.~Gukov, ``Gauge theories labelled by
  three-manifolds,'' \href{http://dx.doi.org/10.1007/s00220-013-1863-2}{{\em
  Commun. Math. Phys.} {\bfseries 325} (2014) 367--419},
\href{http://arxiv.org/abs/1108.4389}{{\ttfamily arXiv:1108.4389 [hep-th]}}.

\bibitem{Dimofte:2011py}
T.~Dimofte, D.~Gaiotto, and S.~Gukov, ``3-manifolds and 3d indices,''
  \href{http://dx.doi.org/10.4310/ATMP.2013.v17.n5.a3}{{\em Adv. Theor. Math.
  Phys.} {\bfseries 17} no.~5, (2013) 975--1076},
\href{http://arxiv.org/abs/1112.5179}{{\ttfamily arXiv:1112.5179 [hep-th]}}.

\bibitem{Yamazaki:2012cp}
M.~Yamazaki, ``Quivers, {YBE} and 3-manifolds,''
  \href{http://dx.doi.org/10.1007/JHEP05(2012)147}{{\em JHEP} {\bfseries 05}
  (2012) 147},
\href{http://arxiv.org/abs/1203.5784}{{\ttfamily arXiv:1203.5784 [hep-th]}}.

\bibitem{Wess:1992cp}
J.~Wess and J.~Bagger, {\em Supersymmetry and supergravity}.
\newblock
1992.
\newblock

\bibitem{Weinberg:1993sg}
E.~J. Weinberg, ``Monopole vector spherical harmonics,''
  \href{http://dx.doi.org/10.1103/PhysRevD.49.1086}{{\em Phys. Rev.} {\bfseries
  D49} (1994) 1086--1092},
\href{http://arxiv.org/abs/hep-th/9308054}{{\ttfamily arXiv:hep-th/9308054
  [hep-th]}}.

\end{thebibliography}\endgroup
\end{document}